\documentclass[onecolumn,prd,letterpaper,unsortedaddress,superscriptaddress,epsfig]{revtex4}
\usepackage{epsfig}
\usepackage{graphicx}
\usepackage{dcolumn}
\usepackage{amsmath}
\usepackage{amssymb}
\usepackage{pslatex}
\usepackage[dvips,hypertex,bookmarks]{hyperref}
\usepackage{html}
\usepackage{url}

\def\deg{$^\circ$~}

\newcommand{\beq}{\begin{equation}}
\newcommand{\eeq}{\end{equation}}

\begin{document}

\title{The Antarctic Impulsive Transient Antenna Ultra-high Energy Neutrino Detector\\
        Design, Performance, and Sensitivity for 2006-2007 Balloon Flight}

\author{
P.~W.~Gorham$^1$,
P.~Allison$^1$,
S.~W.~Barwick$^{2}$,
J.~J.~Beatty$^3$, 
D.~Z.~Besson$^4$,
W.~R.~Binns$^5$,
C.~Chen$^6$,
P.~Chen$^{6,13}$,
J.~M.~Clem$^7$,
A.~Connolly$^8$,
P.~F.~Dowkontt$^5$,
M.~A.~DuVernois$^{10}$, 
R.~C.~Field$^6$,
D.~Goldstein$^2$,
A.~Goodhue$^9$
C.~Hast$^6$,
C.~L.~Hebert$^1$,
S.~Hoover$^9$,
M.~H.~Israel$^5$,
J.~Kowalski,$^1$
J.~G.~Learned$^1$,
K.~M.~Liewer$^{11}$,
J.~T.~Link$^{1,13}$,
E.~Lusczek$^{10}$,
S.~Matsuno$^{1}$,
B.~C.~Mercurio$^3$,
C.~Miki$^{1}$,
P.~Mio\v{c}inovi\'c$^{1}$,
J.~Nam$^{2,12}$,
C.~J.~Naudet$^{11}$,
R.~J.~Nichol$^8$,
K.~Palladino$^3$,
K.~Reil$^6$,
A.~Romero-Wolf$^1$
M.~Rosen$^{1}$,
L.~Ruckman$^1$,
D.~Saltzberg$^9$,
D.~Seckel$^7$,
G.~S.~Varner$^{1}$,
D. Walz$^6$,
Y. Wang$^{12}$,
F.~Wu$^2$
}
\vspace{2mm}
\noindent
\affiliation{
Dept. of Physics and Astronomy, Univ. of Hawaii, Manoa, HI 96822.   
$^2$Univ. of California, Irvine CA 92697.  
$^3$Dept. of Physics, Ohio State Univ., Columbus, OH 43210. 
$^4$Dept. of Physics and Astronomy, Univ. of Kansas, Lawrence, KS 66045. 
$^5$Dept. of Physics, Washington Univ. in St. Louis, MO 63130. 
$^6$Stanford Linear Accelerator Center, Menlo Park, CA, 94025.
$^7$Dept. of Physics, University of Delaware, Newark, DE 19716. 
$^8$Dept. of Physics, University College London, London, United Kingdom.
$^9$Dept. of Physics and Astronomy, Univ. of California, Los Angeles, CA 90095.
$^{10}$School of Physics and Astronomy, Univ. of Minnesota, Minneapolis, MN 55455.
$^{11}$Jet Propulsion Laboratory, Pasadena, CA 91109.
$^{12}$Dept. of Physics, National Taiwan University, Taipei, Taiwan.
$^{13}$Currently at NASA Goddard Space Flight Center, Greenbelt, MD, 20771.
}
\collaboration{ANITA Collaboration}
\noaffiliation

\begin{abstract}
We present a detailed report on the experimental details of the
Antarctic Impulsive Transient Antenna (ANITA) long duration balloon payload,
including the design philosophy and realization, physics simulations, 
performance of the instrument during its first Antarctic flight completed
in January of 2007, and expectations for the limiting neutrino detection sensitivity. 
Neutrino physics results will be reported separately.
\end{abstract}

\maketitle

\section{Introduction}

Cosmic neutrinos of energy in the Exavolt and higher (1 EeV = $10^{18}$~eV) energy range,
though as yet undetected,
are expected to arise through a host of energetic acceleration and interaction
processes at source locations throughout the universe. However, in only one of these
sources--the distributed interactions of the ultra-high energy cosmic ray flux--
does the combination of observational evidence and interaction physics lead to
a strong requirement for resulting high energy neutrinos.
Whatever the sources of the highest energy cosmic rays, their observed presence in
the local universe, combined with the expectation that their sources occur
widely throughout the universe at all epochs, leads to the conclusion
that their interactions with the cosmic microwave background radiation (CMBR)--
the so-called GZK process (after Greisen, Zatsepin, and Kuzmin~\cite{GZK})--
must yield an associated cosmogenic neutrino flux, as first noted by 
Berezinsky and Zatsepin~\cite{BZ}. These neutrinos are often called the
GZK neutrinos, as they arise from the same interactions of the 
ultra-high energy cosmic rays (UHECR) that cause the
GZK cutoff, but they are perhaps more properly referred to as the BZ neutrinos.

In BZ neutrino production scenarios,
current experimental UHECR measurements invariably point to
the presence of an associated ultra-high energy neutrino flux. For UHECR above
several times $10^{19}$~eV, 
intergalactic space is optically thick to UHECR propagation through the CMBR
at a distance scale of several tens of Mpc. Each UHECR source
at all epochs is thus subject to local conversion of its hadronic flux to
secondary, lower energy particles over a distance scale of order 100~MPc in the current
epoch. Neutrinos are the only secondary particle that may freely propagate
to cosmic distances, and the 
resulting neutrino flux at earth is thus related to the integral over the highest-energy cosmic
ray history of the universe, to the earliest epoch at which they occur. 

Although local sources may also contribute to the EeV-ZeV neutrino flux at earth,
the bulk of the flux is generally believed to arise from a much wider spectral convolution,
and will thus be imprinted with the cosmological source distribution in addition to 
effects from local sources. This leads to strong motivations to detect the
BZ neutrino flux: first, it is required by standard model physics, and
thus its absence could signal new physics beyond the standard model. 
Second, it is the only way to
directly observe the UHECR source behavior over cosmic distance scales.
Finally, once established, the spectrum and absolute flux of such neutrinos
may afford a calibrated ``test beam'' for both particle physics
and astrophysics experimentation, providing center-of-momentum
energies on target nucleons of 100-1000 TeV, an energy scale not likely to
be reached by other methods in the near future.

The Antarctic Impulsive Transient Antenna was designed with the goal
of measuring the BZ neutrino flux directly, or limiting it at a 
level which would provide compelling and useful constraints on the early
UHECR source history. The BZ neutrino flux is potentially very low--of order
1 neutrino per square kilometer per week arriving over 2$\pi$ steradians is a
typical estimate. This flux presents an extreme challenge to
detection, since the low neutrino interaction cross section also means that any
target volume will have an inherently low efficiency
for converting any given neutrino. ANITA's methodology centers on 
observing the largest possible volume of the most transparent
possible target material: Antarctic ice, which has
been demonstrated to provide extremely low-loss transmission of radio
signals through its bulk over much of the continent. ANITA then exploits
the Askaryan effect~\cite{Ask62}, coherent, impulsive radio emission from the
charge asymmetry in the electromagnetic component of a high energy
particle cascade in a dielectric medium. 

ANITA searches
for cascades initiated by a primary neutrino interacting
in the Antarctic ice sheet within
its field of view from the Long-Duration Balloon (LDB) altitude of
35-37 km. The observed area of ice from these altitudes is of order
1.5~M km$^2$. Combining this with the electromagnetic 
field attenuation length of ice
which is of order 1~km at ANITA's observation frequency range, ANITA
is sensitive to a target volume of order 1-2~M km$^3$. The acceptance, however,
is constrained by the fact that at any location within the target, 
the allowed solid angle of arrival for a neutrino to be detectable
at the several-hundred-km average distance of the payload is a small
fraction of a steradian. Folding in these constraints, the
volumetric acceptance is still of order hundreds to thousands of km$^3$ steradians
over the range of energy overlap--$10^{18.5-20}$~eV--with the BZ neutrino spectrum. 
This large acceptance, while tempered by the limited exposure in time
provided by a balloon flight, still yields the largest sensitivity
of any experiment to date for BZ neutrinos. In this report we
document the ANITA instrument and our estimates of its sensitivity
and performance for the first flight of the payload, completed
in January of 2007. A separate report will detail the results on
the neutrino flux.

\section{Theoretical Basis for ANITA Methodology.}

The concept of detecting high energy particles through the coherent radio
emission from the cascade they produce can be traced back over 40 years to
Askaryan~\cite{Ask62}, who argued persuasively for the presence of
strong coherent radio emission from these cascades, and even suggested that
any large volume of
radio-transparent dielectric, such as an ice sheet, a geologic saltbed, or
the lunar regolith
could provide the target material for such interactions and radio emission.
In fact all of these approaches are now being pursued~\cite{RICE03,SalSA,GLUE04}.

Although significant early efforts were successful in detecting radio
emission from high energy particle cascades in the earth's
atmosphere~\cite{Jelley_65}, it is important to emphasize that
the cascade radio emission that ANITA detects is {\em
unrelated to the primary mechanism for air shower radio emission.} Particle cascades induced
by neutrinos in Antarctic ice are very compact, consisting of a
``plug'' of relativistic charged particles several cm in diameter and $\sim 1$~cm thick,
which develops at the speed of light over a distance of several meters from the
vertex of the neutrino interaction, before dissipating into residual
ionization in the ice. The resulting radio emission is coherent
Cherenkov radiation with a particularly clean and simple geometry, providing
high information content in the detected pulses.
In contrast, the radio emission from air showers is a complex
phenomenon entangled with geomagnetic and near field effects. 
Attempts to understand and exploit this form of air shower emission for
cosmic ray studies
have been hampered by this complexity since its discovery in the mid-1960's,
although this effort has seen a recent renaissance~\cite{EASradio}.

Surprisingly little work was done on Askaryan's
suggestions that solids such as ice could be
important media for detection until the mid-1980's, when Gusev and
Zheleznykh~\cite{Gusev},
and Markov \&
Zheleznykh~\cite{Markov86} revisited these ideas. 
More recently a host of investigators including 
Zheleznykh~\cite{Zhe88},
Dagkesamansky \& Zheleznykh~\cite{Dag89}, Frichter,
Ralston, \& McKay~\cite{FRM}, Zas, Halzen, \&
Stanev~\cite{ZHS92}, Alvarez-Mu\~niz \& Zas~\cite{Alv97}, and
Razzaque et. al ~\cite{Razz02} among others have taken
up these suggestions and confirmed the basic results through more
detailed analysis. Of equal importance, a set of experiments at
the Stanford Linear Accelerator center have now clearly confirmed the effect
and explored it in significant detail~\cite{Sal01,SalSA,Mio06,slac07}

\subsection{First Order Energy Threshold \& Sensitivity.}

To illustrate the methodology, we consider a specific example.
The coherent radio Cherenkov emission in an electromagnetic $e^+e^-$ cascade arises
from the $\sim 20\%$ electron excess in the shower, which is itself produced
primarily by Compton scattering and positron annihilation in flight. Considering
deep-inelastic scattering charged-current interactions of a high energy neutrino $\nu$ with
a nucleon $N$, given generically by $\nu + N \rightarrow \ell^{\pm} + X$, the 
charged lepton $\ell^{\pm}$ escapes while the hadronic debris $X$ leads to a hadronic cascade.
If the initial neutrino has energy
energy $E_{\nu}$, the resulting hadronic cascade energy will $E_c = yE_{\nu}$, where
$y$ is the Bjorken inelasticity, with a mean of $\langle y \rangle \simeq 0.22$
at very high energies, and a very broad distribution. The average number of electrons and positrons
$N_{e+e-}$ near total shower maximum is of order the cascade energy expressed in
GeV, or
\begin{equation}
N_{e+e-}~\simeq~ {E_c \over 1~{\rm GeV}}~.
\end{equation}

Consider a case with $E_{\nu}=10^{19}$~eV and a slightly positive fluctuation above the mean
giving $y=0.4$.
This leads to $E_c= 4 \times 10^{18}$ eV, giving $N_{e+e-}\sim 4 \times 10^{9}$. The radiating
charge excess is then of order $N_{ex} \simeq 0.2N_{e+e-}$.
Single-charged-particle Cherenkov radiation gives a total radiated energy,
for tracklength $L$ over a frequency band from $\nu_{min}$ to $\nu_{max}$,
of:
\begin{equation}
W_{tot}~=~ \left ({\pi h \over c}\alpha \right )  L
\left ( 1 - {1 \over n^2\beta^2} \right )
\left ( \nu_{max}^2 - \nu_{min}^2 \right )
\end{equation}
where $\alpha\simeq 1/137$ is the fine structure constant, $h$ and $c$ are
Planck's constant and the speed of light, and $n$ and $\beta$ are the medium
dielectric constant, and the particle velocity relative to $c$, respectively.
For a collection of $N$ charged particles radiating coherently (e.g., with mean
spacing small compared to the mean radiated wavelength), the total energy will be
of order $W_{tot} = N^2 w$. In solid dielectrics with density comparable to ice or
silica sand, the cascade particle bunch is compact, with transverse dimensions of
several cm, and longitudinal dimensions of order 1 cm. Thus coherence will
obtain up to several GHz or more.

For a $4 \times 10^{18}$ eV cascade, $N_{ex} \simeq 8 \times 10^{8}$, and
$L\simeq 6$~m in the vicinity of shower maximum in a medium of density $\sim
0.9$ with $n\sim 1.8$ as in Antarctic ice. Taking the mean radio frequency to
be 0.6~GHz with an effective bandwidth of 600 MHz, the net radiated energy is $W_{tot}=
10^{-7}$ J. This energy is emitted into a restricted solid angle defined by
the Cherenkov cone at an angle $\theta_c$ defined by $\cos\theta_c =
(n\beta)^{-1}$, and a width determined (primarily from diffraction
considerations) by $\Delta \theta_c \simeq c \sin \theta_c / (\bar{\nu}L)$.
The implied total solid angle of emittance is $\Omega_c \simeq 2\pi\Delta
\theta_c \sin \theta_c = 0.36$~sr.

The pulse is produced by coherent superposition of the amplitudes of
the Cherenkov radiation shock front, which yields extremely broadband
spectral power over the specified frequency range. The intrinsic pulse width
is less than 100~ps~\cite{Mio06}, and this pulse
thus excites a single temporal mode of the receiver, with
characteristic time $\Delta t = (\Delta \nu )^{-1}$, or about 1.6 ns in our
case here. Radio source intensity in radio astronomy is typically expressed
in terms of the flux density Jansky (Jy), where 1 Jy =
$10^{-26}$~W~m$^{-2}$~Hz$^{-1}$. The energy per unit solid angle derived
above, $W_{tot}/\Omega_c = 2.7 \times 10^{-7}$ J/sr in a 600 MHz bandwidth,
produces an instantaneous peak flux density of $S_c= 1.4 \times 10^7$ Jy at 
the mean geometric distance $D=480$~km of the ice in view, 
after accounting for the fact that at this 
distance the geometry constrains the Fresnel coefficient
for transmission through the ice surface to $\sim 0.12$, since the
radiation emerges from angles close to the total-internal-reflectance
(TIR) angle.

The sensitivity of a radio antenna is determined by its collecting aperture and
the thermal noise background, called the system temperature $T_{sys}$.
The RMS level of power fluctuations in this thermal noise, expressed
in W~m$^{-2}$~Hz$^{-1}$, is given by
\begin{equation}
\Delta S ~=~ { ~k~T_{sys} \over 
A_{eff} \sqrt{\Delta t \Delta \nu}}~
{\rm W ~m^{-2}~ Hz^{-1}}
\end{equation}
where $k$ is Boltzmann's constant and $A_{eff}$ is the effective area of the
antenna.
Note that in our case, because the pulse is band-limited, the term
$\sqrt{\Delta t \Delta \nu} \simeq 1$. For ANITA, a single on-board
antenna has a frequency-averaged effective area of 0.2~m$^2$. For observations
of ice the system temperature is dominated by the ice thermal emissivity 
with $T_{sys} \leq 320$ K, assuming $\sim 140$~K receiver noise temperature.
The implied RMS noise level is thus $\Delta S = 2 \times 10^6$~Jy,
giving a signal-to-noise ratio of 6.3 in this case. 
These simple arguments show
that the expected threshold for neutrino detection is of order $10^{19}$~eV
even to the edges of the observed area viewed by ANITA. In practice,
events may be detected at lower energies due to fluctuations or interactions
closer to the payload, but more detailed simulations of the energy-dependent
acceptance of ANITA do not depart greatly from this first-order example.

\section{Instrument Design}
\subsection{Overview of Technical Approach}
\label{Techoverview}

As indicated by its acronym, ANITA is conceptually an antenna or
antenna array optimized to detect impulsive RF events with a
characteristic signature established by careful modeling and
experimental measurements. The array of antennas should view most
of the entire Antarctic ice sheet beneath the balloon, 
out to the horizon, to retain sensitivity to most of the
potential ice volume available for neutrino event production. It should
have the ability to trigger with high efficiency on events of 
interest, and should have the lowest feasible intrinsic noise levels
in its receivers to maximize sensitivity. It should have broad radio spectral
coverage and dual-polarization capability to improve its ability to 
identify the signals and reject the backgrounds. It must have immunity
to transient or steady radio-frequency interference. It must have
enough spatial resolution of the source of measured pulses to
determine if they match expected signal sources, and to allow for
first-order geolocation and subsequent sky mapping if the event is found to be
consistent with a neutrino. It must make
as many distinct and statistically independent measurements as 
possible of each impulse that triggers the system covering all available
degrees of freedom (spatial, temporal, polarization, and spectral), 
because the number of potential neutrino events among 
these triggers may be close to
zero, and this potential rarity of events demands that the information
content of each measured event be maximized.

These guiding principles have led to a technical approach that
centers around dual-polarization, broadband antenna clusters with 
overlapping fields-of-view, combined with 
a trigger system based on a heritage of RF impulse detection
instruments, both space-based (the FORTE satellite~\cite{Jacobson_99,FORTE04}) 
and ground-based (the GLUE and RICE experiments~\cite{GLUE04,RICE03}). 
The need for direction determination, combined with the constraints on usable
radio frequency range dictated by ice parameters, 
leads to an overall geometry for both
individual antennas and the entire array, that is governed by the
requirements for radio pulse interferometry over the spectral band
of interest.

The key challenges for ANITA are in the area of background rejection
and management of electromagnetic interference (EMI). 
Impulsive interference events are likely to be primarily from
anthropogenic sources, and in most cases do not
mimic real cascade Cherenkov radio impulses because they lack many of the 
required properties such as polarization-, spectral-, and phase-coherence. 
A subset of impulsive anthropogenic interference, primarily
from systems where spark gaps or rapid solid-state switching relays are
employed, can produce events which are difficult to distinguish from
events of interest to ANITA, and thus the task of pinpointing the origin
of any impulsive event is of high importance to the final selection of
neutrino candidates. If, after rejection of all impulsive events associated
with any known current or prior human activity, there remains a
class of events which are distributed across the integrated field
of view of the payload and in time in a way that is inconsistent with
human origin, we may then begin to consider this event class
as containing neutrino candidates. Whether they survive with that
designation will ultimately depend on our ability to exclude all other
known possibilities.

In a previous balloon experiment~\cite{Anitalite}, we
found Antarctica to be relatively radio quiet at balloon altitudes once the 
payload leaves the vicinity of the largest bases.
What continuous wave (CW) interference is present can be
managed with careful trigger configuration and threshold adjustment
to servo-adjust for the temporary increases in narrow-band power that occasionally
are seen. With regard to impulsive interference, we found triggers due to
it to be relatively infrequent away from the main bases though even the
smaller bases did occasionally produce bursts of triggers. 

A less well-understood background may arise from ultra-high
energy air showers which can produce a tail of radio emission out to
ANITA frequencies, but these events, though they may produce triggers,
are eliminated on the basis of their direction, arising from above
the horizon, and their loss of coherence at VHF and UHF frequencies.
However, in all cases above, ANITA may be presented with
unexpected challenges.

\subsection{Background Interference Issues.}
\label{EMIdisc}
Because ANITA operates with extremely high radio bandwidth over
frequencies that are not reserved for scientific use, the problem of radio
backgrounds, both anthropogenic and natural, is crucial to the development
of a robust mission design. We have
noted previously that the thermal noise power 
\footnote{$P_N= k_B T_{sys} \Delta f$, for $k_B=1.38 \times 10^{-23}$J/K, $T_{sys}=$ the
system noise temperature in Kelvin, and $\Delta f=$ the bandwidth in Hz.}
provides the ultimate background limitation, for both impulsive and 
time-averaged measurements, 
in much the same way that photon noise provides one of
the ultimate limits to optical imaging systems. 

Electromagnetic interference may take different forms: near-sinusoidal
"Carrier Wave" (CW) interference can have very high narrow-band power
and saturate the system, or it can appear at a low level, sometimes
as a composite of contributions from many bands, and effectively
act to raise the aggregate system noise. Impulsive EMI often arises from
electronic switching phenomena, and may trigger the system even if it cannot be
mistaken for signals of interest, since the trigger should be as
inclusive as possible. ANITA has only one chance per true neutrino event to detect
and characterize the radio wavefront as it passes by the payload; thus
it must be as efficient as possible at triggering on anything
similar to the events of interest.
In the end, it is the information content of a given triggered 
measurement that will determine the confidence with which we can
ascribe it to a neutrino origin. This conclusion is the primary
mission design driver for the type of payload and the number of antennas.

The design of the mission, payload,
ballooncraft, and all ancillary instrumentation must 
therefore be evaluated in
the light of whether it produces EMI, mitigates it, responds appropriately
to it, or facilitates rejection of it. In the end, when all background
interference has been rejected, what is left becomes the substance for
ANITA science.

\subsubsection{Anthropogenic Backgrounds.}
\label{AnthropoEMI}
Backgrounds from man-made sources do not in general pose a risk of being
mistaken for the signals of scientific interest, unless they arise from
locations where no human activity is previously known. As we will show
later in this report, ANITA's angular reconstruction ability for
terrestrial interference events gives accuracies of order 1 degree
or better, enabling ground location of event sources to a level
more than adequate to remove events that originate from known camps or
anthropogenic sources. Human activity in Antarctica
is highly controlled and positions and locations for all such activity are
logged with high reliability during a season.
However, man-made sources can still pose a significant 
risk of interfering with the operation of the instrument.
Interference from man-made terrestrial or orbital sources is
a ubiquitous problem in all of radio astronomy. In this
respect ANITA faces a variety of potential
interfering signals with various possible impacts on the
data acquisition and analysis.

\paragraph{Satellite signals.}
Orbiting satellite transmitter power is generally
low in the bands of interest. For example, the
GPS constellation satellites at an altitude of 21000 km,
have transmit powers of order 50W in the 1227 MHz and
1575 MHz bands, with antenna gains of 11-13 dBi. The
implied power at the earth's surface is -127 dB W m$^{-2}$
maximum in the 1227 MHz band. The implied RMS noise voltage
for ANITA, given the antenna's effective area at this
frequency, is of order 0.7 $\mu$V, far below the RMS thermal noise
voltage ($\sim 10-15\mu$V RMS) referenced to the receiver inputs. 
Current satellite systems do not
typically operate in ANITA's band, however there are some legacy
systems that can produce detectable power within ANITA's band.
As we will discuss in a later section, ANITA has encountered 
some satellite interference in the 200-300 MHz range, but it
has not caused significant performance degradation to date.

Satellites do not in general intentionally produce nanosecond-scale
impulsive signals; however, such signals may be produced by
solid-state relay or actuator activity on a satellite that
is changing its configuration. Such signals would appear to come
from above the horizon, but might also show up in reflection off
the ice surface. In this latter case, the Fresnel coefficient
for such a reflection will in general signficantly boost the horizontal
polarization of such a reflection, and this characteristic 
provides a strong discriminator, if the initial above-the-horizon
impulse was for some reason not detected.

\paragraph{Terrestrial signals.}
The primary risk for terrestrial signals is not that they
trigger the system. Terrestrial sources often do produce
significant impulsive interference, and will trigger our system at significant rates
anytime the payload is within view of such anthropogenic sources. 
However, such triggers are easily selected against in post-analysis since
their directions can be precisely associated with known sources in
Antarctica. The greater issue for ANITA occurs if there
is a strong transmitter in the field of view which saturates
the LNA, causing its gain to decrease so that the sensitivity in
that antenna is lost. The present LNA design tolerates
up to about 1 dBm output before saturation, with an input
stage gain of 36 dB. Thus a signal of $0.25\mu$W coupled
into the antenna would pose a risk of saturation and
temporary loss of sensitivity. 

Since the antenna effective area is of order 0.6 m$^2$ at
the low end of the band, ANITA therefore tolerates
up to a 0.2 MW in-band transmitter at or near the horizon, or
a several kW in-band transmitter near the nadir, accounting for the
off-axis response of the ANITA antennas. Most of the higher
power radar and other transmitters in use in Antarctica
are primarily at the South Pole and McMurdo stations. 
Such systems did reduce
our sensitivity when the payload was in close proximity to
McMurdo station, and to a lesser degree, when in view of the
South Pole station.

\subsubsection{Other possible backgrounds.}

\paragraph{Lightning.}
Lightning is known to produce intense bursts of electromagnetic energy, but
these have a spectrum that falls steeply with frequency, with very little
power extending into the UHF and microwave regimes. 
Although lightning does occur over the Southern
Ocean~\cite{Jacobson_01,Jacobson_00}, it is unknown
on the Antarctic continent. We do not expect lightning to comprise
a significant background to ANITA.

\paragraph{Cosmic Ray Air Shower backgrounds.}
Cosmic ray extensive air showers (EAS) at EeV energies also produce an
electromagnetic pulse, known from observations since the late 1960's.
The dominant RF emission comes from synchrotron radiation in the
geomagnetic field. This emission is coherent below about 100~MHz,
transitioning to partial coherence above about 200~MHz in the ANITA band.
Although there has been a recent increase in activity to measure
the radio characteristics of EAS events 
in the coherent regime below 100~MHz~\cite{Huege_Falcke05}, 
there is still little reliable information regarding the partially coherent
regime where ANITA is sensitive to such events, although in fact several
of the early detections of such events were at 500~MHz~\cite{Fegan}.
The radio emission from EAS is highly beamed, so the acceptance for
such events is naturally suppressed by geometry. They are also expected
to have a steeply falling radio spectral signature, and thus an inverted
spectrum compared to events originating from the Askaryan process, which
has an intrinsic rising spectrum over the frequency region that coherence obtains, and
a slow plateau and decline above those frequencies.

ANITA may detect such events either by direct signals or reflected signals
off the ice surface, in a manner similar to that mentioned above for 
posible impulses from satellites. The EAS signals are known to be linearly polarized,
with the plane of polarization determined by the local geomagnetic field
direction. Since the field is largely vertical in the polar regions, there
is a tendency for the EAS radio emission to be horizontally polarized
for air showers with large zenith angles. ANITA's field-of-view,
which has maximum sensitivity near the horizon, thus favors EAS events
with these large zenith angles. Such events when observed directly
arrive from angles above the horizon, but under the right circumstances 
they may also be seen in reflection,
thus appearing to originate from below the horizon. They might thus be
confused with neutrino-like events originating from under the ice,
if their radio-spectral and polarization signature was not considered.
In an appendix we will address this possible physics background and
show why it is straightforward to separate it from the events of
interest.

\subsection{CSBF Support Instrumentation Package.}

\begin{figure}[htb!]
\begin{center}
\epsfig{file=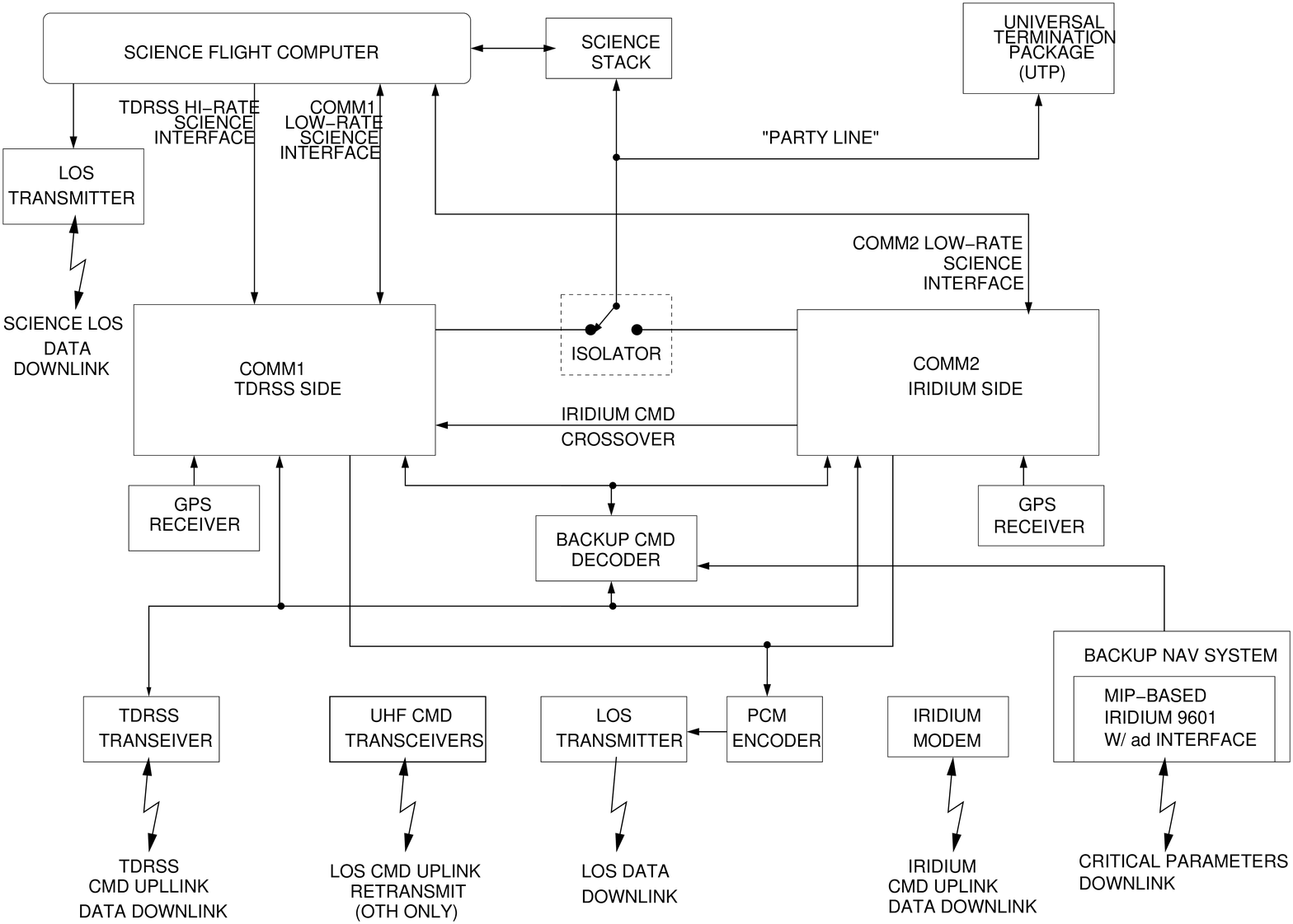,width=3.95in}
\caption{\it Block diagram of the NSBF SIP. \label{SIPfig}}
\end{center}
\end{figure}	

Support for NASA long-duration balloon payload launches and in-flight
services is provided through the staff of the Columbia Scientific Balloon
Facility (CSBF), based in Palestine, Texas, USA. 
CSBF has developed a ballooncraft Support Instrument Package (SIP),
an integrated suite of computers, sensors, actuators, relays, transmitters, and
antennas, for use with all LDB science instruments.  The CSBF SIP
is controlled by a pair of independent flight computers which handle science
telemetry, balloon operations, navigation, ballast control and
the final termination and descent of the payload.  
A system diagram of the SIP is provided in Figure~\ref{SIPfig}.  
A {\em Science Stack}, a configurable set of block modules, 
is also available as an option to the SIP providing such functions as a
simple science flight computer, analog-to-digital conversion, and open-collector command outputs
for additional instrument command and control.

The SIP also
provides the telemetry link between the ANITA flight computer and data
acquisition system and ground based operations.  Data from the ANITA
computer is sent over serial lines to the SIP package which handles
routing and transmission over line-of-sight (LOS),
Tracking and Data Relay Satellite System 
(TDRSS), and IRIDIUM communication
pathways.  ANITA utilizes the NSBF SIP Science Stack to provide
the ability to command the flight CPU system off and on and reboot the
computer during flight.

With regard to computational resources of the SIP, these are designed to
fulfill existing LDB requirements, including preserving a full archive of
all telemetered data that is passed through the SIP from the science
instrument. This function thus provides an additional redundant 
copy of the telemetered data that can be used if there is telemetry
loss or corruption.

One important characteristic of the SIP relevant to ANITA is that it is
not highly shielded from producing local EMI, at least at the extremely
low level required for compatibility with ANITA science goals. Of necessity
the SIP was thus enclosed in an external Faraday housing, 
with connectors, and penetrators designed in a manner similar to what was done for the ANITA primary
electronics instrumentation.

\subsection{Gondola Structure.}
	\label{gondola}

\begin{figure*}[!htb]
\centerline{
\includegraphics[width=3.95in]{Payl_LV.eps}~~\includegraphics[width=2.5in]{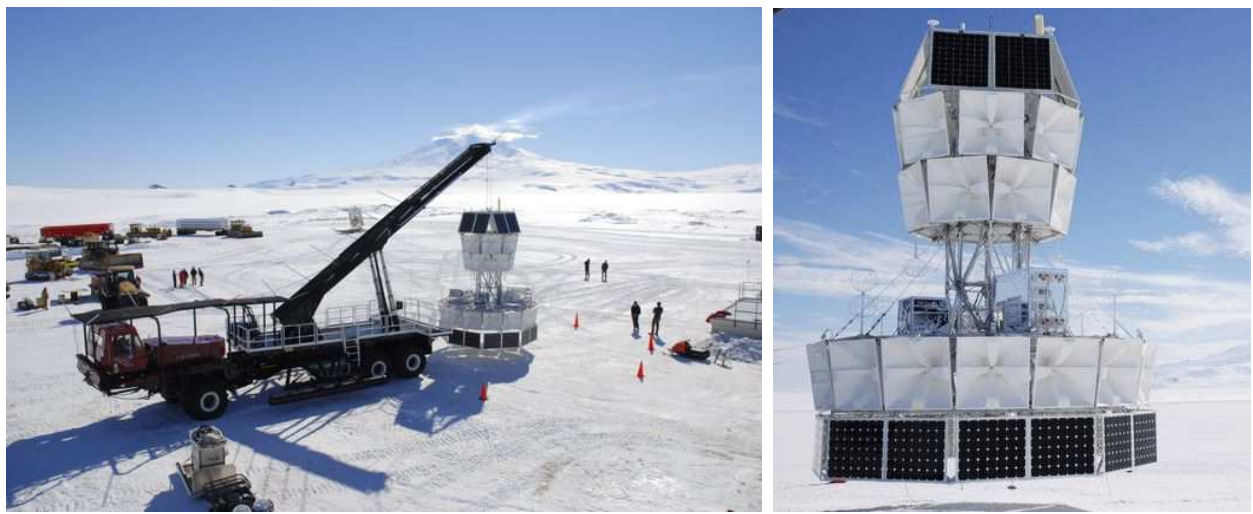}}
\begin{small}
\caption{ ANITA payload in flight-ready configuration with launch vehicle.\label{payload06a}}
\end{small}
\end{figure*}

The gondola structure consists primarily of an aircraft-grade aluminum alloy
frame.  A matrix of tubular components  is pinned together via a
combination of socket joints, tongue and clevis joints, and quarter-turn
cam-lock fasteners.  Views just prior to the launch of the payload show the
structural elements of the gondola in Fig.~\ref{payload06a}.
The frame is based on octagonal symmetry where eight vertical members,
plus cross bracing, provide an internal backbone that allows for the
attachment of spoke-like trusses to which the horn antennas fasten.  
Three ring-shaped
clusters of quad-ridged horn antennas constitute the primary ANITA sensors.  The top two
antenna clusters have eight antennas each.
Positioned around the perimeter of the base is a sixteen horn cluster.
All of the eight horn clusters have a 45\deg azimuthal offset angle between
adjacent antennas, with a 22.5\deg azimuthal offset between the top two
rings.  The antennas in the sixteen horn ring are offset from each other by 22.5\deg.
All of the antennas in the upper two eight horn rings and the ones in the
sixteen horn ring are canted down 10\deg below horizontal to optimize their
sensitivity, based on Monte Carlo studies of the effects of the tapering
of the antenna beam when convolved with the neutrino arrival directions and
energy spectrum.

The nearly circular plane that is established by the sixteen antenna ring,
near the base of the gondola, provides a large deck area for most of the other
payload components.  This region is covered by lightweight
panels made of dacron sailcloth on the topside and a reflective layer on the underside
to maintain thermal balance.  The ANITA electronics housing,
the NASA/CSBF SIP, and the battery packs are mounted
on the structural ribs of the deck. Most external metallic structure is painted
white to avoid overheating in the continuous sunlight, and critical components
such as the instrument housings and receivers are covered with silver-backed teflon-coated
tape to provide high reflective rejection of  
solar radiation and high emissivity for internal heat dissipation.

\subsection{Power subsystem.}
	
The ANITA power system is composed of a photovoltaic (PV) array, a charge controller, 
batteries, relays, and DC-to-DC converters. The PV array is an omni-directional array 
consisting of eight panels configured in an octagon, with the panels hanging vertically 
(see instrument figure). Although PV panels flown on high-altitude balloons are typically 
oriented at $\sim 23$\deg to the horizontal, in Antarctica, the large solar albedo from the ice 
results in more irradiance incident on the panels (for most conditions) if they hang 
vertically. Each panel consists of 84 solar cells  electrically connected in 
series. They were mounted on frames made of aircraft-grade spruce wood with a coarse 
webbing (Shearweave style1000-P02) stretched on the frames. 

The PV arrays were designed and fabricated by SunCat Solar. The solar 
cells used were Sunpower A-300 cells with a rated efficiency of 21.5\% and dimensions 
12.5 cm square and thickness 260 um. Bypass diodes were placed in parallel with 
successive groups of 12 cells within a panel (7-diodes/panel) to mitigate the effect of a 
possible single cell open circuit failure during flight. Additionally, a blocking diode was 
placed between each panel output and the charge controller to prevent cross-charging of 
panels with different output voltages resulting from different illuminations and 
temperatures. To reduce Fresnel reflection losses for high-refractive index silicon (n=3.46 
at 700 nm), the silicon cells had two anti-reflective (AR) coatings applied. An AR coating 
with refractive index n=1.92 was applied by the solar cell manufacturers. Additionally, 
during fabrication of the panels by SunCat Solar, a second AR coating with refractive 
index 1.47 was applied. This results in calculated Fresnel losses of 13-14\% for incidence 
angles from 0 to 40$^{\circ}$. 

The maximum power point (MPP) voltage and current generated 
by these cells under standard conditions (STC) are 0.560V and 5.54A respectively. 
However, the actual V and I vary considerably depending upon the irradiance and cell 
temperature. The single-cell temperature coefficient for the voltage is -1.9 mV/C. PV 
panel temperatures varied over the range of -10C to +95C, depending upon the irradiance 
incident upon the cells. The temperatures were measured by semiconductor temp sensors 
(AD590) glued to the back of cells.  PV array circuit components (diodes) also introduce 
losses in the output voltage and power. The actual measured PV voltage input to the 
charge controller during the flight ranged from 42.5 to 47 V (in good agreement with 
estimates using the cell temperature and temperature coefficient) and the current was 
about 9~A giving a total power of 400 W. 

The omni-directional array is inherently an unbalanced system; i.e. the irradiance incident on each 
panel differs. For a given orientation of the gondola, some panels are directly irradiated 
by sunlight plus solar albedo from the ice and others are irradiated only indirectly from 
solar albedo. Additionally, for those that are directly irradiated, the solar incidence angle 
is different. This results in individual panels that generate very different currents at any 
given time. Because of the differing temperatures, the individual panels also have 
significantly different output voltages (the voltage differences are small compared to the 
current differences) that feed into the charge controller. As mentioned above, the 
blocking diodes prevent cross-charging of panels generating different voltages.

When using an unbalanced array, to achieve the maximum power output, it is important to 
use a charge controller that senses and operates at the actual MPP as opposed to one that 
operates at a constant offset voltage from the array open-circuit voltage. We used an 
Outback MX-60 charge controller to supply power to the ANITA instrument. Conductive 
heat sinks were installed on the power FETs and transistors and the heat was conducted 
to the instrument radiator plate. We operated in the 24 
V mode and flew nine pairs of 12 V Panasonic LC-X1220P (20 AH) lead acid batteries 
that were charged by the charge controller and would have provided 12 hours of power 
in case of PV array failure.

The Instrument Power box consisted of the MX-60 charge controller, solid-state power 
relays, and Vicor DC/DC converters for the external 
radio-frequency conditional module (RFCM) amplifiers. The main power 
relays for the cPCI crate were controlled by discrete commands from the SIP. All other 
solid-state relays were controlled by the CPU, either under software control or by 
commands from the ground. The DC/DC box consisted of 
Vicor DC/DC converters which provided the +5, +12, -12, 
+3.3, +1.5, and 5 voltages required by the cPCI crate and peripherals. All voltages and 
currents were read by the housekeeping system.

\begin{figure}[!htb]
\centering
\centerline{\includegraphics[width=3.75in]{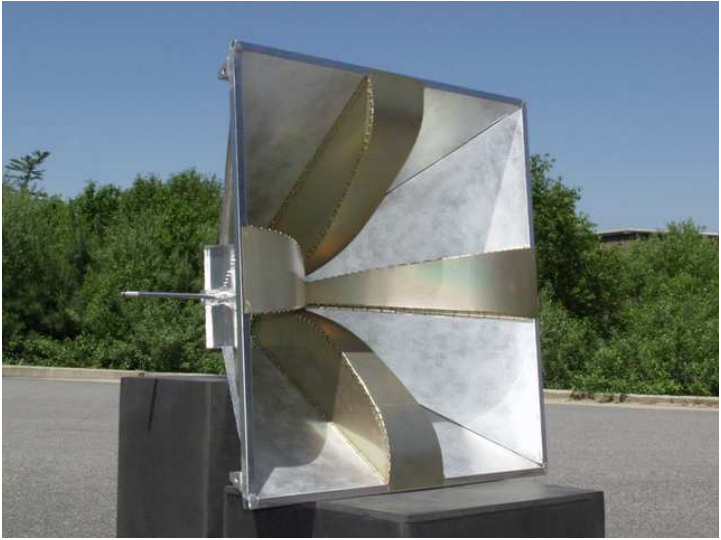}~~\includegraphics[width=3.0in]{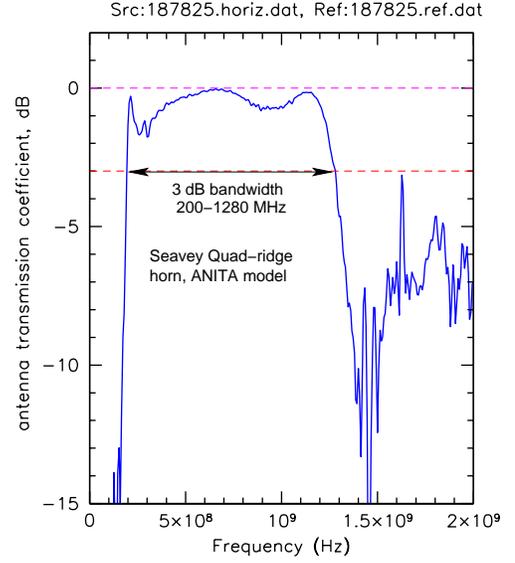}}
\begin{small}
\caption{ Left: A photograph of an ANITA quad-ridged dual-polarization horn. 
Right: Typical transmission coefficient for signals into the quad-ridged horn as a function
of radio frequency. \label{qrhorn}}
\end{small}
\end{figure}

\subsection{Radio Frequency subsystem}
\subsubsection{Antennas.}

Figure~\ref{payload06a} shows the ANITA payload configuration just
prior to launch in late 2006 at Williams Field, Antarctica.
The individual horns are a custom design produced for ANITA by
Seavey Engineering, Inc., now a subsidiary of Antenna Research Associates, Inc.
These horns are the primary ANITA antennas, and
may be thought of as a flared quad-ridged waveguide section;
the back of the horn does in fact terminate in a short section of waveguide.
The dimension of the mouth is of order 0.8~m across, and the horns can
be close-packed with minimal disturbance of the beam response since
the fringing fields outside the mouth of the horn are small.
Figure~\ref{qrhorn} shows an individual antenna prior to painting, and a corresponding
typical transmission curve indicating the efficiency for coupling power into the
antenna, as a function of radio frequency.

\begin{figure}[!htb]
\centering
\centerline{\includegraphics[width=3.5in]{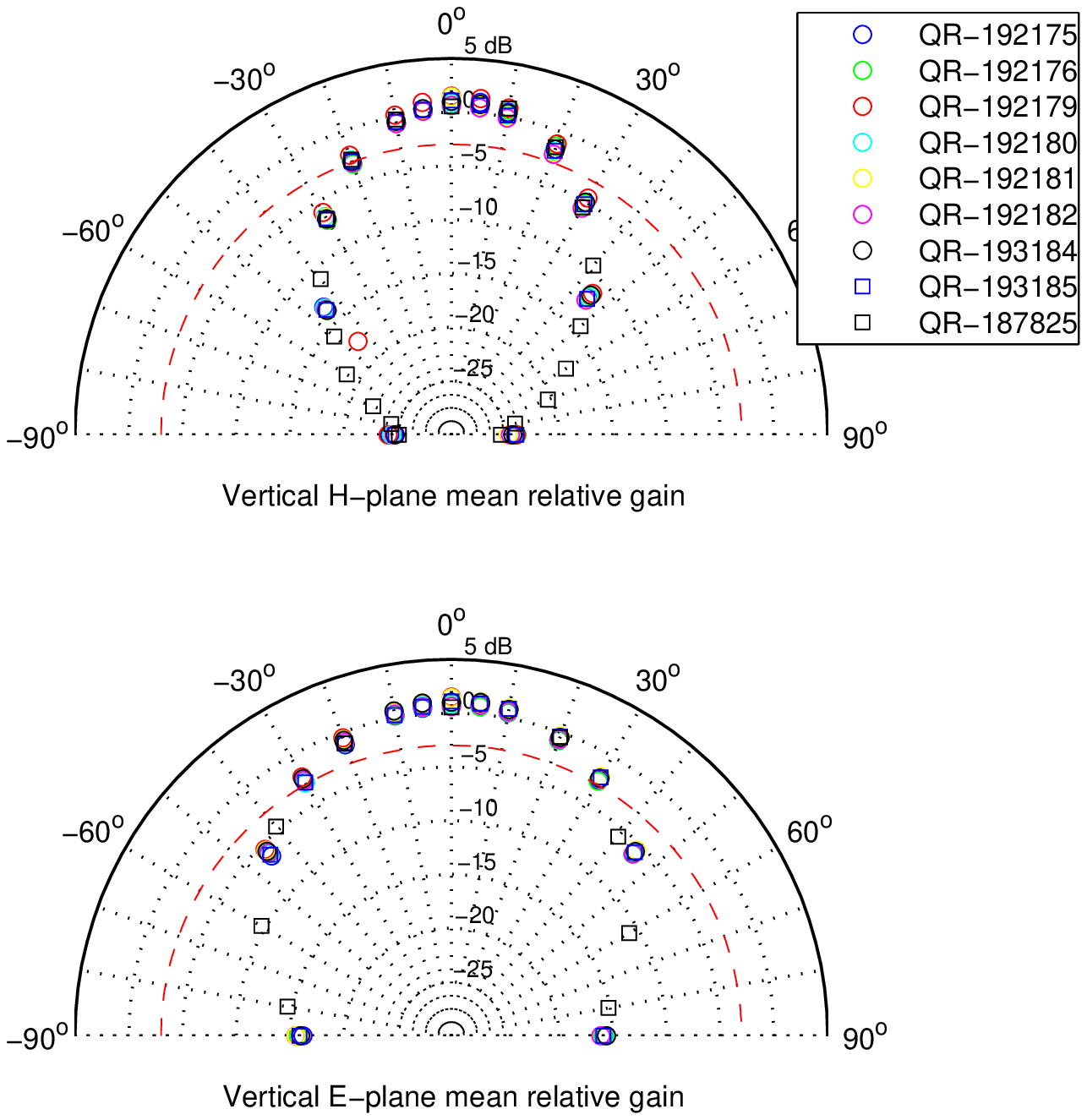}~~\includegraphics[width=3.5in]{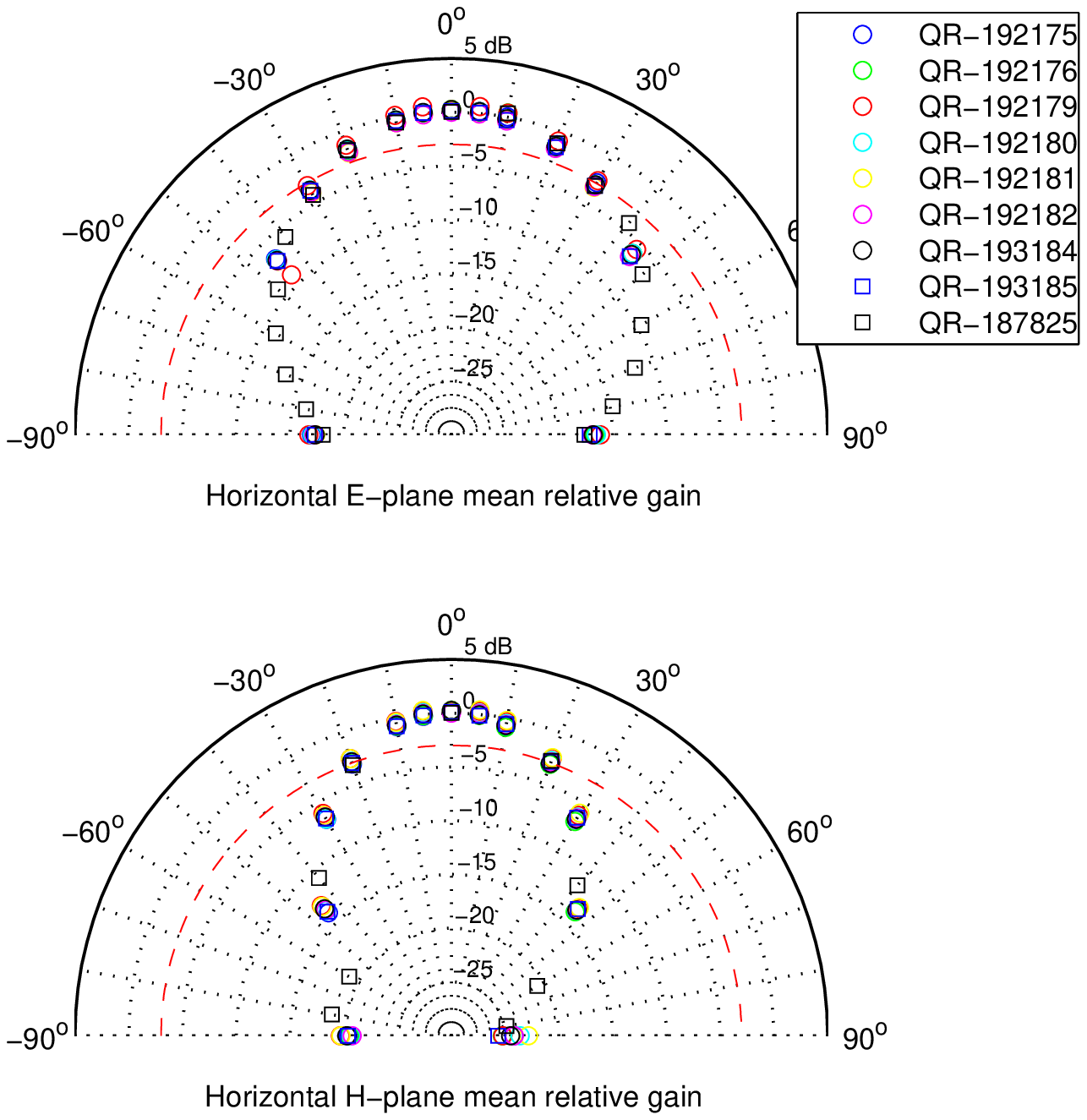}}
\begin{small}
\caption{ Left: Antenna vertical-polarization relative directivity in dB relative to the peak gain
for both E and H-planes. Right: the same quantities for the horizontal polarization. Nine different
antennas are shown. Gain is frequency-averaged for a flat-spectrum impulse across the band
for different angles in \label{VHgain}}
\end{small}
\end{figure}

The average full-width-at-half-maximum (FWHM) beamwidth 
of the antennas is about $45^{\circ}$ with a corresponding directivity
gain (the ratio of $4\pi$ to the main beam solid angle) of
approximately 10 dBi average across the band. Fig.~\ref{VHgain}
illustrates this for nine different ANITA antennas, showing the frequency-averaged
response relative to peak response along the principal
antenna planes (E-plane and H-plane for both polarizations) as a function of angle.
The choice of beam pattern for these antennas also determined the 
$22.5^{\circ}$ angular offsets in azimuth, as this was chosen to provide
good overlap between the response of adjacent antennas, but still maintaining
reasonable directivity for determination of source locations.

By arranging an
azimuthally symmetric array of 2 cluster groups of 8+8 (upper) 
and 16 (lower) antennas, each
with a downward cant of about $10^{\circ}$, we achieve complete
coverage of the horizon down to within $40^{\circ}$ of the nadir,
virtually all of the observable ice area. The antenna beams in this
configuration overlap within their 3 dB points, giving redundant
coverage in the horizontal plane. The $\sim 3$~m separation between
the upper and lower clusters of 16 antennas provides
a vertical baseline for establishing pulse direction in elevation angle.
Because the pulse from a cascade is known to be highly linearly-polarized,
we convert the two linear polarizations of the antenna into dual circular
polarizations using standard $90^{\circ}$ hybrid phase-shifting combiners. 
This is done for two reasons: first, a linearly polarized pulse 
will produce equal amplitudes in
both circular polarizations, and 
thus some background rejection is gained  by
accepting only linearly-polarized signals; and second, 
the use of circular polarizations removes any bias in the
trigger toward horizontal or vertically polarized impulses.

\begin{figure}[htb!]
\centering
\centerline{\includegraphics[width=3.55in]{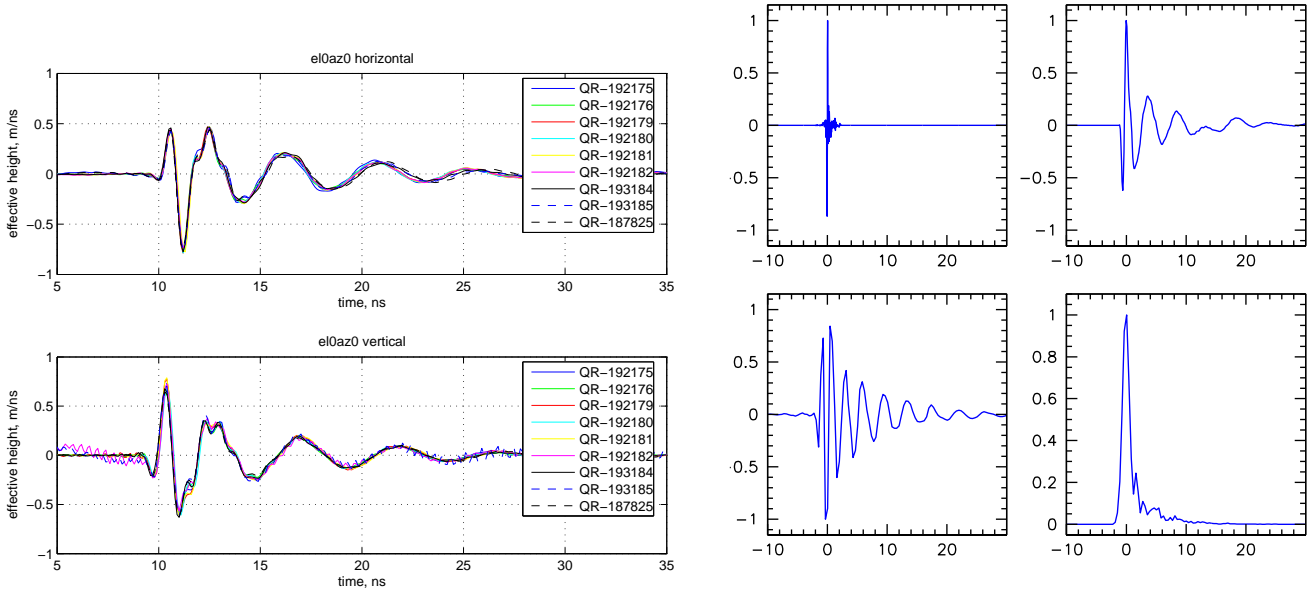}~~\includegraphics[width=3.25in]{Impulse07.eps}}
\vspace{3mm}
\caption{ Left: Antenna impulse response as measured for nine ANITA antennas, here in units that also give the instantaneous
effective height~\cite{Mio06}. Right: ANITA impulse response as it appears various stages of the signal chain.\label{impulse}}
\end{figure}

Because ANITA's sensitivity to neutrino events depends crucially
on its ability to trigger on impulses that rise above the
intrinsically impulsive thermal noise floor,
ANITA's antennas and receiving system must preserve the narrow
impulsive nature of any signal that arrives at the antenna. 
Fig.~\ref{impulse} shows the measured behavior of
the system impulse response at various stages. On the left, we
show details of the measured impulse response of nine of the flight antennas,
in units that give the instantaneous effective height $h_{eff}(t)$. The actual voltage time
response $\mathcal{V}(t)$ at the antenna terminals, assuming they are attached to a matched load,
is then just the convolution of this function with the incident field $\mathcal{E}(t)$:
$$\mathcal{V}(t) ~=~ \frac{1}{2} \mathcal{E}(t) \otimes h_{eff}(t)$$
where the convolution operator is indicated by the symbol $\otimes$. This
equation can also be expressed as an equivalent frequency domain form,
though in that case the quantities are in general complex.

On the right, we show the evolution of an Askaryan impulse through the
ANITA system. The initial
Askaryan impulse (a) is completely unresolved by the ANITA system,
since its intrinsic width is of order 100~ps (reference~\cite{Mio06}
provides the actual measured data that form the basis for this). The horn antenna
response (b) includes group delay at the edges of the frequency band
which leads to the low-frequency tail, and such group delay variation
is more pronounced after the system bandpass filters are applied (c).
However, the voltage envelope is slightly misleading, as the total
power (or intensity) response (d) of the system is still
confined to 1.5~ns FWHM.

\begin{figure*}[htb!]
\centering
\includegraphics[width=6.5in]{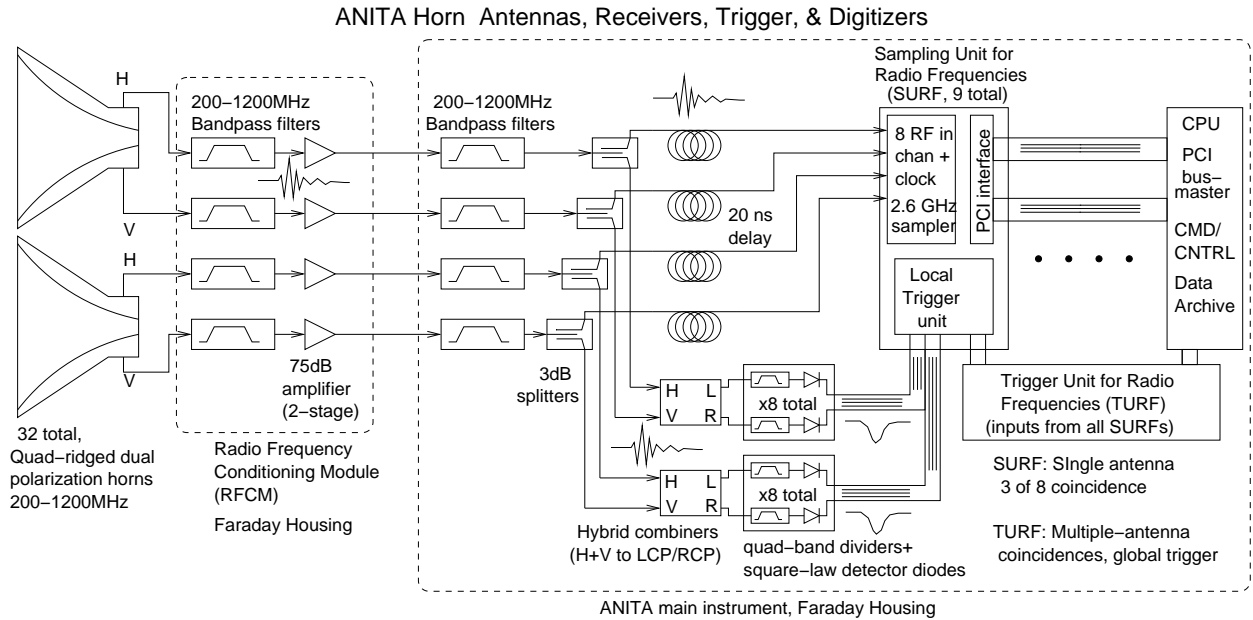}
\begin{small}
\caption{Block diagram of the primary RF subsystems for ANITA.}
\label{Block07}
\end{small}
\end{figure*}

	\subsubsection{Receivers.}
	
\begin{figure}[htb!]
\centering
\centerline{\includegraphics[width=3.85in]{rfcm_photo.eps}~~\includegraphics[width=3.05in]{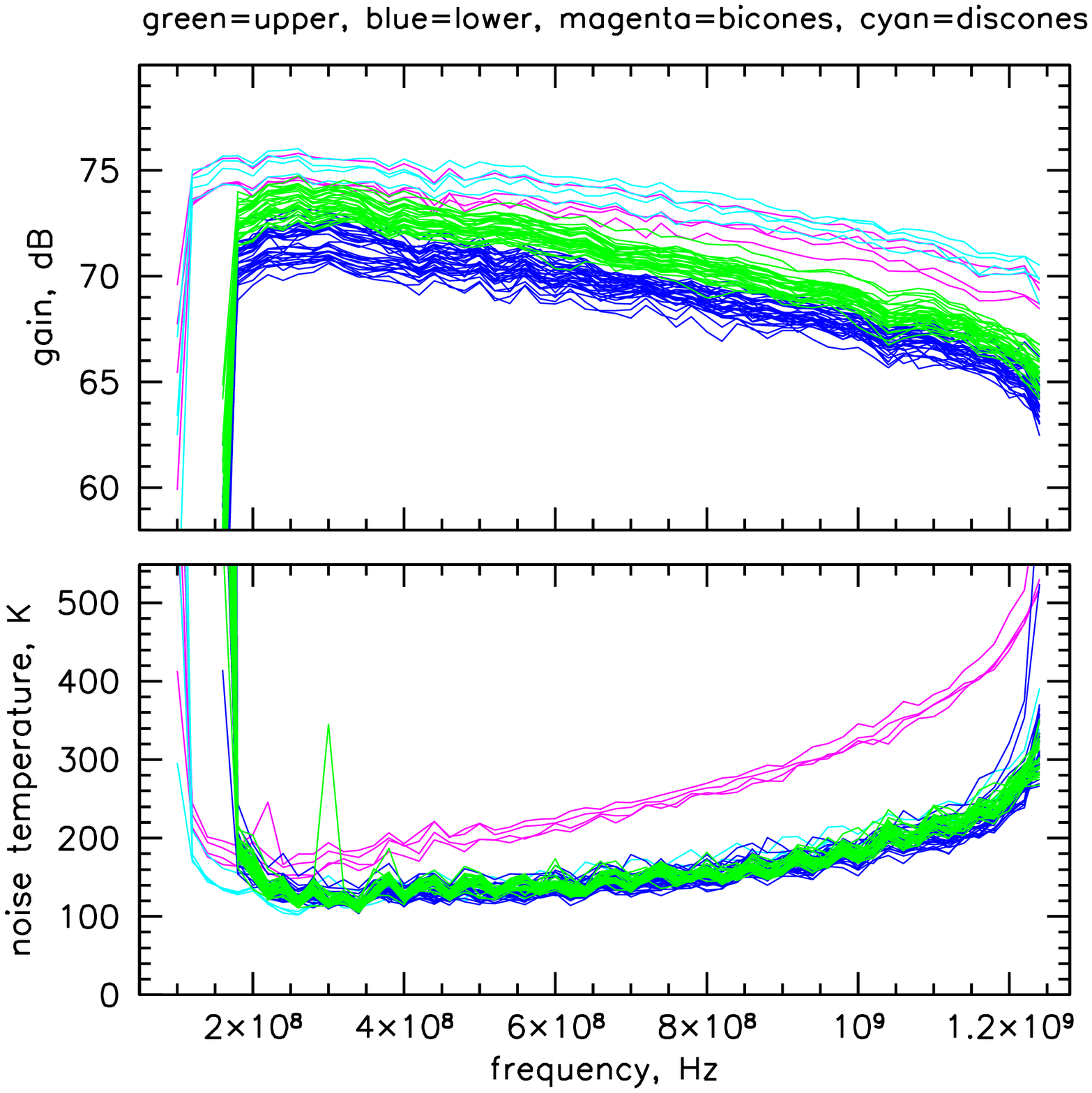}}
\begin{small}
\caption{ Left: Photo Layout of ANITA receiver module. 
Right: Gain and noise measurements for all ANITA receivers}
receivers.
\label{RFCM}
\end{small}
\end{figure}

The RF front end for ANITA consists of a bandpass filter, followed by
a low-noise-amplifier (LNA)/power limiter combination, then followed by a 
2nd stage booster amplifier. An example of one of the receivers is shown in
Fig.~\ref{RFCM}(left), with its enclosure cover removed to show the internal components. 
These elements are all in close proximity to the horn antennas to ensure no
transfer losses through cables, and are enclosed in a Faraday box for additional
EMI immunity. Once the signals are boosted by the 2nd stage
amplifier, they are transmitted via coaxial cable to the receiving section of 
the trigger and digitizer, which is contained in a large shielded Faraday 
enclosure. Once the signals arrive at this location, a second bandpass filter is 
applied to remove the out-of-band noise from the LNA, and the signals are then 
ready for insertion into the trigger/digitizer system. 

Figure~\ref{RFCM}(right) shows a composite overlay of measurements of
the total gains and 
noise temperature for all ANITA channels, including cable
attenuation losses up to the inputs of the digitizers. 
The gain slope is dominated by the intrinsic amplifier response. The noise 
temperature arises from the LNAs, with about 90~K intrinsic
amplifier noise, combined with the front-end bandpass filter, which creates
most of the additional $\sim 50$~K due to emissivity along
the signal path. The combined average noise temperature of
$\langle T_{sys}\rangle \simeq 140$~K was 
found to contribute at most about 40-45\% of the total noise
while at float, as we describe in a later section. Several of the receivers
also had a coupler section added to them to allow for insertion of a calibration
pulse into their respective antennas; these have a higher noise figure, but these
channels were not used for the primary signal triggering.

\subsection{Trigger and Digitizer Subsystem}
\subsubsection{Trigger System.}

In a long-duration balloon flight, primary power is solar
which places tight restrictions on the payload power budget.  Just as
severe is the need to eliminate the heat generated by the
payload electronics, which places
practical limits of a kW or less on the entire payload instrumentation.
ANITA seeks to digitize a large number of radio-frequency channels at
rates of at least Nyquist frequencies for a bandwidth that extends
up to 1.2~GHz, implying Nyquist sampling of at least 2.4 Gsamples/second. 
Commercial digitizers that run continuously at such rates are generally
high-power devices, typically 5 watts per channel at the time of
ANITA electronics development. For 80 channels, the required power for
only the digitizers alone would be several hundred Watts, and the
downstream electronics then required to parse and decimate the huge data rate
(several Terabits per second) would use a comparable amount of power.
Folding in the requirements for amplifiers and other analog and digital power needs,
the payload power budget was not viable using commercial digitizers.

\begin{figure}[h]
\includegraphics[width=6.35in]{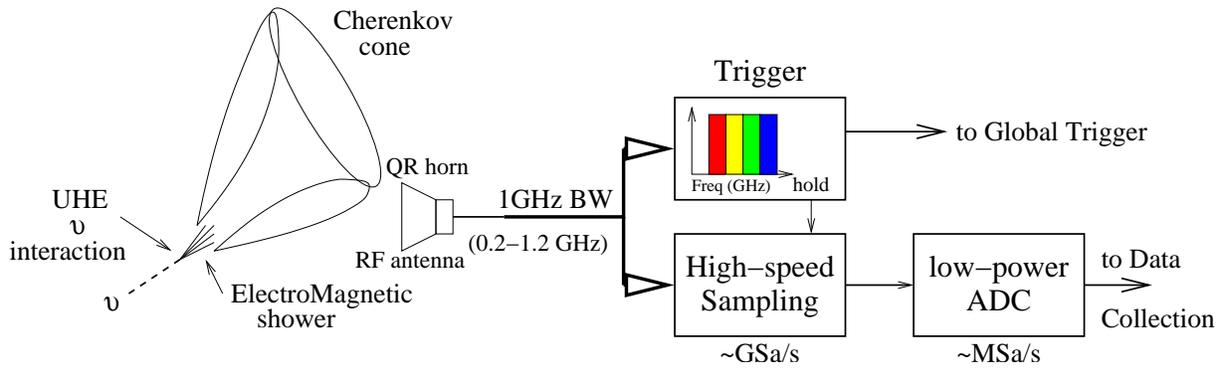}
\caption{In order to minimize the power required, signals from the
antennas are split into analog sampling and trigger paths.  To provide
trigger robustness, the full 1GHz bandwidth is split into 4 separate
frequency bands, which serve as separate trigger inputs.
\label{block_diag} }
\end{figure}

To make this problem tractable within the power, space, and weight budget
of an LDB payload, we elected to develop a separate analog trigger
system which would detect the presence of an incoming plane-wave
impulse, then only digitize a time window around this pre-detected
signal.  The as-built design for the low-powered RF trigger and digitization
electronics is summarized in Table~\ref{specs}.
A divide-and-conquer strategy to address the power and performance
issues raised by these specifications is shown in Fig.~\ref{block_diag}.

\begin{table}[h]
\begin{center}
\caption{ANITA Electronics Specifications.}
\begin{tabular}{|l|c|l|}
\hline \textbf{Design Parameter} & \textbf{As-built Value \& Comments} 
\\
\hline \# of RF channels & 80 = 32 top, 32 bottom, 8 monitor \\  
\hline Sampling rate & 2.6 GSa/s,  greater than Nyquist \\
\hline Sample resolution & $ \geq$ 9 bits = 3 bits noise + dynamic range \\
\hline Samples in window & 260 for a 100 ns window \\
\hline Buffer depth & 4 to allow rapid re-trigger \\
\hline Power/channel & $ <$ 10W including LNA \& triggering \\
\hline
\hline \# of Trigger bands & 4, with roughly equal power per band \\
\hline \# of Trigger channels & 8 per antenna (4 bands x 2 pols.) \\
\hline Trigger threshold & $\leq 2.3\sigma $ above Gaussian thermal noise \\
\hline Accidental trigger rate & $\leq 5$ Hz, gives 'heartbeat' rate \\
\hline Raw event size & $\sim 35$ kB, uncompressed waveform samples \\
\hline
\end{tabular}
\label{specs}
\end{center}
\end{table}

\paragraph{Triggering at Thermal Noise Levels}

Actual neutrino signals are not expected to be observed at a rate
greater than of order several per hour at most, if all previous bounds
are saturated. However, to avoid creating too restrictive a trigger
condition, the trigger was designed not to depend on 
exact time-alignment (or phase) of the incoming signal, over a
time window of order 10 ns. This allows
accidentals from random thermal noise to also trigger the system,
so that a continuous stream of events would be recorded, allowing
a continuous sampling of the instrument health. Since the thermal
noise floor in radio measurements is very well defined by the
overall system temperature, this ensures that the sensitivity of the
instrument remains high. As long as these thermal noise triggers do
not saturate the data acquisition system, causing deadtime to actual
events of interest, this methodology is effective. The thermal 
noise events so recorded have negligible probability of time
alignment to mimic an actual signal. Appendix~\ref{Thermal_App}
gives further results demonstrating this conclusion.

\paragraph{Trigger Banding}

In order to provide optimal robustness in the presence of unknown but
potentially incapacitating Electro-Magnetic Interference (EMI)
backgrounds, a system of non-overlapping frequency bands has been
adopted.  Typical anthropogenic backgrounds are narrow-band, and while
a strong emitter in a given band would likely raise the trigger
threshold (at constant rate) such that it would be effectively
disabled, the other trigger bands could continue to operate at thermal
noise levels.

\begin{figure*}[thb!]
\centerline{\epsfig{file=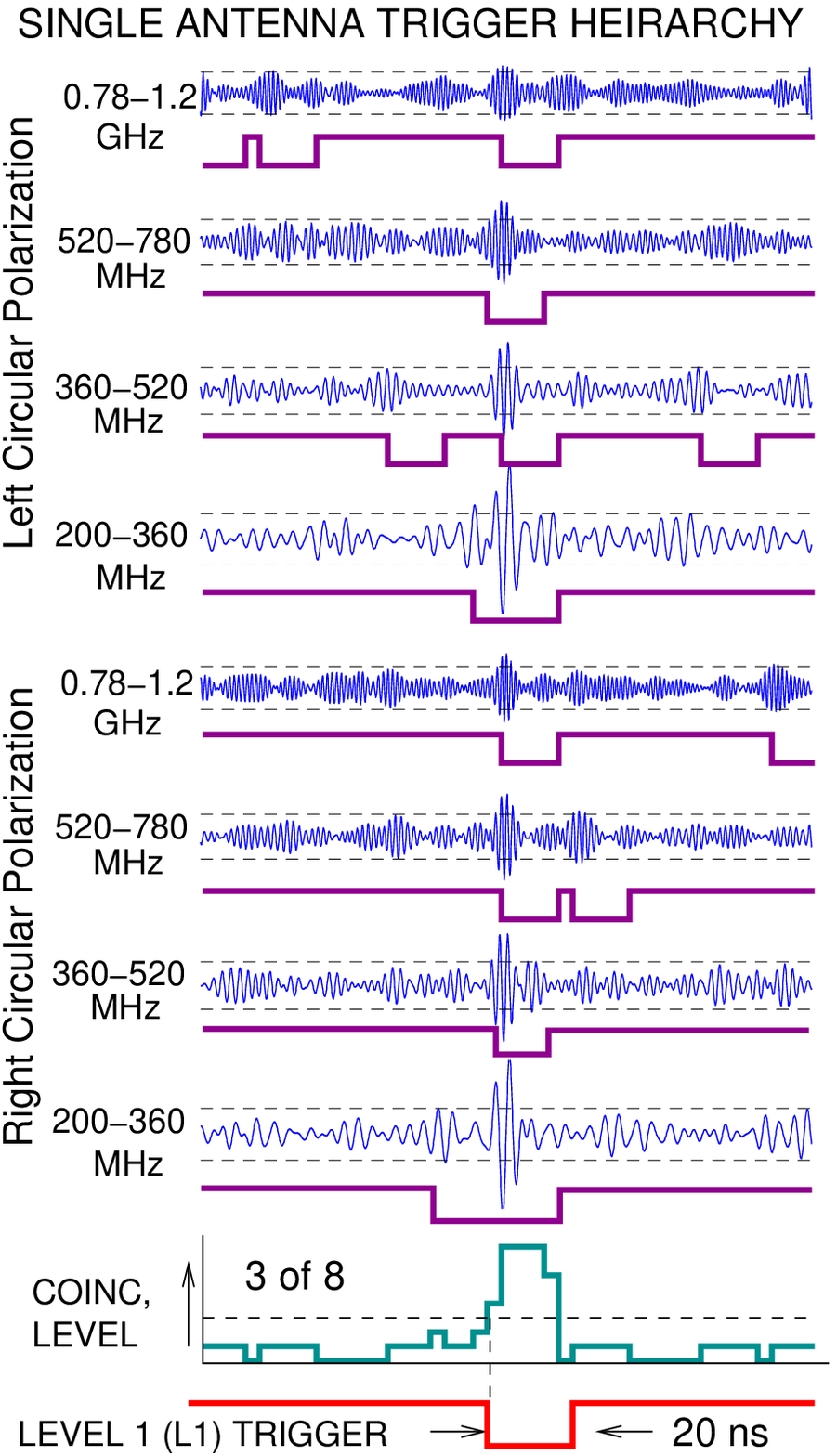,width=2.65in}\includegraphics[width=3.85in]{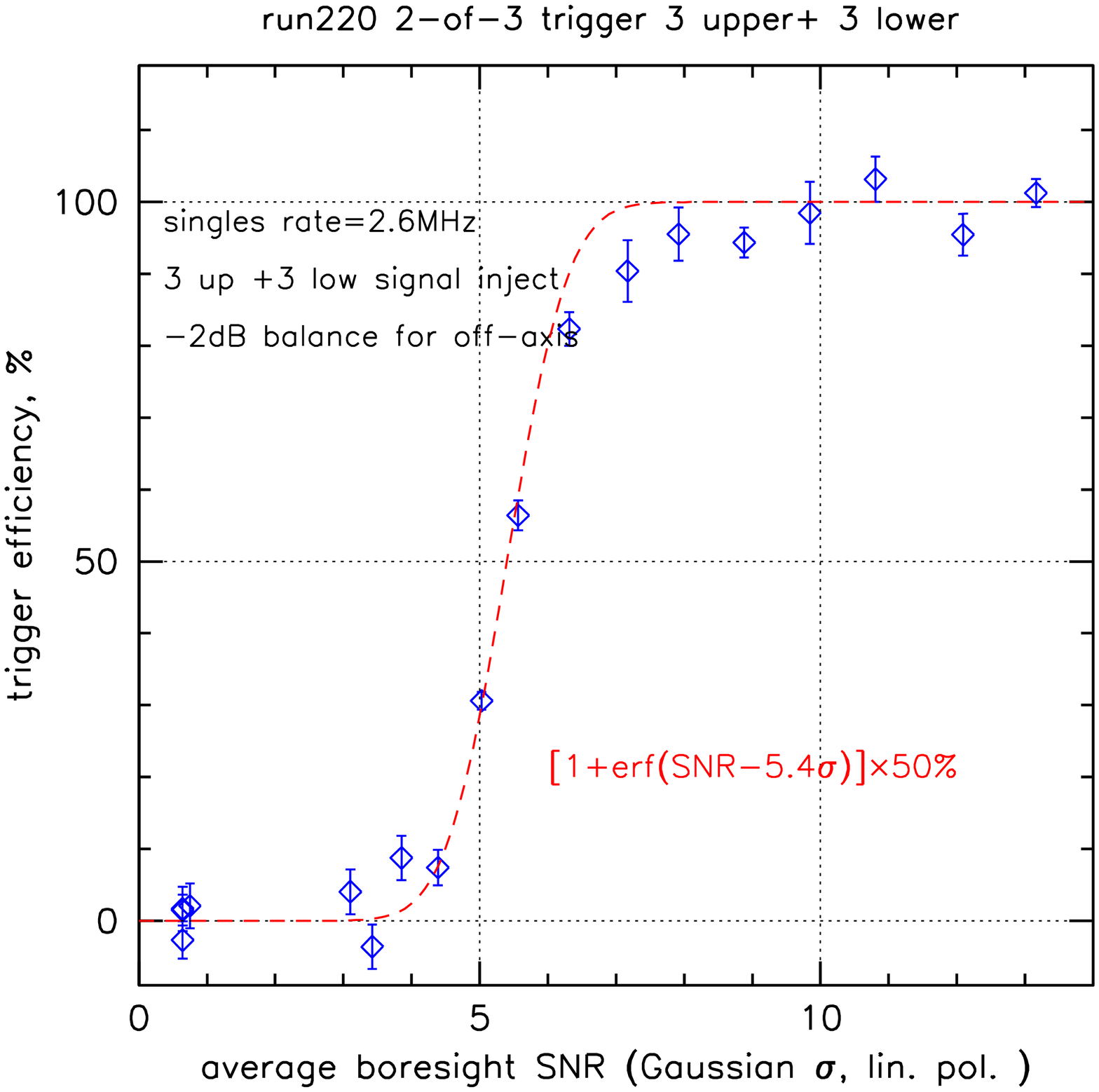}}
\caption{ Left: The baseline single-antenna (L1) triggering is illustrated
schematically. In practice the double-sided voltage thresholds
shown are implemented by first using a tunnel-diode square-law
detector to rectify the band voltage signals into a
unipolar pulse.  Right: Plot of measured global (L3) trigger efficiency 
vs. threshold in standard deviations above the RMS thermal noise
voltages for ANITA, using a shaped Askaryan-like pulse 
from a pulse generator under pure thermal noise conditions.\label{trigger}}
\end{figure*}

Signals from the vertical and horizontal polarizations of the quad-ridge
horn antennas are amplified and conditioned in the RF chain described
in the preceding subsection.  These RF signals are then split
into two paths, a trigger (lower) and digitizer (upper) path as
indicated in Fig.~\ref{Block07}. The trigger path first passes
through the hybrid 90\deg combiner which converts the H \& V
polarizations to LCP and RCP. The signals then enter a
sub-band splitter section where they are
divided into frequency bands with centers at
$\nu_c = $265, 435, 650, and 990~MHz and bandwidths of
$\Delta\nu =$130, 160, 270, 415~MHz, 
or fractional bandwidths $\Delta\nu/\nu_{c} \simeq 44\%$ on average.
This partitioning is performed in order to
provide rejection power against narrow-bandwidth
anthropogenic backgrounds.  In contrast, as true Askaryan
pulses are temporally compact and coherent, significant RF power is
expected across several bands.  In addition the thermal noise in each of
the bands are statistically independent, and requiring a multi-band
coincidence thus permits operation at a lower effective thermal noise
floor. This is illustrated schematically in Fig.~\ref{trigger}.

\paragraph{Level 1 trigger.}
The frequency sub-band signals are then led into a square-law-detector
section which uses a tunnel-diode, a common technique in radio
astronomy. The resulting output unipolar pulses (negative-going in this
case, typically 7~ns FWHM) are filtered and sent to a local-trigger unit (a 
field-programmable gate array or FPGA) on one of the nine 8-input-channel Sampling
Unit for Radio Frequencies (SURF) boards, within the
compact-PCI ANITA crate, and therefore attached to the host computer and
global trigger bus. Within the SURF FPGA, the square-law detector outputs
are led through a discriminator section with a programmable
threshold. The single-band thresholds are set in a noise-riding mode
where they servo on their rate, with a time constant of several seconds,
maintaining typical rates of 2.6-2.8~MHz under pure thermal noise
conditions, corresponding approximately to the 
$2.3\sigma_V$ ($\sigma_V = V_{rms}$, the root-mean-square received voltage) 
level mentioned above.
The SURF FPGA also then applies the
single-antenna trigger requirement: the eight sub-bands generate a 10~ns
logic level for each signal above threshold, and when
any three of these logic gates overlap, a single-antenna trigger is generated. 
These triggers are denoted Level 1 (L1) triggers, and
occur at typical rates of 150~kHz per antenna for thermal-noise conditions.

To determine the expected ANITA L1 accidental rate $R_{L1}$ of $k$-fold coincidences among the
$n=8$ sub-band single-antenna channels, consider a trial event, defined by a hit in
any one of the $n$ channels, which then triggers a logic transition out of the discriminator
to the logic TRUE state for a duration $\tau$. 
Then consider the probability during this trial that $k-1$ or more ($k=3$ for ANITA) 
additional sub-band discriminator logic signals arrive while the first is still
in TRUE state, corresponding to a hit above threshold for that channel.
The rate of TRUE states per channel is $r$. We do not for now assume $r\tau<<1$.

The probability to observe exactly $k-1$ out of $n-1$ additional channels in the TRUE state 
after one channel has changed its state 
is given by the binomial (e.g., the $k$ out of $n$ `coin toss') probability:
$$ P(k-1:n-1)  = \frac{(n-1)!}{ (k-1)!(n-k)!}  p^{k-1} (1-p)^{n-k} ~.$$
The single channel `coin-toss' probability $p$ is just
given by the fractional occupancy of the TRUE state per second per channel:
$p~=~r\tau$.

The probability per trial to observe greater than $k-1$ out of $n-1$
channels is then just the cumulative probability density of the binomial distribution
times the observation interval:
$$ P(\geq k-1:n-1) = \sum_{j=k-1}^{n-1}   \frac{(n-1)!}{ (j)!(n-1-j)!}  (r\tau)^j [1-r\tau]^{n-j}  ~dt $$

For $r\tau<<1$ as it often is in practice, this simplifies to
$$ P(\geq k-1:n-1) \simeq \frac{(n-1)!}{ (k-1)!(n-k)!} (r\tau)^{k-1}  $$
since only the leading term in the sum contributes significantly and the term
$1-r\tau \simeq 1$.

The rate is then determined by multiplying the single-trial probability 
by the number of ensemble trials per second,
which is just equal to the total number of channels times the singles rate per channel.
The singles rate
per channel is given simply by $r$, and the total singles rate across all channels is $nr$. 
Thus the total rate in the limit of $r\tau<<1$, is:
$$ R_{L1} = nr P(\geq k-1:n-1) \simeq n \frac{(n-1)!}{ (k-1)!(n-k)!} r^k\tau^{k-1}  
=  \frac{(n)!}{ (k)!(n-k)!} k r^k\tau^{k-1}~.  $$

\paragraph{Level 2 trigger.}
When a given SURF module detects an L1 trigger, indicating either a 
possible signal or (most probably) a thermal noise fluctuation, it
reports this immediately to the Trigger Unit for Radio Frequencies (TURF)
module, which occupies a portion of the compact-PCI backplane
common to all of the SURF modules. The TURF contains another
FPGA, which determines whether a level 2 (L2) trigger, which corresponds to
two L1 events in any adjacent antennas of either the upper or lower ring, within a 20~ns
window, has occurred. L2 triggers occur at rate of about 2.5~Khz
per antenna pair, or about 40~kHz aggregate rate for thermal noise.

\paragraph{Level 3 trigger.}
If a pair of L2s occur in the upper and lower rings within a 30~ns
window and any up-down pair of the antennas share the same azimuthal sector 
(known as a ``phi sector''), 
a level 3 (L3) global trigger is issued, and the digitization
of the event proceeds. These occur at a rate of about 4-5~Hz for
thermal noise. Fig.~\ref{trigger}(right) shows a measurement of the 
effective threshold for L3 triggers in terms of the peak pulse
SNR above thermal noise. Here three upper and three lower antennas were stimulated
with a shaped pulse and the L3 rate was measured as a function of the
input pulse SNR to estimate the effective global threshold,
here in Gaussian standard deviations above the thermal noise 
RMS voltage.  ANITA begins to respond at about $4\sigma_V$, reaches
of order 50\% efficiency at $5.4 \sigma_V$ and is fully efficient
at of order $\sim 7\sigma_V$.

In Appendix~\ref{ThermalApp}, 
we further analyze the rate of accidentals in terms of their ability to
reconstruct coherently to mimic a true signal event, and we find that
the chance probability for this is of order 0.003 events for the ANITA flight,
presenting a negligible background.

\begin{figure}[htb!]
\centerline{\includegraphics[width=3.25in]{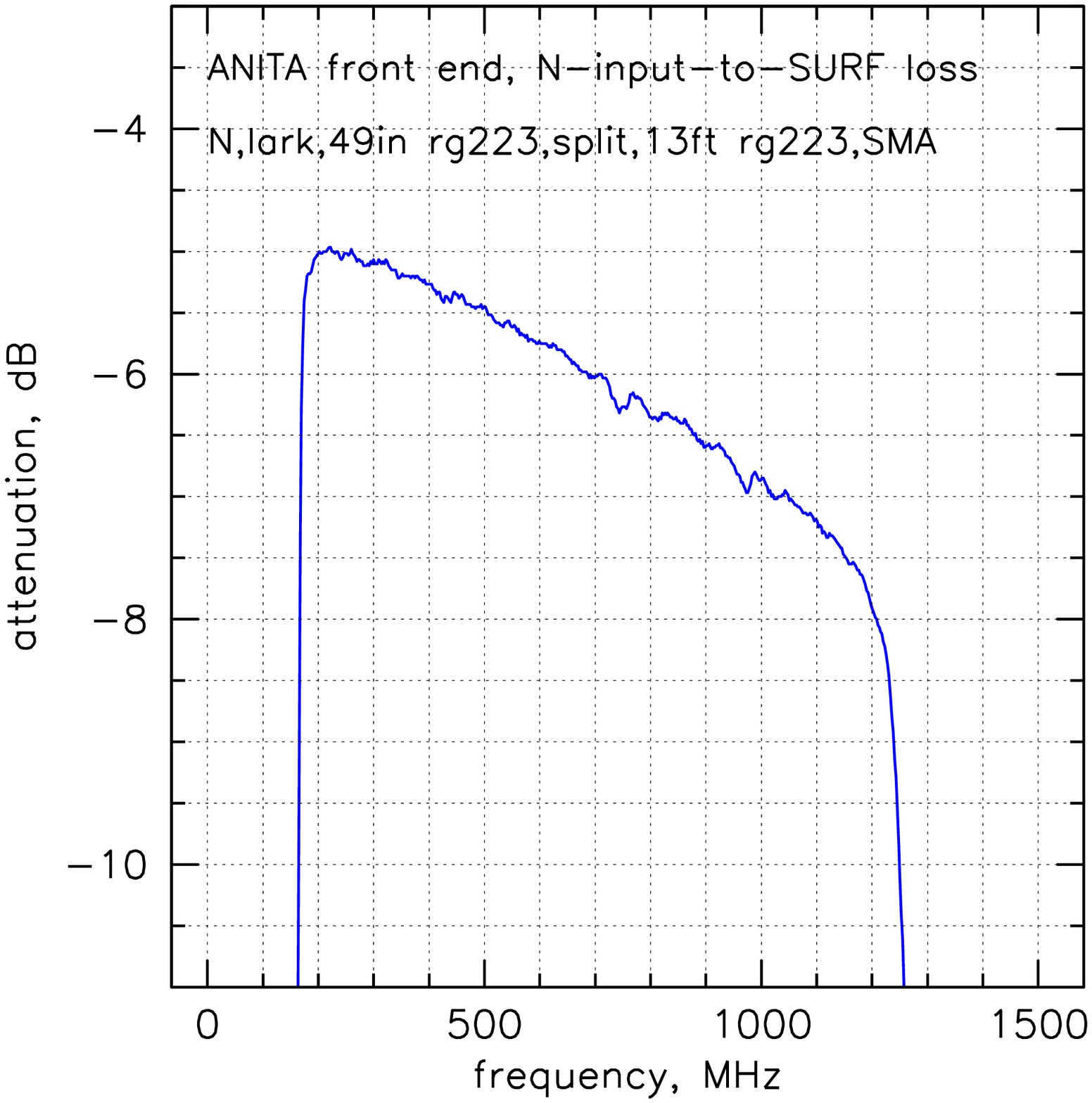}~~\includegraphics[width=3.4in]{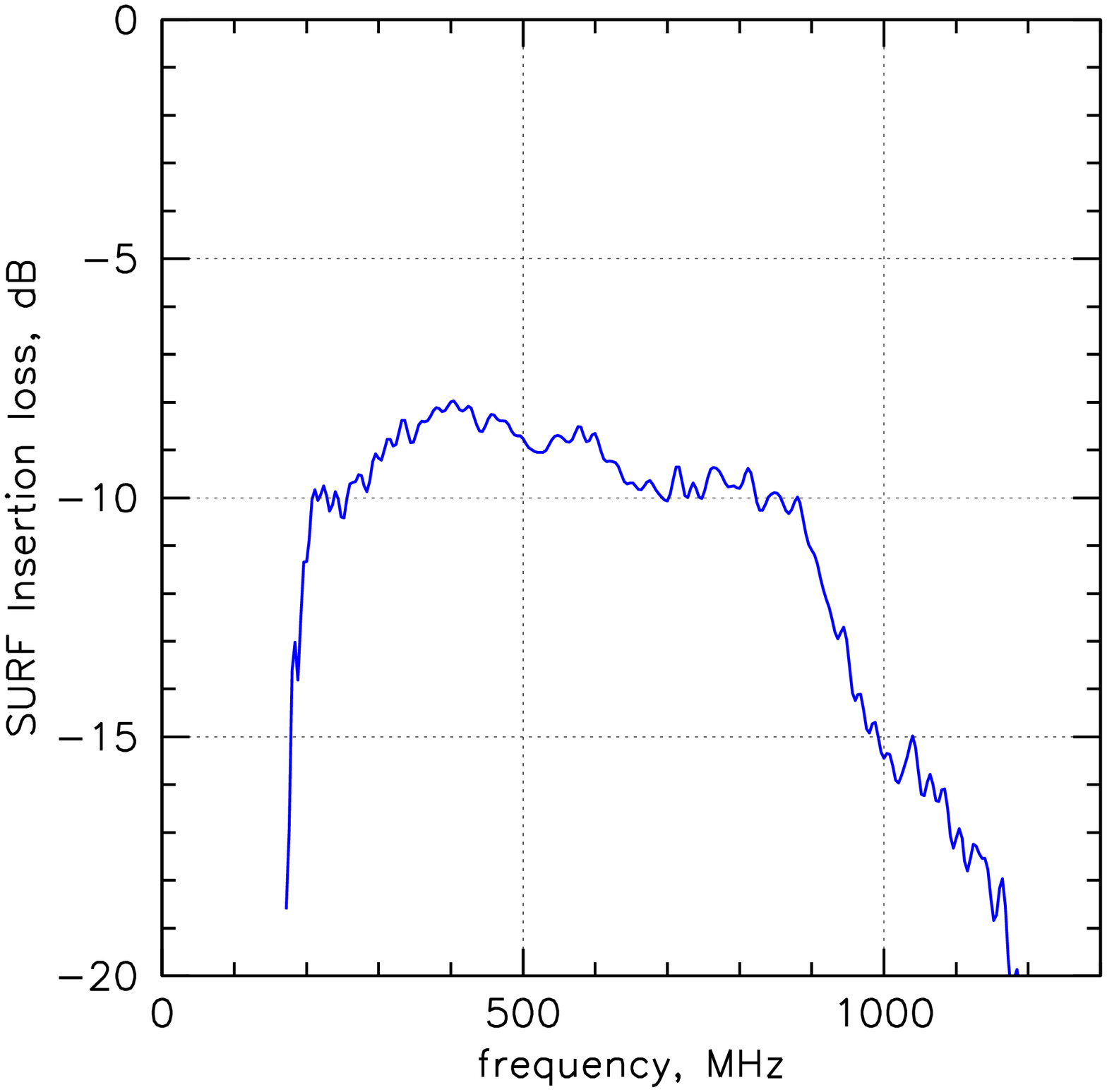}}
\caption{Left: Detailed view of input signal chain insertion losses, up to
the input to the SURF. The losses in the trigger path are slightly lower since the 20~ns
delay cable is omitted.  Right:Typical insertion losses for the SURF digitizer inputs used to record the
waveforms. The rolloff above about 850~MHz leads to a loss of SNR of the
recorded waveforms compared to the actual RF trigger path.
\label{SURFloss} }
\end{figure}

\subsubsection{Digitizer System.}

The upper path in Figure~\ref{Block07} is the digitization path.  A
low-power, high sampling-speed Switched Capacitor Array (SCA) continuously
samples the raw RF inputs over the entire 1 GHz of analog
bandwidth defined by the upstream RF conditioning, at a
sample rate of 2.6~Gsamples/s. Sampling is halted
and the analog waveform samples held for readout 
upon the fulfillment of a trigger
condition. The SCA sampler, which does not actually digitize its stored samples until
commanded to do so for a trigger, uses far lower power than 
traditional high speed continuous-digitizing samplers (such as
oscilloscopes). Without the custom development of this technology
by G. Varner of the ANITA team, the power budget for ANITA
would have grown substantially, from  of order 1 W/ch to perhaps
10~W per channel for a commercial digitizer as was used for
ANITA-lite~\cite{Anitalite}. In addition, the continuously sampled data would have
added a processing load of order 200~Gbyte/sec to the trigger
system.

Fig.~\ref{SURFloss} shows measurements of the signal chain and SURF channel insertion loss vs. radio
frequency. On the left, the losses up to the input of the SURFs are shown; these are primarily cable
and second-stage bandpass filter losses. Similar losses apply to the trigger path as well, though
slightly lower since there is no 20~ns delay cable in that path. On the right, the SURF insertion losses are
shown, and these are unique to the waveform recording path. 
The loss above about 850 MHz tend to significantly reduce the
intrinsic peak voltages in the most impulsive waveforms compared to what is seen by the analog trigger
inputs. These amplitude losses can be corrected
to some degree as shown in a later section, but there is still a net loss of SNR in the 
deconvolved waveforms compared to what is seen by the trigger path.

It is evident from this plot that ANITA's digitizers did not fully achieve the 
design input bandwidth span of 200-1200 MHz; the last quarter of this band has reduced
response compared to the design goal. The main impact of this is not to reduce the
trigger sensitivity of the instrument, since the digitizer does not constrain the input bandwidth
of the trigger system. Rather, the primary impact is in evaluation of potential neutrino candidates,
for which we would like to be able to reconstruct an accurate spectral density for the
received radio signal. For the current digitizer system, the reconstructed spectral content above
900~MHz will thus be subject to errors that increase with frequency; however, this
frequency region is also the region where ice attenuation is rising quickly~\cite{icepaper}.

\subsection{Navigation, attitude, \& timing}

	\paragraph{Absolute Orientation.}

In order to geometrically reconstruct neutrino events, accurate position, 
altitude, absolute time, and pointing information are required. To provide such 
data on an event-by-event basis, a pair of Global Positioning System (GPS) 
units was used. They provide more than sufficient accuracy to fulfill the science 
requirements, see Table~\ref{GPS}. In addition these units provide the ability to
synchronously trigger and read out the system 
on an absolute timing mark (such as the nearest second), a feature which is essential to
the ground-to-flight calibration sequence, where a ground transmitter needs to
be globally synchronized to the system during flight, including a propagation
delay offset.

ANITA had a mission-critical
requirement for accurate payload orientation
knowledge, to ensure that the free-rotation of the payload would
not preclude reconstruction of directions for events at the
sub-degree level of accuracy. 
Such measurements were accomplished
with a redundant system of 4 sun-sensors, a magnetometer, and
a differential GPS attitude measurement system (Thales Navigation/Magellan ADU5).
These systems performed well in flight and met the mission design
goals. In calibration done just prior to ANITA's 2006 launch, we
measured a total $(\Delta \phi)_{RMS} = 0.071^{\circ}$, very close to the limit of
the ADU5 sensor specification, and well within our allocated
error budget.

The ADU5 is connected to the flight computer with a pair of 
RS-232 interfaces; one carries the attitude information packets for the 
housekeeping readout, and the second provides readout when 
during the UTC second a digital trigger line was set at the ADU5. 
The second GPS unit, a G12 sensor from 
Thales, is used to get a second trigger timing piece of information over one 
serial line and timing information for the flight computer's NTP (Network Time 
Protocol) internal clock. Position and attitude information 
is updated every second. Also every second the NTP server gets an update to keep 
good overall clock time at the computer. The trigger is also connected directly 
to GPS time: GPS second from the 1-second readout and fraction of a second from 
the phototiming data block. Additional pointing information is derived from the 
sun sensors and a tip-tilt sensor mounted near the experiment's center of mass. 
They provide a simple crosscheck to the attitude data with very different 
systematics and also offer a measure of redundancy.

\begin{table*}[bt]
\caption{ Navigation, attitude, and timing sensor requirements and provided accuracy.}
\label{GPS}
  \begin{center}
     \begin{tabular}{|l|l|l|l|} \hline
Parameter &Determination method & Required Accuracy & System Accuracy \\ \hline \hline
Position/Altitude & Ordinary GPS & 10m horiz./20m vert.& 5m Horizontal/10m Vertical \\
UTC  &  GPS Phototiming pulse & 20ns & $<10$ns \\
Pointing & Short-baseline differential GPS & 0.3$^{\circ}$ rotation/0.3$^{\circ}$ tip & $<0.07^{\circ}$/$<0.14^{\circ}$ \\
Pointing Sun-sensor & Tip-tilt sensor & 0.3$^{\circ}$ rotation/0.3$^{\circ}$ tip & 1$^{\circ}$/1$^{\circ}$ \\ \hline
     \end{tabular}
   \end{center}
\end{table*}

\section{Flight Software \& Data Acquisition}
The ANITA-I flight computer was a standard c-PCI single board computer, based on the Intel Mobile Pentium III CPU. 
The operating system was Red Hat 9 Linux, which was selected due to driver availability.

\begin{figure}[hbt]
    \centering
    \resizebox{\textwidth}{!}{\includegraphics{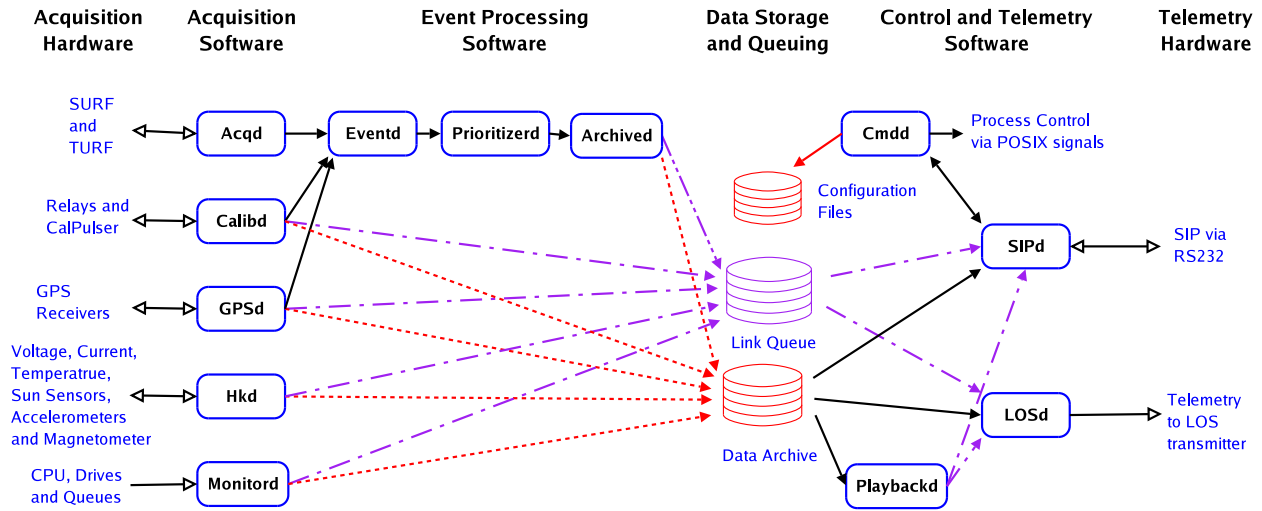}}
    \caption{A schematic overview of the flight software processes. The open arrows show software interaction with 
hardware components, and closed arrows indicate data transfer between processes. The telemetry data flow across 
the ramdisk is indicated by the dot-dashed (purple) line, and permanent data storage to the on-board drives is 
shown by the the dotted (red) lines.}
    \label{f:overview}    
  \end{figure}

The design philosophy behind the ANITA flight software was to create, as far as possible, 
autonomous software processes. Inter-process data transfer consists of FIFO queues of data 
files and links on the system ramdisk, these queues are managed using standard Linux 
filesystem calls. Process control is achieved using configuration files and a small 
set of POSIX signals (corresponding to: re-read config files, restart, stop, and kill). 
A schematic overview of the flight software processes is shown in Figure~\ref{f:overview}.

The flight software processes break down into three main areas, which will be discussed in following sections. 
These areas are:
\begin{enumerate}
\item Data acquisition -- processes which control specific hardware and through which all data from that hardware is obtained.
\item Event processing -- processes which augment or analyze the waveform data in the on-line environment.
\item Control and telemetry -- processes which control the telemetry hardware and those which are responsible for process control.  
\end{enumerate}

\subsection{Data Acquisition}
The bulk of the data, around 98\%, acquired during the flight is in the form of digitized waveform data 
from the SURF boards. The remaining 2\% consists of auxiliary information necessary to process and 
interpret the waveform data (payload position, trigger thresholds, etc.) and data that is used to 
monitor the health of the instrument during flight (temperatures, voltages, disk spaces, etc.).

\subsubsection{Waveform Data}
The process that is responsible for the digitization and triggering hardware, the SURF and TURF, is 
Acqd (the Acquisition Daemon). The Acqd process has four main tasks:
\begin{itemize}
\item Acquiring waveform data from the SURFs
\item Acquiring trigger and timing data from the TURF (via TURFIO).
\item Acquiring housekeeping data from the SURF (scaler rates, read back thresholds, and RF power).
\item Setting the thresholds and trigger band masks to control the trigger, dynamically adjusting 
the thresholds to maintain a constant rate.
\end{itemize}

Once the TURF has triggered an event and the SURFs' have finished digitizing, the event data 
is available for transfer across the c-PCI backplane to the flight computer. The flight 
computer polls the SURFs to check when an event has finished digitization and the data is 
ready to be transferred across the c-PCI backplane.

An event consists of 260 16-bit waveform data words per channel, there are 9 channels per 
SURF and 9 SURFs in the c-PCI crate. A complete raw event is approximately 41KB. To achieve 
better compression, see Section~\ref{s:compression}, the raw waveform data is pedestal 
subtracted before being written to the queue for event processing.

\subsubsection{Trigger Control}
In addition to acquiring the waveform and trigger data, Acqd is also responsible for 
setting the thresholds and trigger band masks that control the trigger.  There are 
three handles through which Acqd can control the trigger:
\begin{itemize}
\item The single channel trigger thresholds (256 channels).
\item The trigger band masks (8 channels per antenna).
\item The antenna trigger mask (32 antennas in total).
\end{itemize}

The default mode of operation is to have all of the masks off, such that every trigger 
band and every antenna can participate in the trigger. In the thermal regime, i.e. away 
from anthropogenic RF noise sources such as camps and bases, the trigger control 
operates by dynamically adjusting the single channel thresholds to ensure that each 
trigger channel triggers at the same rate (typically 2-3MHz). The dynamic adjusting 
of the thresholds is necessary as even away from man-made noise sources the RF power 
in view varies with the temperature of the antenna and its field of view, i.e with 
the position of the sun and galactic center with respect to the antenna. The thresholds 
are varied using a simple PID (proportional integral differential) servo loop that 
was tuned in the laboratory using RFCMs with terminated inputs.

During times when the balloon is in view of large noise sources, such as McMurdo station, 
a different triggering regime is necessary to avoid swamping the downstream processes with an 
unmanageable event rate. To allow for this all of the trigger control options are commandable 
from the ground, see Section~\ref{s:commanding} for more details on commanding. Using these 
commands some of the available options for controlling the trigger rate are:
\begin{itemize}
\item Adjust the global desired single trigger channel rate.
\item Adjust individual single channel rates independently. 
\item Remove individual trigger channels (i.e. frequency bands) from the antenna level (L1) trigger.
\item Remove individual antennas from the L2 and L3 triggers.
\end{itemize} 

\subsubsection{Housekeeping Data}
In addition to the waveform data, housekeeping data is also continuously captured, both for 
use in event analysis and also for monitoring the health of the instrument during flight. 
Table~\ref{t:housekeeping} is a summary of the various types of housekeeping data acquired 
by the flight software processes.

\begin{table}[hbt]
  \centering
  \begin{tabular}{| c | l | c |}
\hline
    Process & Housekeeping Data & Rate \\
\hline
Acqd & Trigger rates and average RF power & up to 5\,Hz \\
Calibd & Relay status & 0.03\,Hz \\
GPSd & GPS position, velocity, attitude, satellites, etc. & up to 5\,Hz  \\
Hkd & Voltages, currents, temperatures, pressures, etc. & up to 5\,Hz \\
Monitord & CPU and disk drive status  & 0.03\,Hz\\
\hline
  \end{tabular}
  \caption{The types of housekeeping data acquired by flight software processes.}
  \label{t:housekeeping}
\end{table}

\subsection{On-line Event Processing}
At altitude the bandwidth for downloading data from the payload to the ground systems 
is very limited, see Section~\ref{s:datadown}. In order to maximize the usage of this 
limited resource the events are processed on-line to determine the event priority and 
they are compressed and split into a suitable format for telemetry. 

\subsubsection{Prioritization}
The Prioritizerd daemon is responsible for determining the priority of an event. 
This priority is used to determine the likelihood of a given event being telemetered 
to the ground during flight. The prioritizer looks at a number of event characteristics 
to determine priority. The hierarchical priority determination is described below:
\begin{itemize}
\item Priority 9 --  If too many waveforms (configurable) have a peak in the FFT 
spectrum (configurable), the event is given this low priority, to veto events 
from strong narrow band noise sources.
\item Priority 8 -- If two many channels peak simultaneously, determined via 
matched filter cross-correlation, the  event is assumed to be generated on-payload and is rejected.
\item Priority 7 -- Compares the RMS of the beginning and end of waveforms to veto non-impulsive events.
\item Priority 6 -- This is the default priority. Thermal noise events will 
be assigned this priority if they are not demoted for one of the above reasons, or promoted for one of the below reasons.
\item Priority 5 -- An event is promoted to priority five if it passes the 
test of N of M (configurable) neighboring antennas peaking within a time 
window (configurable). Events must satisfy this condition to be c
onsidered for promotion to priorities 1-4.
\item Priority 4 -- A cross-correlation is performed with boxcar 
smoothing, and there are peaks in 2-of-2 antennas in one ring and 1-of-2 in the other.
\item Priority 3 -- A cross-correlation is performed with boxcar 
smoothing, and there are peaks in 2-of-3 antennas in both rings.
\item Priority 2 -- A cross-correlation is performed with boxcar 
smoothing, and there are peaks in 2-of-2 antennas in both rings.
\item Priority 1 -- A cross-correlation is performed with boxcar 
smoothing, and there are peaks in 3-of-3 antennas in both rings.
\end{itemize}

\subsubsection{Compression} \label{s:compression}
Several encoding methods were investigated for the telemetry of waveform data. 
Of the methods investigated, the optimum method was determined to be a 
combination of binary and Fibonacci encoding. The steps involved in the compression are described below:
\begin{itemize}
\item The waveform is pedestal subtracted and zero-meaned.
\item The data is truncated to 11-bits (the lowest bit in the data is shorted to the next to lowest)
\item A binary bit size is determined based upon the smallest power of 
two above the three standard deviation point of the waveform.
\item All samples that lie within the range of values that can be 
encoded with a binary number of this dimension are encoded as their 
binary value, those outside are assigned the maximal value.
\item The difference between the maximal value and the actual sample value 
are then encoded using Fibonacci coding in an overflow array. Fibonacci 
coding is useful for this purpose as it efficiently encodes large numbers, 
and it has some built in immunity to data corruption as each encoded number ends with 11.
\item The full encoded waveform then consists of 260 $n$-bit binary numbers 
followed by $M$ Fibonacci encoded overflow values.
\end{itemize}

\begin{figure}[hbt]
  \centering
  \resizebox{0.6\textwidth}{!}{\includegraphics{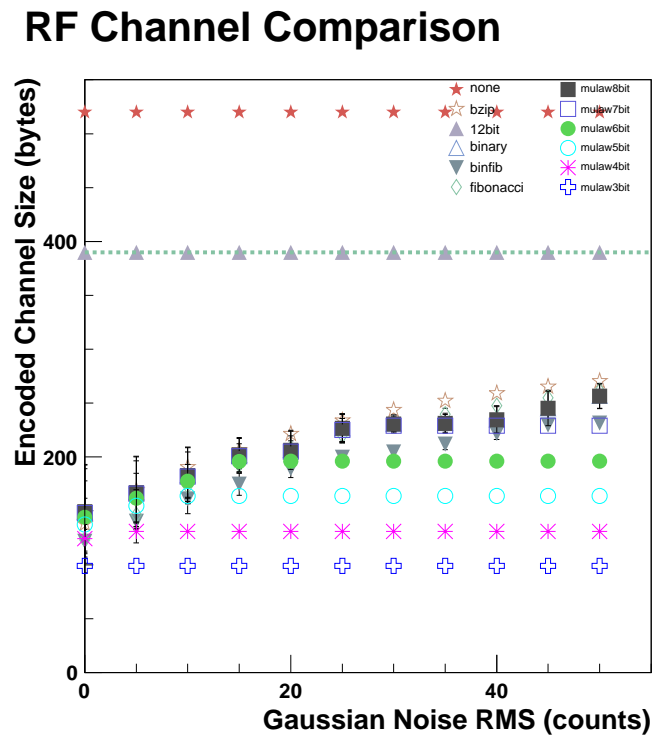}}
  \caption{A comparison of the performance of the different loss-less and lossy 
compression methods tested on ANITA waveform data with adjustable Gaussian 
noise levels. The binary/Fibonacci method detailed in the text proved to be 
the most effective method for encoded the telemetry data. The mu-law methods 
were the only lossy methods that were considered.}
  \label{fig:compress}
\end{figure}

Figure~\ref{fig:compress} shows the performance of the binary/Fibonacci 
encoding method in comparison with the other methods considered. This 
method was chosen as it proved to give the best compression of the loss-less methods considered.

\subsection{Control and Telemetry}
One critical aspect of the flight software for a balloon experiment is the 
control telemetry software. This software represents the only link between 
the experiment and the scientists on the ground. As such the software needs 
to be both very robust, to withstand a long flight away from human interaction, 
and also flexible enough to cope with unexpected failures during the flight.

\subsubsection{Data Downlink} \label{s:datadown}
During the flight it is critical to telemeter enough information that the 
scientists on the ground can determine whether the instrument is operating, 
and acquiring sensible data. All of this information needs to be pushed 
through one of two  narrow (bandwidth limited) pipes to the ground. When 
close to the launch site the data is sent over a Line of Sight (LOS) 
transmitter with a maximum bit rate of 300kbps. Once over the horizon 
the only data links available are satellite links: i) a continuous 6\,kbps 
using the TDRSS satellite network; and ii) a maximum of 248 bytes every 30 
seconds using the IRIDIUM network. The ODIUM link is only used to monitor 
payload health during times when the other, higher bandwidth, links are 
unavailable (due to satellite visibility and other issues).

There are several data streams that are telemetered, these are summarized 
in Table~\ref{tab:telem}. Clearly, only a tiny fraction of the total data 
written to disk is able to be telemetered to the ground using the satellite links.
\begin{table}[hbt]
  \centering

  \begin{tabular}{|c|c|c|c|}
    \hline
   Data Type & Size (bytes) & TDRSS (packets/min) & LOS (packets/min) \\
    \hline 
    Header & 74 & 150 & 300 \\
    Waveform & 14,000  & 2 & 120 \\
    GPS & 88 & 6 & 300 \\
    Housekeeping & 150 & 6 & 60 \\
    RF Scalers & 1380 & 1 & 30 \\
    CPU Monitor & 64 & 1 & 1 \\
    \hline
  \end{tabular}
  \caption{A summary of the data types, sizes and telemetry rates over the LOS and TDRSS downlinks.}
  \label{tab:telem}

\end{table}

\begin{table}[hbt!]
  \centering

  \begin{tabular}{|c|c|c|}
    \hline
    Command Name & Description & Number \\
    \hline 
    Send Config Files & Telemeter config file & $\sim 300$ \\
    Set PID Goal & Adjust the single band trigger rate & $\sim 250$ \\
    Send Log Files & Telemeter log file  & $\sim 250$ \\
    Kill Process & Tries to kill the selected process & $\sim 150$ \\
    Restart Process & Tries to restart the selected process & $\sim 100$ \\
    Set Band Mask & In/Exclude frequency bands  & $\sim 100$ \\
    TURN RFCM On/Off & Turn on/off the power to the amplifiers & $\sim 45$ \\
    Set Antenna Mask & In/Exclude antennas from the trigger & $\sim 20$ \\
    GPS Trigger Flag & Enable/Disable GPS triggered events & $\sim 20$ \\
    Start New Run & Starts a new run & $\sim 20$ \\
    Set Channel Scale & Change the goal rate of an individual band & $\sim 15$ \\
    \hline
  \end{tabular}
  \caption{The most common commands used during the ANITA-I flight, with a 
brief description of the intended command outcome, and an indication of 
the number of times the command was used during the ANITA-I flight.}
  \label{tab:commands}
\end{table}

\subsubsection{Commanding} \label{s:commanding}
During the flight it is possible to send commands from the ground to the payload 
over one of the three available links: line of sight, TDRSS and IRIDIUM. 
Each command contains up to 20 bytes of information. The most common commands 
used during the ANITA-I flight are shown in Table~\ref{tab:commands}.

\section{Monte Carlo Simulation Sensitivity Estimates}

ANITA's primary goal is the detection of the cosmogenic neutrino flux. All
first-order modeling of such fluxes assume they are quasi-isotropic; that
is to say, although they originate from sources of small angular extent and retain
their directional information throughout their propagation trajectory to earth, the
sources are presumed to be distributed in a spatially uniform manner as projected on
the sky and viewed from Earth. Thus to a good approximation we model the incoming
neutrino flux as sampling a parent distribution that is uniform across the spherical
sky.

Because of the complexity of the neutrino event acceptance for the target volume 
of ice that ANITA's methodology employs, a complete analytic estimate of the
neutrino detection sensitivity is very difficult, and we have relied on detailed
Monte Carlo simulation methods to perform the appropriate 
multi-dimensional integration which determines the acceptance as a function of
neutrino energy.

\begin{figure}[htb!]
\includegraphics[width=5.85in]{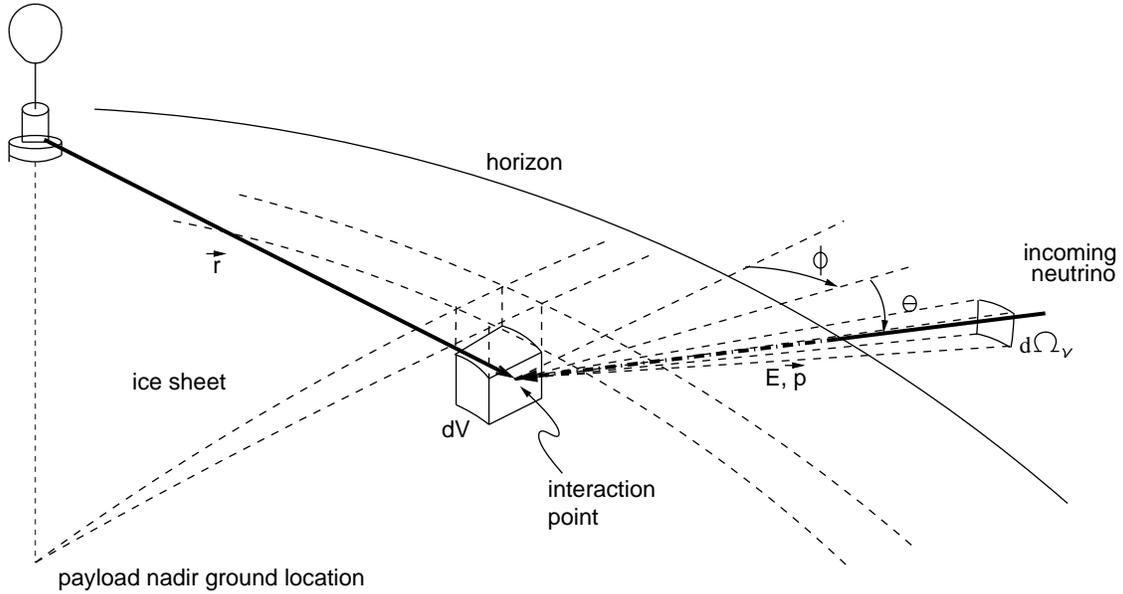}
\caption{ Schematic view of the geometry for a neutrino interaction in
the field-of-view of ANITA.
\label{Anitageom} }
\end{figure}

One may visualize the acceptance by first considering a volume element 
located somewhere within the horizon of the ANITA payload at some instant
of time, at a depth within several electric field attenuation lengths of the
ice surface. In the absence of the surrounding ice and earth below this volume element, 
the neutrino flux passing through such a volume is isotropic. However,
the attenuation due to the earth and surrounding ice modifies the flux
and resulting acceptance, so that, at EeV energies where neutrino attenuation
lengths are several hundred km water equivalent for the standard model
cross sections, most of the neutrino acceptance below the horizon is lost.
A schematic view of the geometry is shown in Fig.~\ref{Anitageom}. The volume
element $dV$, and position $\vec{r}$ with respect to the payload's
instantaneous location, encloses an interaction location which is subject to a neutrino
flux over solid angle element $d\Omega_{\nu}$, at offset angles $\phi,\theta$ with respect
to the radial azimuth from the payload and local horizontal directions.
Note that the payload angular acceptance $d\Omega$ is related to
$d\Omega_{\nu}$ by convolution with the detectability of neutrinos at every
location within the field of view. The geometry of the Cherenkov cone and
refraction effects then complicate the convolution even further, and Monte Carlo
integration becomes an attractive way to deal with this complexity.

We may still define an average volumetric acceptance $\langle \mathcal{V}\Omega \rangle$, 
in km$^3$-water-equivalent steradians (km$^3$we sr), 
as the physical target density-times-volume $\rho V$ of the detector
multiplied by the weighted solid angle $\Omega_\nu$ over which 
initially isotropic neutrino fluxes produce interactions within the detector,
and then multiplied again by the fraction $N_{det}/N_{int}$ 
of such interactions that are detected. For any given volume element of the
target, the latter term will depend on the convolution of the emission solid angle
for detectable radio signals with the arrival solid angle of the neutrinos. Although the
symbol $\mathcal{V}\Omega$ appears to imply that the two parts of the acceptance
(volume and solid angle) can be factored, this is generally not true
in practice since they tend to be strongly convolved with one another as a function of
energy. That is to say,
any given volume element may provide events detectable over a certain solid angle
centered on some portion of the sky, but a different volume element will 
in general have both a different net solid angle for detection and also
a different angular region over which those detections occur, all as a function of energy. 
However, we can define an average acceptance solid angle $\langle \Omega \rangle$ and
this factor is a useful quantity in this calculation.

The differential number of neutrino interactions per unit time, per unit neutrino arrival solid angle,
per unit volume element in the detector target,  can be written as:
\begin{equation}
\frac{d^3 N_{int}}{dt~d\Omega_{\nu}~dV} ~=~ 
\int_{E_{thr}}^{\infty} dE_{\nu}~
F_{\nu}(E_{\nu})~
\sigma_{\nu N,e}(E_{\nu})~\rho(\vec{r})~N_{A}~P_{surv}(E_{\nu},\vec{r},\theta_{\nu},\phi_{\nu})
\end{equation}
where $E_{thr}$ is the threshold energy for the detector, $\sigma_{\nu N,e}$ the
neutrino total cross section on nucleons (or electrons, for $\nu_e+e$ scattering),
$F_{\nu}(E_{\nu})$ is the neutrino flux as a function of energy,
$\rho(\vec{r})$ is the ice density at interaction position $\vec{r}$, $N_{A}$ is 
Avogadro's number, and $P_{surv}(E_{\nu},\vec{r},\vec{p}_{\nu})$ is the survival probability for a
neutrino of energy and 3-momentum $E_{\nu}, \vec{p}_{\nu}$ to arrive at position 
$\vec{r}$, which is the location of the volume element under consideration. This
probability can be further written:
\begin{equation}
P_{surv}(E_{\nu},\vec{r},\theta_{\nu},\phi_{\nu}) = 
e^{-L_{int}(E_{\nu}) X(\vec{r},\theta_{\nu},\phi_{\nu})}
\end{equation}
with the neutrino interaction length $L_{int} = (\sigma_{\nu N,e}N_A)^{-1}$,
and the function $X(\vec{r},\theta_{\nu},\phi_{\nu})$ the total column density
along direction $\theta_{\nu},\phi_{\nu}$ from point $\vec{r}$ within the detector. The
function $X$ thus contains the earth-attenuation dependence.

The number of neutrino interactions is thus
\begin{equation}
N_{int}~=~ \int_0^T dt \int_0^{4\pi} d\Omega_{\nu} \int_0^{V} dV \frac{d^3 N_{int}}{dt~d\Omega_{\nu}~dV}
\label{nint}
\end{equation}

To determine the average acceptance solid angle $\langle \Omega \rangle$,
imagine the detector target volume completely isolated from any 
exterior target matter of the earth, bathed in an isotropic flux of
neutrinos from $4\pi$ solid angle, with the assumption that the interaction length is
large enough compared to the target thickness that we may neglect self-attenuation within the
target. The number of interactions for this
idealized case is 
\begin{equation}
N_{int,0}~=~ \int_0^T dt \int_{4\pi} d\Omega_{\nu} \int_{V} dV 
\int_{E_{thr}}^{\infty} dE_{\nu}~F_{\nu}(E_{\nu})~\sigma_{\nu N,e}(E_{\nu})~\rho(\vec{r})~N_{A}
\end{equation}
or the same integral as equation~\ref{nint} except that the survival probability
$P_{surv}=1$ for all arrival directions.  With this prescription, the
average acceptance solid angle is given by
\begin{equation}
 \langle \Omega \rangle ~=~ 4\pi~\frac{N_{int}}{N_{int,0}}
\end{equation}

The corresponding number of detections depends both on the interactions and the probability 
$P_{det}$ of detection of the resulting shower. To first order, deep inelastic 
neutrino charged-current interactions lead to an immediate local hadronic shower and a
single charged lepton which escapes the vertex and can subsequently interact. For electron
neutrinos at $10^{18-20}$~eV, this lepton interaction usually takes place rapidly,
and produces an immediate secondary electromagnetic shower, which will be elongated due
to Landau-Pomeranchuk-Migdal (LPM) suppression of the bremsstrahlung and pair-production
cross sections. For
other neutrino flavors, the secondary lepton will propagate long distances through the
medium but can produce detectable secondary showers through electromagnetic (bremsstrahlung or
pair production) or photo-hadronic processes. Since the average Bjorken inelasticity 
$\langle y(E_{\nu}) \rangle \simeq 0.22$ at these energies, the secondary lepton in these charged current
interactions leaves with most of the energy on average, so a secondary shower with
any appreciable fraction of the lepton's energy may exceed the hadronic vertex in
shower energy. Accounting for this, the number of detections can be written
\begin{equation}
N_{det}=\int_0^T dt \int_{4\pi} d\Omega_{\nu} \int_{V} dV \frac{d^3 N_{int}}{dt~d\Omega_{\nu} dV} 
\left [ P_{det,h}(yE_{\nu},\vec{r}_{h},\theta_{\nu},\phi_{\nu}) +
\alpha_{cc}P_{det,\ell}((1-y)E_{\nu},\vec{r}_{\ell},\theta_{\nu},\phi_{\nu}) \right ] 
\end{equation}
where $P_{det,h}(yE_{\nu},\vec{r_{h}},\theta_{\nu},\phi_{\nu})$ is the detection probability for
the hadronic showers as a function of shower energy $yE_{\nu}$, shower centroid position $\vec{r}_h$,
and shower momentum angles $\theta_{\nu},\phi_{\nu}$. The corresponding detection probability for
subsequent lepton showers is written as 
$$\alpha_{cc}P_{det,\ell}((1-y)E_{\nu},\vec{r}_{\ell},\theta_{\nu},\phi_{\nu})$$ 
where $\vec{r}_{\ell}$ is the centroid of the leptonic shower and  
$\alpha_{cc} = 1,0$ for charged- or neutral-current interactions respectively 
(the latter have zero detection probability since the outgoing lepton is a neutrino; 
for simplicity in this present treatment we do not
include subsequent possible neutrino interactions from neutral-current events 
within the detector and their contribution is small in any case).

Once the two integrals given by $N_{int}$ and $N_{det}$ are determined, the volumetric 
acceptance is given by 
\begin{equation}
\langle \mathcal{V}\Omega \rangle = \frac{N_{det}}{N_{int}}~V_0 \Omega
~=~ 4\pi V_0 \frac{N_{det}}{N_{int,0}}
\end{equation}
where $V_0$ is the total physical volume over which the trial neutrino flux arrives.

\subsection{Monte Carlo Implementation}

We have developed two largely independent simulation codes, one at the University of Hawaii,
hereafter denoted the Hawaii code, and a second originally developed at UCLA, but now maintained at
University College London, hereafter denoted the UCL code. Although some empirical
parameterizations (for example, ice refractive index vs. depth) may use common code or
source data, the methodologies of the two codes are entirely independent.

\begin{figure}[htb!]
\centerline{\includegraphics[width=4.5in]{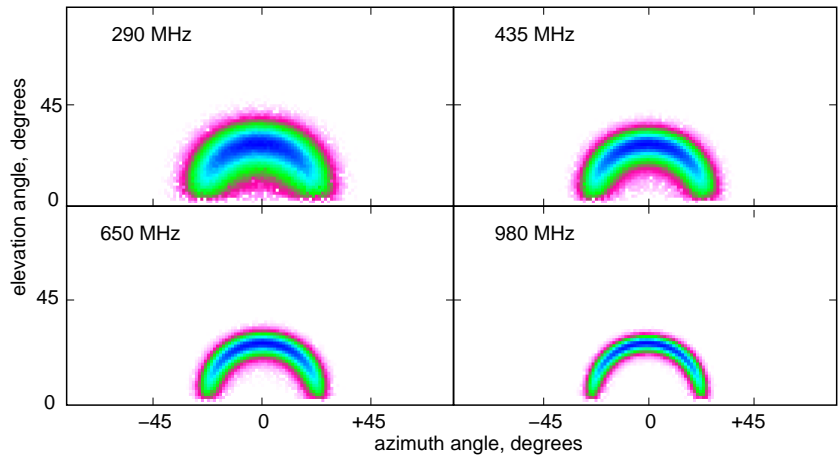}}
\caption{Simulated false-color images of relative E-field strength 
of the emerging Cherenkov cone
as projected onto the sky due to a neutrino event at 
energy of order $3 \times 10^{18}$~eV, with 
an upcoming angle of several degrees relative
to the local horizon. The different panes show the
response at different radio frequencies as indicated. The color
scale is normalized to the peak (blue) in each pane; in fact
the higher frequencies peak at higher field strengths. 
\label{skyimg} }
\end{figure}

Both the Hawaii and UCL simulations of the ANITA experiment estimate the experiment's sensitivity
from a sample of simulated ultra-high energy neutrino interactions in Antarctic ice. 
The programs use variations of ray tracing methods to propagate the 
signal from the interaction to the payload antennas, and then
through the three-level ANITA instrument trigger. The neutrino events can be
drawn from a parent sample that match any given source spectrum, or, for more
general results, the acceptance can be estimated as a function of energy by
stepping through  a series of monoenergetic neutrino fluxes to map out
the energy dependence. This latter approach then allows for estimates
of the acceptance which can be convolved with any input neutrino flux.

\begin{figure}
\begin{center}
\centerline{\epsfig{figure=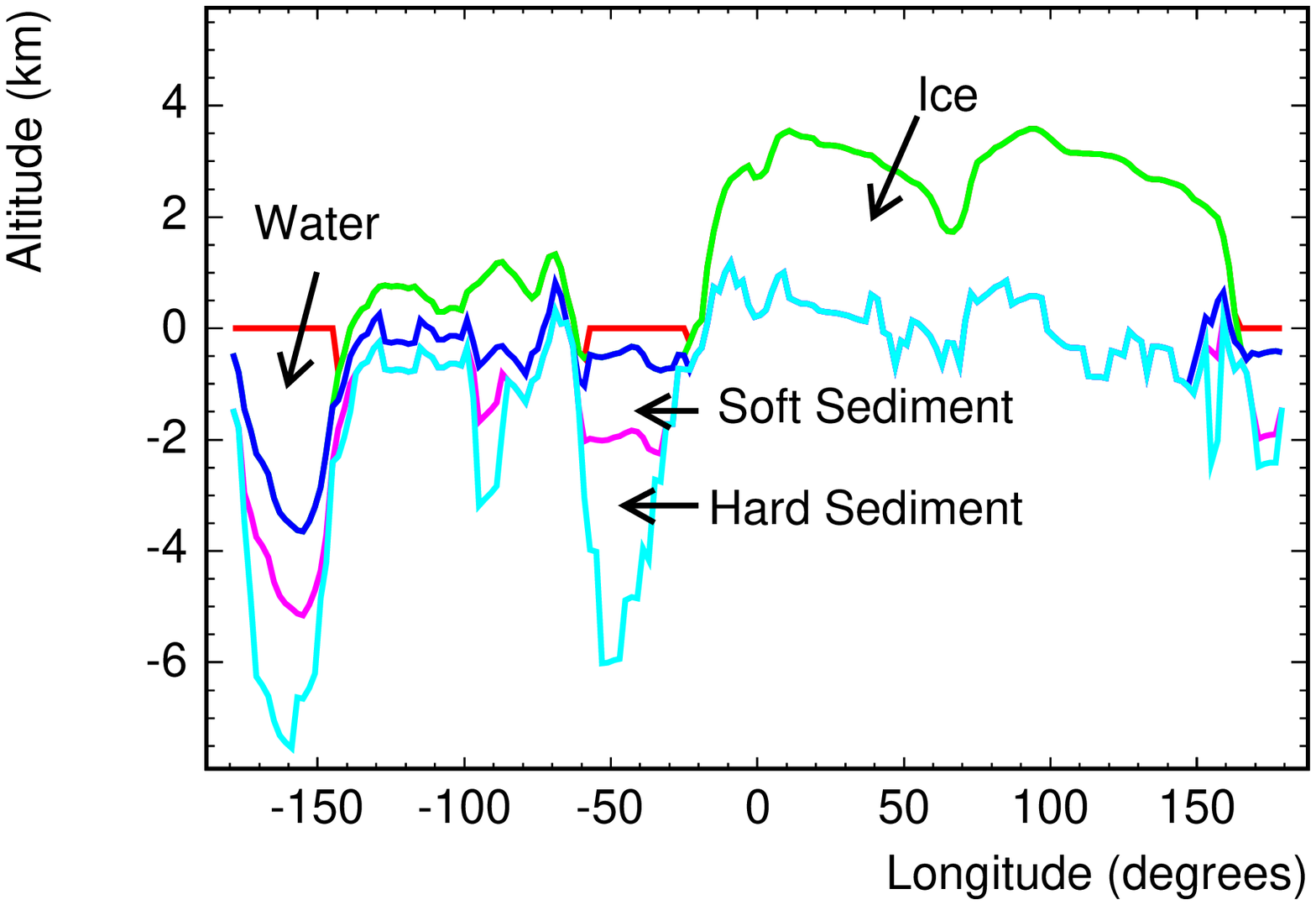,width=3in}~~\epsfig{figure=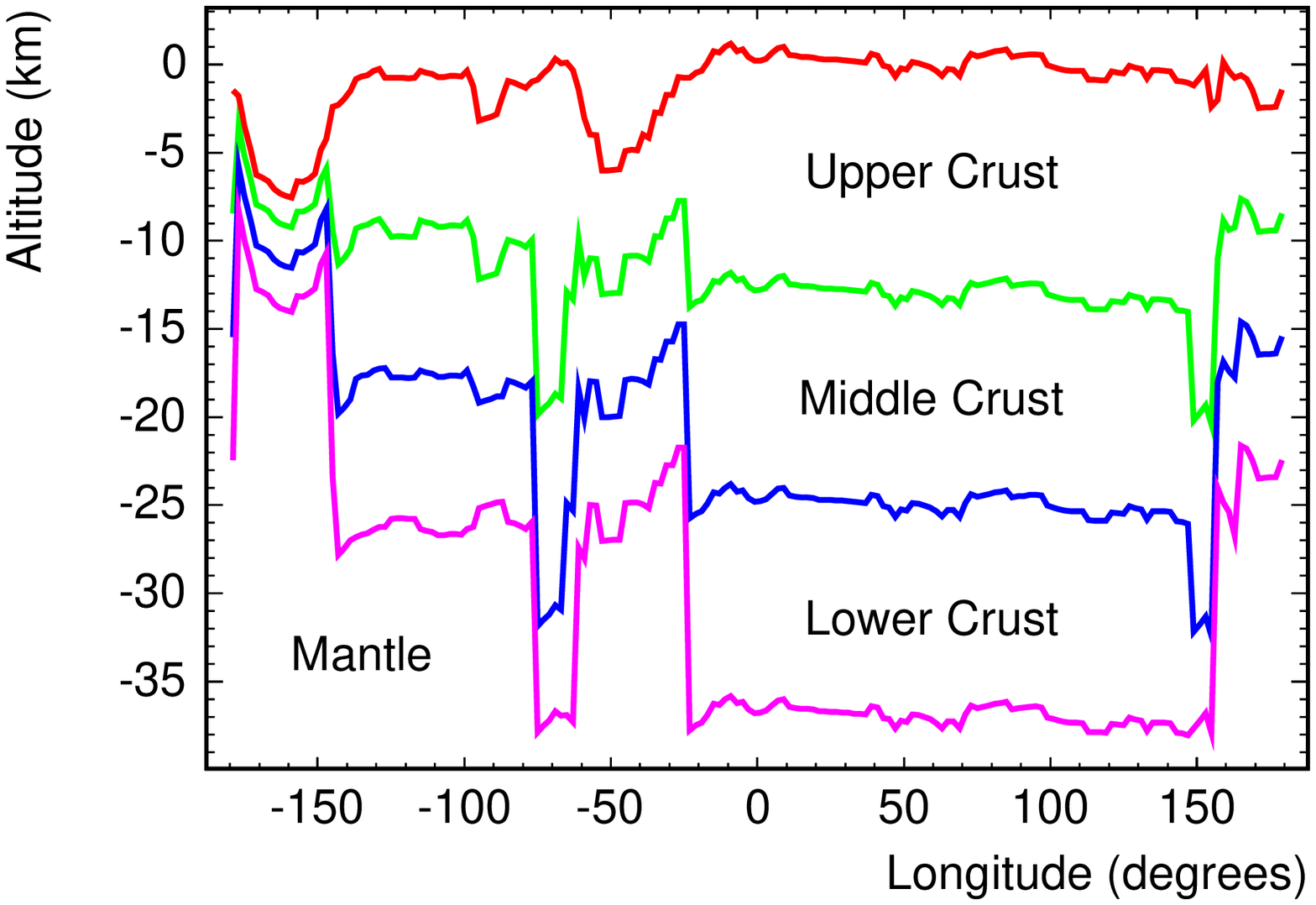,width=3in}}
\caption{Left: Altitude of the upper four layers given in Crust 2.0 along the
the $75^{\circ}$ S latitude line.  The horizontal axis is degrees in
longitude. Right:  Altitude of the lower three layers given in Crust 2.0 along the
the $75^{\circ}$ S latitude line.
\label{fig:layers}}
\end{center}
\end{figure}

The two simulations also differ in the areas in which they focus their 
highest fidelity. The Hawaii code attempts to model the entire pattern of the sky brightness
of radio emission
produced by a neutrino interaction, creating a library of these events which
is later sampled for different relative sky positions of the payload with respect to
the events. Fig.~\ref{skyimg}(left) shows
one example of this type of modeling, in this case for a smooth ice surface.
This sky-brightness modeling includes also a first-order model for
the wavelength-scale surface roughness of the ice. The method uses a facet-model of
the surface, and probes it with individual ray traces that are distributed in their
number density according to the Cherenkov cone intensity 
(see Fig.~\ref{ERays} included in an Appendix). 
This surface ray-sampling thus gives a first-order
approximation that has been found to give reasonable fidelity for such effects
as enhanced transmission coefficients beyond the total-internal-reflectance angle,
for example. Additional detail regarding surface roughness effects are included 
in an Appendix. 

The Hawaii code does not attempt
detailed modeling of the physical topography of Antarctica but instead uses a standard ice
volume with homogeneous characteristics, and estimates of the actual
ANITA flight acceptance use a weighted-average based on the piecewise geography
of the flight path.
 
The UCL simulation, in contrast, does include a detailed modeling of the Antarctic
ice sheets and subcontinent based on the CSEDI Crust 2.0 model of
the Earth's interior, and the BEDMAP model for Antarctic ice
thicknesses. Figure~\ref{fig:layers} shows the 
elevations of the seven crustal layers included in Crust 2.0; the BEDMAP database
provides similar resolution.
This approach provides higher fidelity estimates of the instantaneous
ice volume below the payload for any given time during the flight. This approach
toward event modeling, with its limited sampling of ray propagation, does not easily lend itself to treatment of surface
roughness however, but is optimized for speed, using a principal-ray approximation
for the neutrino interaction-to-payload radiation path. Thus each event is 
observed along a single ray refracted through the surface from the cascade location.

\begin{figure}[htb!]
\includegraphics[width=3.85in]{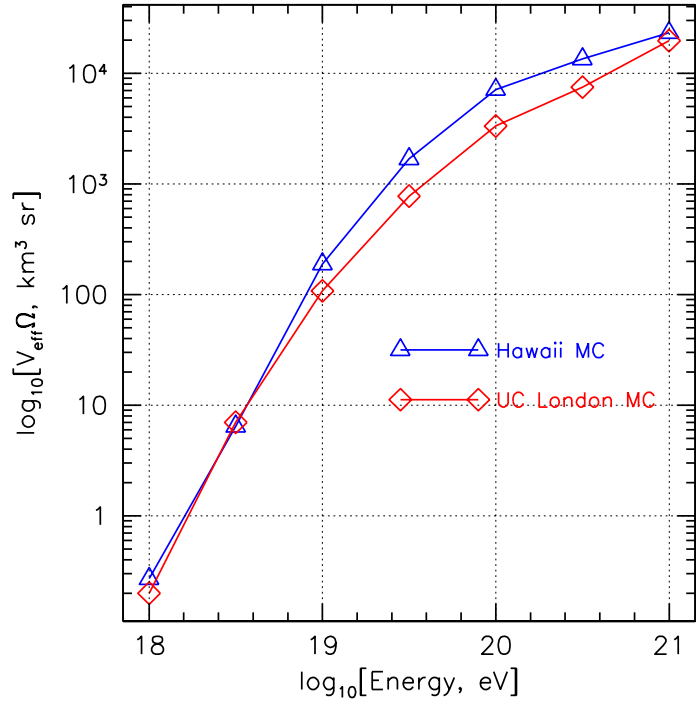}
\caption{Curves of simulated acceptance vs. energy for the two 
independent ANITA Monte Carlos.
\label{Veffsr} }
\end{figure}

Despite their distinct differences in approach, the two methods have 
converged to good agreement in the total neutrino aperture, when compared for
a given standard ice volume. Details of the two methods are described in two
appendices.

In Fig.~\ref{Veffsr} we show the results of the two estimates of total effective
volumetric acceptance (volume times acceptance solid angle for an isotropic source
such as the cosmogenic neutrinos) as estimated by the UH and UCL simulations.
The values may be thought of as the total water-equivalent volume of material 
that has 1 steradian of acceptance to a monoenergetic flux of neutrinos at 
each plotted energy. It is evident that there is reasonably good agreement
for ANITA's acceptance, with values that differ by of order a factor of 2
at most.

In practice the solid angle of acceptance for any volume
element of the ice sheet surveyed by ANITA is in fact very small, and the 
acceptance is also quite directional, covering only a relatively small band of sky at
any given moment. The actual sky coverage will be shown in the next section.

\section{Pre-flight Calibration}

In June of 2006 prior to the ANITA flight, the complete ANITA payload and flight system
was deployed in End Station A at the Stanford Linear Accelerator Center (SLAC)
about 13~m downstream of a prepared target of 7.5 metric tons of pure carving-grade ice
maintained at -20C in a refrigerated container. Using the 28 GeV electron
bunches produced by SLAC, electromagnetic showers were created in the ice
target, resulting in Askaryan impulses that were detected by the payload 
with its complete antenna array, trigger, and data acquisition system. This beam test
experiment thus provided an end-to-end calibration of the flight system,
and yielded on- and off-axis antenna response functions for ANITA as well as a separate
verification of the details of the Askaryan effect theory in ice as
the dielectric medium. 

\begin{figure}[htb!]
\includegraphics[width=5.5in]{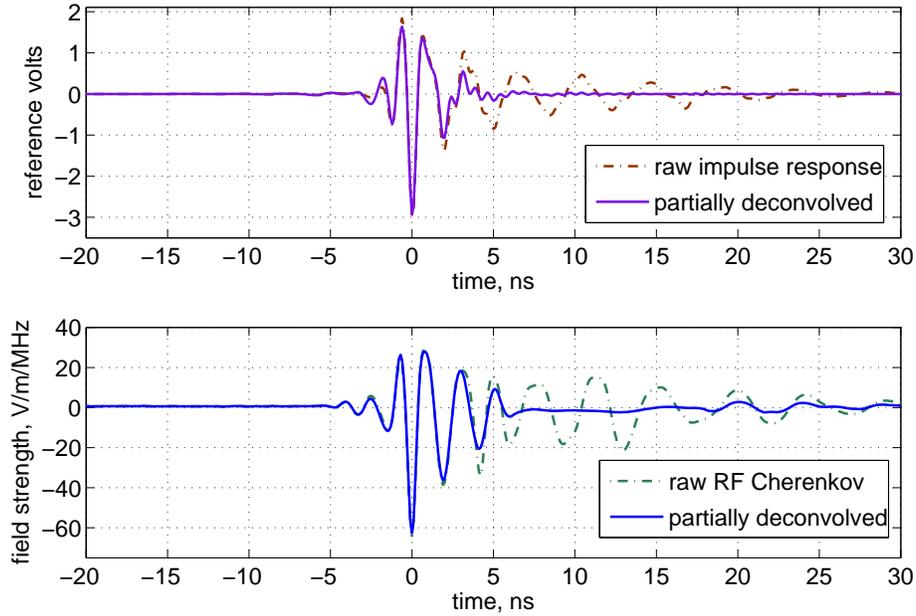}
\caption{ Top: measured ANITA quad-ridge horn impulse response. Dashed lines 
include the strong
nonlinear group-delay component near the horn low-frequency cutoff, and the
solid line removes this low-power component using wavelet deconvolution, for
better clarity. Bottom: waveform measured by ANITA antennas during SLAC
calibration using direct Askaryan impulse signals from ice; dashed and solid lines
are the same as in the top pane of the figure.
\label{wfm_ref} }
\end{figure}

A complete report of the results for the Askaryan effect
in ice has been presented in a separate report~\cite{slac07}, which
we denote as paper I, and we 
do not reproduce that material here. However, antenna response details were
not reported in paper I, so we provide those here since they are relevant
to the ANITA-1 flight. Fig.~\ref{wfm_ref}(top) shows an antenna impulse response
function, 
and (bottom) the same set of response functions for the actual Askaryan signals
recorded during the SLAC beamtest. The dashed and solid lines show
both raw and partially deconvolved to remove the lowest-frequency component,
which gives the long tail of $\sim 200$~MHz ringing that is seen in the plots.
This tail is due to the rapid rise in group-delay as one approaches the
low-frequency cutoff of the antenna, and although it can be useful as
a kind of ``fingerprint'' of the antenna response in the presence of noise,
it does not contribute much power to the actual trigger. 
The solid lines show the response function with the low-frequency
component removed by wavelet deconvolution, showing the dominant power
components. These data may be compared to the laboratory measurements in
Fig.~\ref{impulse} above.

\section{In-flight Performance}

ANITA's primary operational mode during the flight was to maintain 
the highest possible sensitivity to radio impulses with nanosecond-scale
durations, but the trigger was designed to be as loose as possible to
also record many different forms of impulsive interference. The
single-band thresholds were allowed to vary relative
to the instantaneous radio noise. The system that
accomplished this noise-riding behavior utilized
a Proportional-Integral-Differential (PID) servo-loop
with a typical response time of several-seconds based on several Hz
sampling of the individual frequency band singles rates. 

\subsection{Antenna temperature}

\begin{figure*}[htb!]
\includegraphics[width=6.95in]{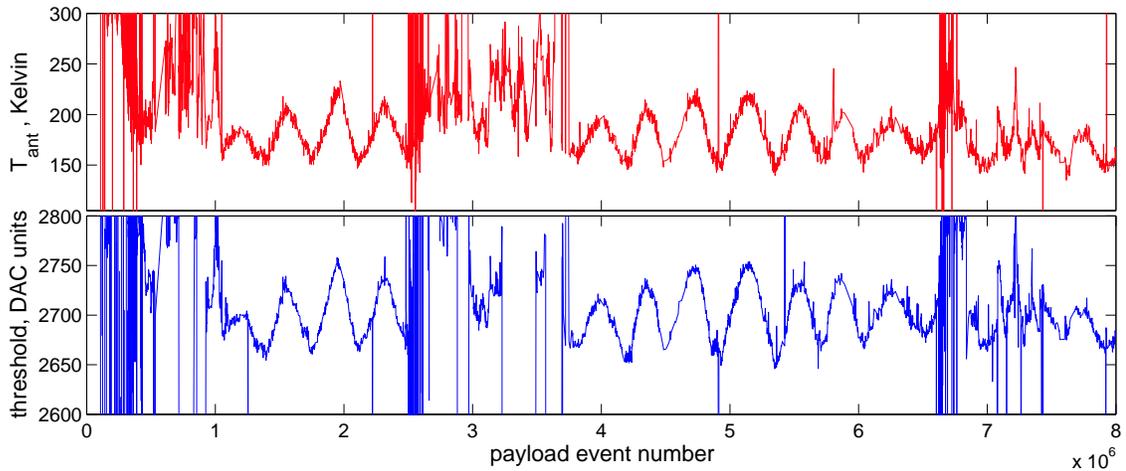}
\caption{Top: Antenna effective temperature vs. event number (at an average
rate of about 4.5~Hz) during the ANITA flight.
The effective temperature is the average over the antenna beam including 230K ice and
5-100K sky, with varying contributions from the Sun and Galactic Center
which produce the diurnal modulation. Modulation due to free rotation of
the payload over several minute timescales has been averaged out.
Bottom: The noise-riding antenna average threshold is shown on
the same horizontal scale, showing the servo response between the instrument
threshold and the apparent noise power. This noise-riding approach was used to
retain a relatively constant overall trigger rate, giving stability to the
data acquisition system.
\label{rfpower} }
\end{figure*}

The power received by the antennas in response to the ambient
RF environment directly determines the limiting sensitivity of the
instrument. This received system power $P_{SYS}$, referenced to the
input of the low-noise amplifier (LNA)
is often expressed in terms of the
effective temperature, via the Nyquist relation~\cite{Nyquist},
which can be expressed in simple terms as:
\begin{equation}
P_{SYS} ~=~ kT_{SYS} ~=~ k(T_{ANT}~+~T_{RCVR})
\end{equation}
where $k$ is Boltzmann's constant, $T_{ANT}$ is the effective antenna
temperature, which includes contributions from the apparent sky 
(or sky+ice in our case) temperature in the field of view, along with
antenna mismatch losses; $T_{RCVR}$ is the additional noise added 
by the front-end LNA, cables, and filters, and second-stage amplifiers
and other signal-conditioning devices, all of which constitute the
receiver for the incident signal. $T_{RCVR}$, which is an
irreducible noise contribution, is usually fixed by the
system design, and is calibrated separately (c.f. Fig.~\ref{RFCM}). It 
is typically stable during 
operation and can be used as a reference to estimate additive noise. 
In contrast $T_{ANT}$ may vary significantly during a flight
and must be monitored to ensure the best possible sensitivity of the
instrument.

Absolute monitoring of the sensitivity for ANITA was accomplished by scaling
the measured total-band power
from the calibrated noise figure of the front-end receivers. This
power was separately sampled at several Hz for each of the input channels,
using a radio-frequency power detector/amplifier based on the MAXIM Integrated Circuits 
MAX4003 device,
calibrated with external noise diodes. The absolute noise power
was also cross-calibrated by comparison with known noise temperatures
for the galactic plane and sun which were in the fields-of-view of the 
antennas throughout the flight, modulated by the payload rotation
and by diurnal elevation changes.

\begin{figure*}[htb!]
\includegraphics[width=5.5in]{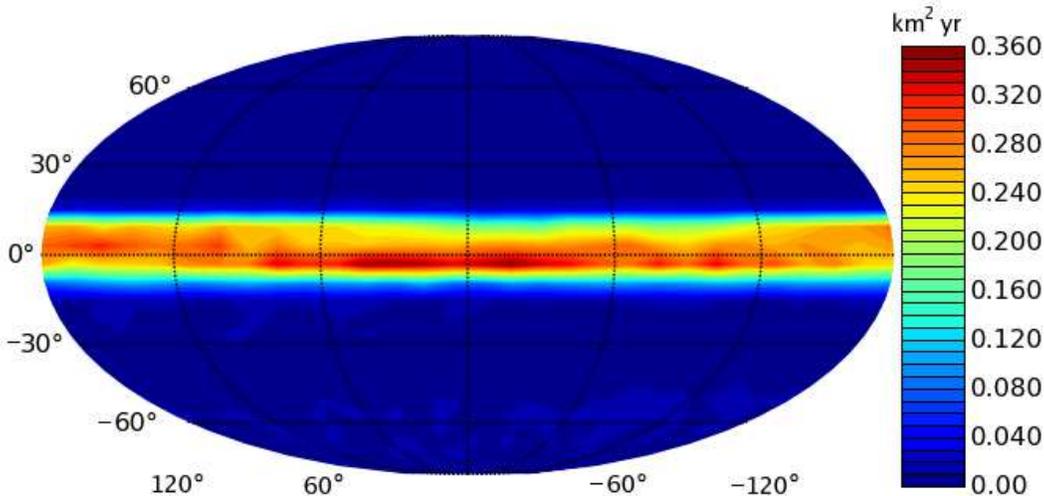}
\caption{A false-color map of sky exposure quantified by
total neutrino aperture (here given for 
$E_{\nu}=10^{20}$~eV) as a function of right-ascension and declination for
the ANITA-1 flight. 
\label{skymap} }
\end{figure*}

A plot of the measured antenna temperature, derived from the RF power 
measurements, is shown in Fig.~\ref{rfpower} as a function of the 
payload event number recorded during flight, which may be converted to
payload time in seconds using a mean trigger rate of about 4~Hz. 
Several features of this plot require explanation.
A firmware error in a commercial Global-Position System (GPS) 
unit used for  flight synchronization caused a series of in-flight timekeeping errors
which led to a drift in payload time vs. real time; this was corrected 
in the data after the flight. Starting on day 21 (roughly event number 7M) of the flight,
the flight computer began to experience anomalies which led to a series of
reboots of the system, and increased deadtime. This behavior persisted
until the termination of the flight after 35 days aloft. Beyond day 21,
although we continued to take triggered data with an effective livetime
of about 40\%, the environmental sensor data is fragmented and 
only a portion is shown here.

Values which rise off-scale on the vertical axis indicate periods when 
EMI dominated the system noise, as ANITA's trajectory came within
the horizon of either McMurdo Station or Amundsen-Scott Station, both
of which had strong transmitters in the ANITA bands. Other periods are
largely quiescent as indicated by the plot; in particular the period
between events 4M to 6.6M. During these periods the observed antenna temperature is dominated by
the 230K ice in the lower portion of the antenna fields of view, 
averaged with the  cold sky (3-10 K over our band) in the upper portion.
The plotted values are averaged over all payload rotation angles (ANITA
was allowed to freely rotate around its balloon suspension point).
The observed modulation of the antenna temperature is thus due the
diurnal changes in elevation angles of the Galactic Center and Sun,
and has the expected amplitude for these sources over our frequency band,
dominated by the nonthermal rise at the lower end of our band near
200~MHz, where the antenna fields of view are also the largest.

\subsection{Navigation performance}

Figure~ref{GPS} summarizes the navigation performance during the flight. Reconstruction
of the ground position of any source that is detected during flight depends
on accurate knowledge of the payload position and attitude. There are six degrees
of freedom: latitude, longitude, altitude, orientation, pitch, and roll, and these
are all determined via the ADU5 system on ANITA. Redundant information with somewhat
lower precision is also obtained by a sun-sensor and magnetometer suite.

\begin{figure*}[htb!]
\centerline{\includegraphics[width=6.5in]{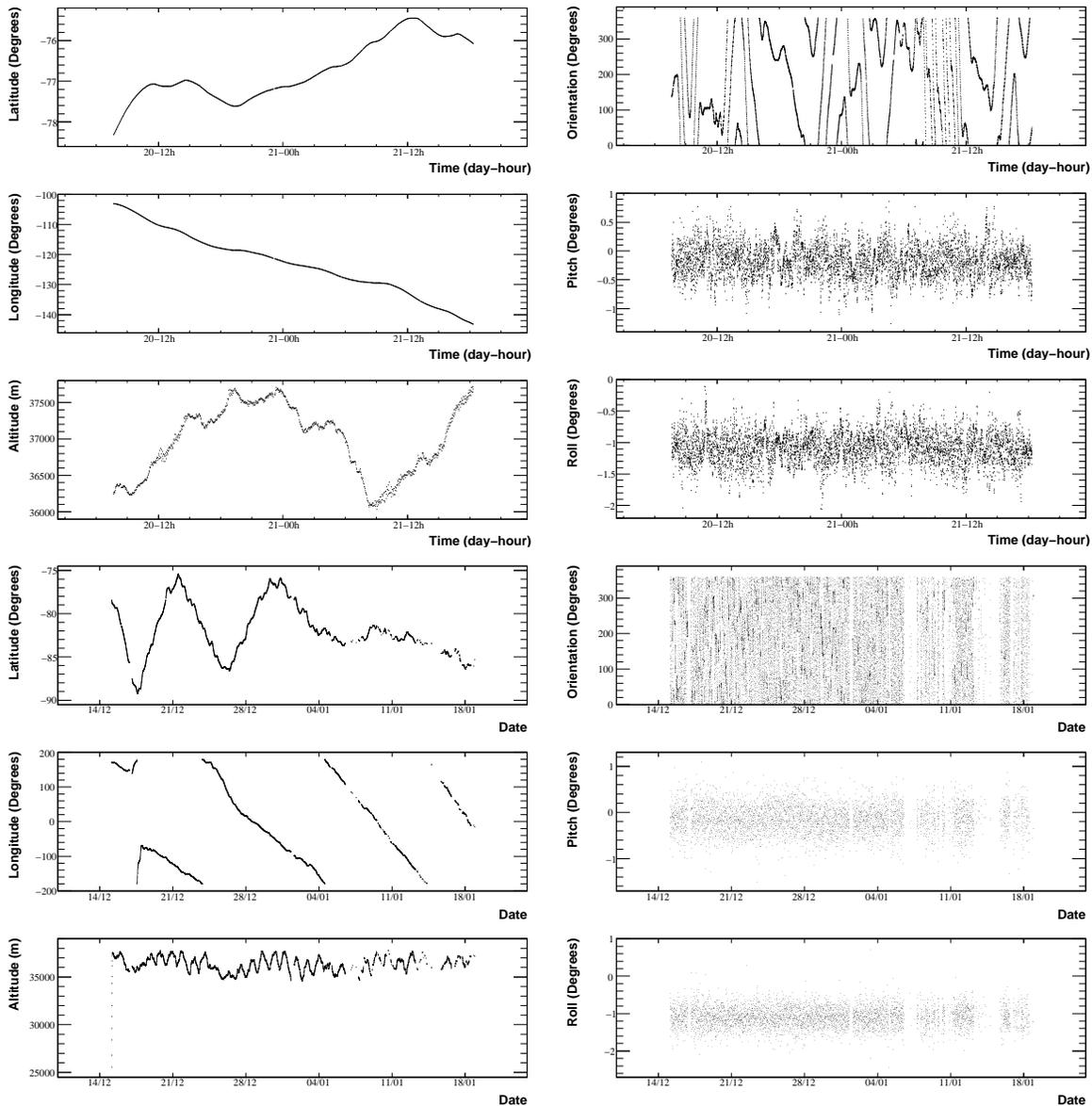}}
\caption{ A composite set of plots representing the in-flight performance of ANITA's
navigation system. The left side plots are for geodetic position and altitude,
and the right-side plots show the attitude parameters. The top 6 plots zoom in on a single
data run, covering several flight days. The corresponding plots on the bottom half of the
figure show the total time series for the full flight. Data for the orientation changes rapidly
as the payload is allowed to freely rotate; thus the individual rotation cycles are difficult
to see in the entire flight's data but show up clearly for the single data run. 
\label{GPS} }
\end{figure*}

Fig.~\ref{GPS} shows several features that illustrate important aspects of the
navigation system. The payload was freely-rotating; that feature is illustrated by the
orientation plots in the figure, which show the often rapid change of azimuth for the
fixed reference coordinate system on the payload as it rotated with periods varying
from minutes to hours during the flight. The payload pitch and roll are somewhat
ill-defined for an azimuthally symmetric payload, but they do indicate that the
reference plane of the GPS was slightly tilted, by about $1^{\circ}$ in the
``roll'' direction with respect to the horizontal. Overall, the performance of
this navigation subsystem was within the flight requirements, and helped to ensure
the accuracy of the direction reconstruction of impulsive radio sources detected throughout the flight.

\subsection{Flight path}

\begin{figure*}[htb!]
\centerline{\includegraphics[width=3.1in]{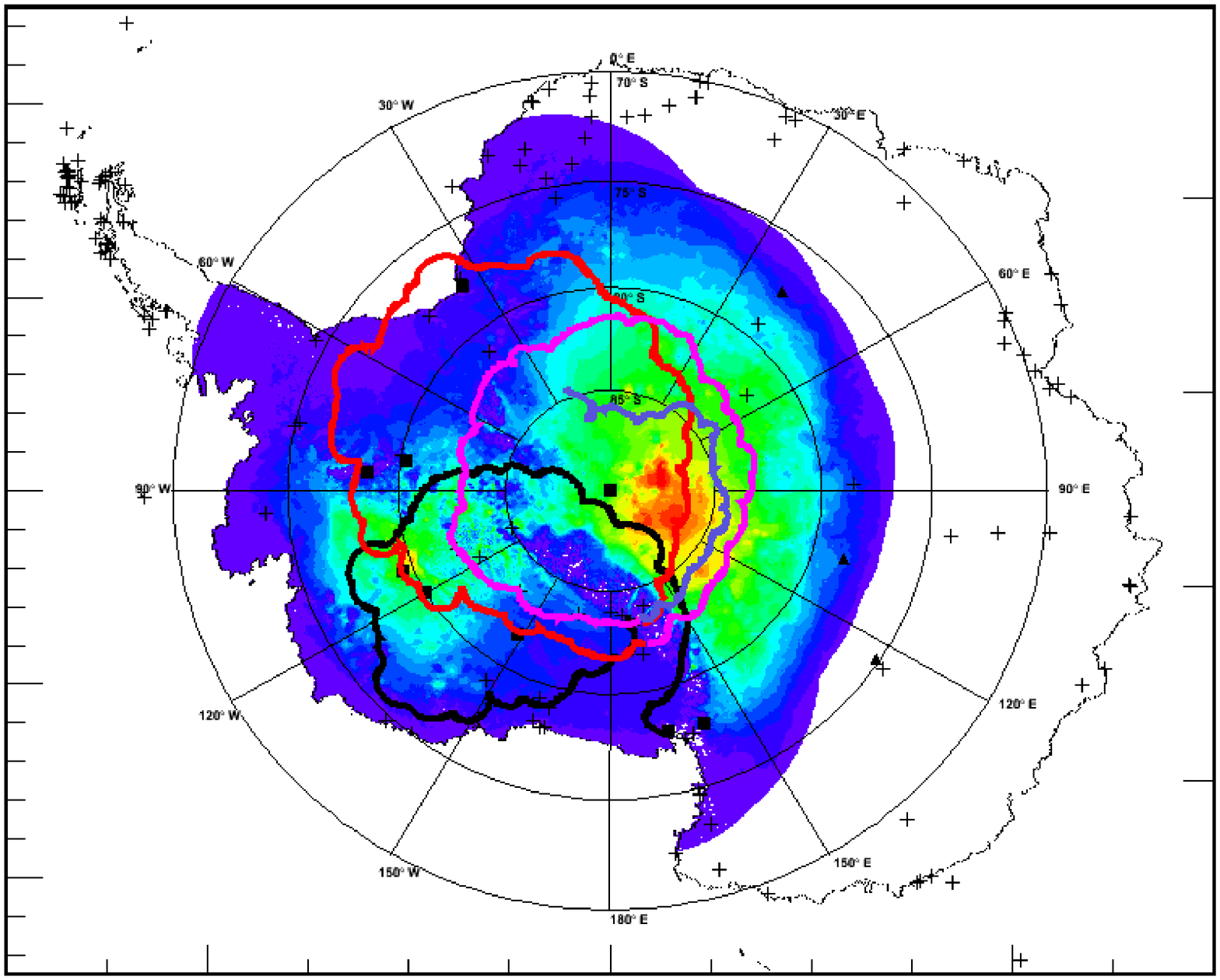}~~\includegraphics[width=3.5in]{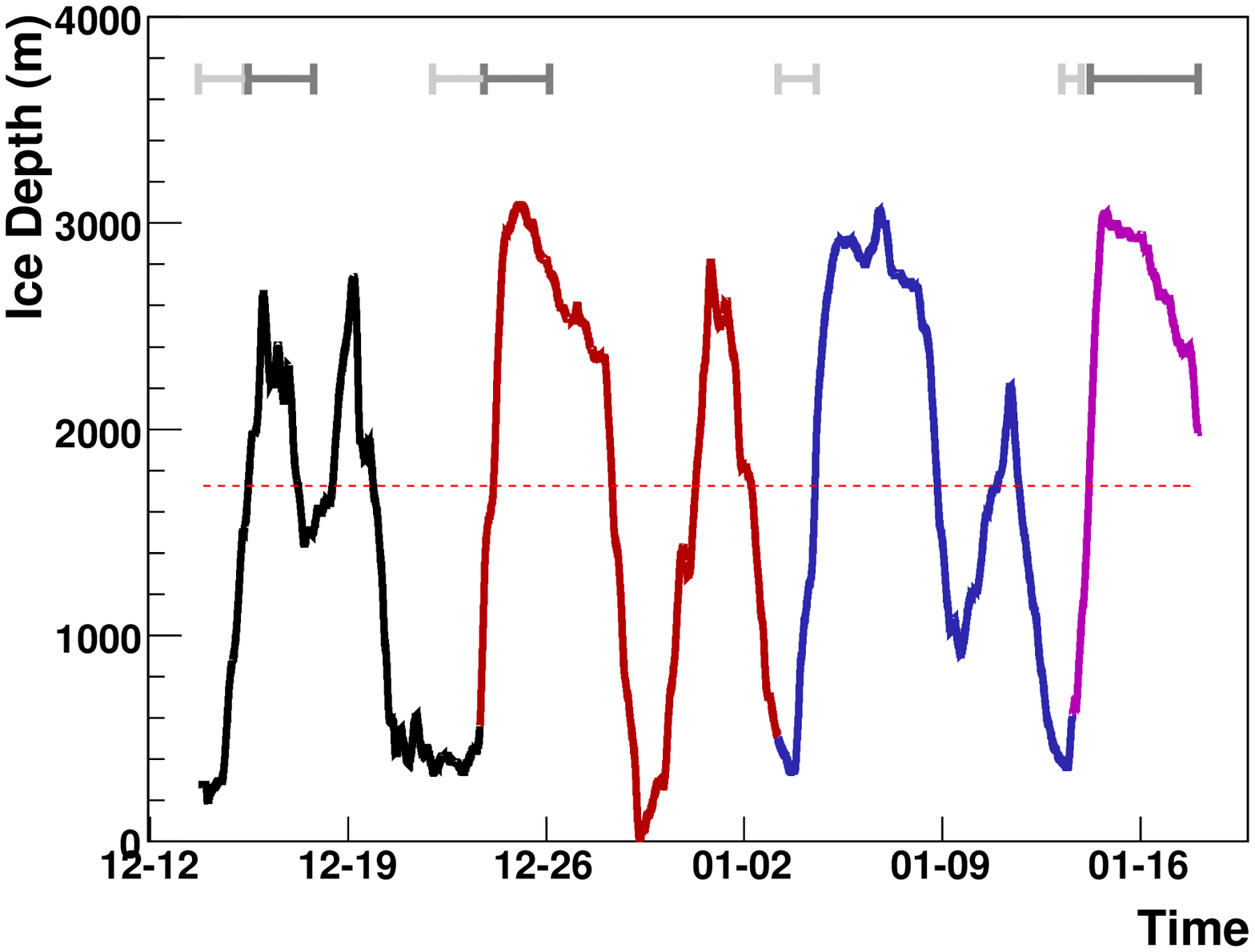}}
\caption{Left: Flight path for ANITA-1, with depth of ice plotted. Only ice within the actual field-of-view of the payload is
colored. Ice depths: red: $>4$km; yellow: 3-4km; green:2-3km; light blue 1-2km, blue: $<1$km. Right: average
depth of ice within the payload field of view during the flight. Because the average includes all ice within
the field of view, the minima and maxima of this curve do not correspond to the minima and maxima in the
actual localized ice depths.
\label{flightpath} }
\end{figure*}

During the austral summer in the polar regions, the polar vortex usually establishes a nearly 
circumpolar rotation of the upper stratosphere, and payload trajectories are often
circular on average, even over several orbits. During the austral summer of 2006-2007 however
the polar vortex did not stabilize in a normal configuration, but remained rather weak and
centered over West Antarctica. This led to an overall offset of the centroid of the
orbital trajectories for ANITA-1, as seen in Fig.~\ref{flightpath}(left).

This anomalous polar vortex had two consequences for ANITA's in-flight 
performance: First, it led to a much larger dwell time in regions where there
were higher concentrations of human activity, primarily McMurdo and Amundsen-Scott
South Pole Stations. The presence of these and other bases within the
field-of-view of the payload led to a higher-than expected rate of noise
triggers and compromised the instrument sensitivity over a significant portion of the
flight.  Second, the higher-than-normal dwell time over the smaller and thinner
West Antarctic ice sheet,
rather than the main East Antarctic ice sheet, led to a lower overall exposure of
ANITA to deep ice. The first consequence is apparent in Fig.~\ref{rfpower}, where
it is evident that as much as 40\% of the flight suffered some impact on 
sensitivity due to the presence of noise sources in the field of view. The 
second consequence is quantified in Fig.~\ref{flightpath}(right) where we
plot the field-of-view-averaged depth of the ice along the flight trajectory, along
with the overall average. Despite the large dwell time over West Antarctica, ANITA
still achieved an average depth of ice during the entire flight of more than 1500~m.

\subsection{Sky coverage}

In Fig.~\ref{skymap} we show a false-color map 
of the relative neutrino sky exposure in
Galactic coordinates for ANITA's flight. The exposure is determined both
by the geometric constraints of the flight path, and by the neutrino
absorption of the earth, which limits the acceptance to within several
degrees of the horizon at any physical location over the ice. It
is evident from this plot that ANITA had significant exposure both to
the galactic plane and out-of-plane sky, and although the total sky solid
angle covered is only of order 10\%, the exposure is adequate to constrain
a diffuse neutrino flux component, as well as provide some constraints
on point sources within the exposed regions.

\subsection{Data quality and volume}

ANITA recorded approximately 8.2~M RF-triggered events during the 35 day
flight, and for each of these events all of the antennas and different
polarizations were read out and archived to disk. The system was able
to maintain a trigger rate of between 4 and 5 Hz during normal operation
without any significant loss, and this rate, which was determined by
the individual antenna thresholds, met the pre-flight sensitivity goal
for periods when the payload was not in view of strong 
anthropogenic noise sources. In practice, less than 1\% of all recorded
triggers do actually arise from a coherent plane-wave impulse arriving from
an external source. The vast majority are random thermal-noise coincidence
events, which are recorded to provide an ongoing measure of the 
instrument health, and separate validation of the sensitivity.

\begin{figure*}[tb!]
\includegraphics[width=6.5in]{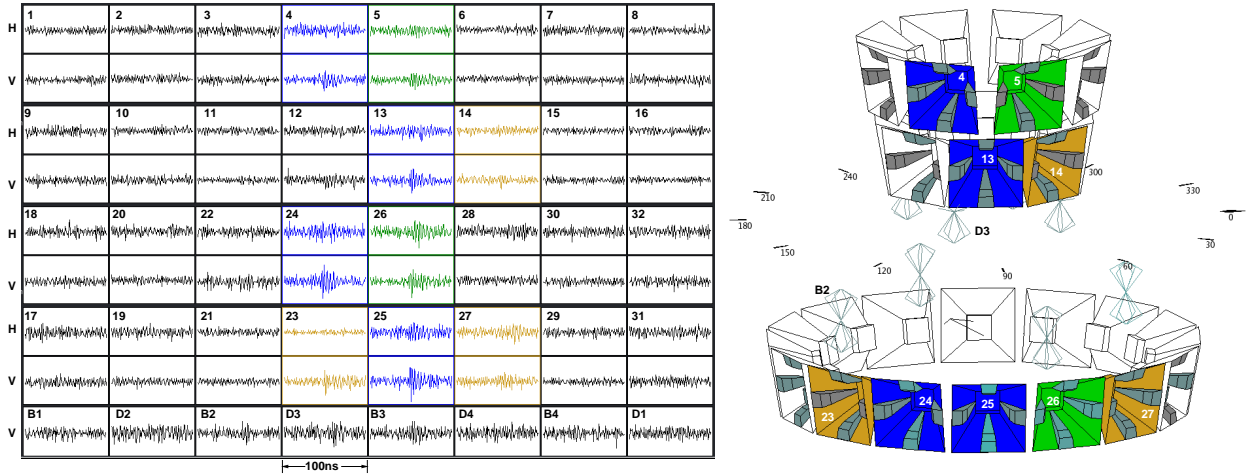}
\caption{Plot of an example triggered event (attributed to interference from a West Antarctic encampment)
during the ANITA flight, as viewed with the realtime data viewers, both as a 
set of the entire payloads waveforms (left) and the trigger antenna geometry (right).
Colors indicate the trigger hierarchy: yellow indicates L1 trigger, green L2, and
blue a global L3 trigger.
\label{7767328} }
\end{figure*}

The performance of the data acquisition system for non-thermal events
is best summarized by showing
a typical impulsive event as recorded by the system and displayed by
the in-flight monitor on the ground receiving computer system. A partial
display of event 7767328 is shown in Fig.~\ref{7767328}. On the left are
the recorded waveforms for this event, which clearly show an impulsive signal
superposed with thermal noise on the left, where all 72 antenna signals are
displayed. Vertical columns are organized in azimuth around the payload,
with horizontal rows providing both the vertical and horizontally polarized
signals from the vertically offset antenna rings of the payload. The
physical geometry of the arriving signal is indicated in the payload
view to the right, where colors indicate
which ones produced the actual triggers and at what level. These colors also
match up with the colored waveforms on the left. 

These raw-data waveforms, while appearing somewhat noisy, actually have
quite high information content, with up to ten usable signals detectable above
thermal noise even in the raw data in this event. After processing, which
typically includes a matched-filter correlation designed to optimize the SNR of
broadband impulsive events, such signals are used to provide
strong geometric constraints on the arrival direction of the event, as
we will detail in a later section.

\subsection{Ground-to-payload calibration pulsers.}

The ANITA instrument requires the highest possible precision in reconstructing
the arrival direction of any incoming source event so that it may 
reject any impulses that might mimic neutrino signals but arise from
anthropogenic electromagnetic interference. In addition, ANITA must
verify the in-flight sensitivity of the trigger system to external events.
To accomplish this, we developed a ground-based calibration approach which
relied on pulse transmitters which could be directed at the payload from
several sites during the flight, and used to establish both pointing accuracy
and sensitivity. Here we describe the performance and results using these
systems.

\label{sec:ground_pulser}

\subsubsection{Description}
\label{sec:gp:description}

We deployed four ground-to-payload transmitter 
antennas at two distinct sites during the 2007 -- 2008 ANITA flight.
The antennas transmitted radio impulses to the payload in order to verify
instrument health and provide a sample of neutrino-like signals for testing our 
analysis codes.  A field team traveled to Taylor Dome ($77^{\circ}$,52' S; $150^{\circ}$,27' E)
and operated a surface 
quad-ridged horn antenna and a borehole discone-type 
antenna in a 100 meter deep borehole, about 30~m below the
firn-ice boundary at Taylor Dome.  Results from concurrent
Taylor Dome experiments by the ANITA team
on ice properties at this site are presented in a separate report~\cite{Besson08}.

The team at 
Williams Field at the LDB launch site 
operated a surface quad-ridged horn antenna and a discone in a 26 meter deep
borehole.  All four antennas transmitted a kilovolt-scale impulse $\approx 500$~ps
wide with a flat frequency spectrum.  The impulse was chosen to be as close as 
possible to theoretical predictions of a neutrino Askaryan signal.  The surface 
antenna at McMurdo could adjust the amplitude and polarization of the impulse and 
occasionally transmitted brief continuous wave (CW) signals.  The surface antennas at both McMurdo 
and Taylor Dome were quad-ridged horn antennas identical to those on the ANITA 
payload.  The borehole antennas were both discone antennas designed and built 
at the University of Hawaii for use in ice.

The payload used two methods of detecting impulses: the standard RF trigger
and a forced trigger using a preset time offset relative to the GPS pulse-per-second (PPS).  
Anthropogenic EMI from McMurdo restricted our ability to operate in RF trigger mode when the 
base was above the horizon.  Despite this noise, we were able to use RF triggers 
for several hours during which noise levels were low.  Both as a supplement to the RF triggered 
events and also for easy event identification, we timed the McMurdo antennas' transmissions 
such that the signals always arrived at the payload at a constant time offset from the GPS
PPS.  These PPS triggers allowed us to record calibration signals during periods that 
would have otherwise been unusable due to anthropogenic noise.

\begin{figure}[thb!]
 \centerline{\epsfig{figure=Cal_rf_map.eps,width=3.3in}\epsfig{figure=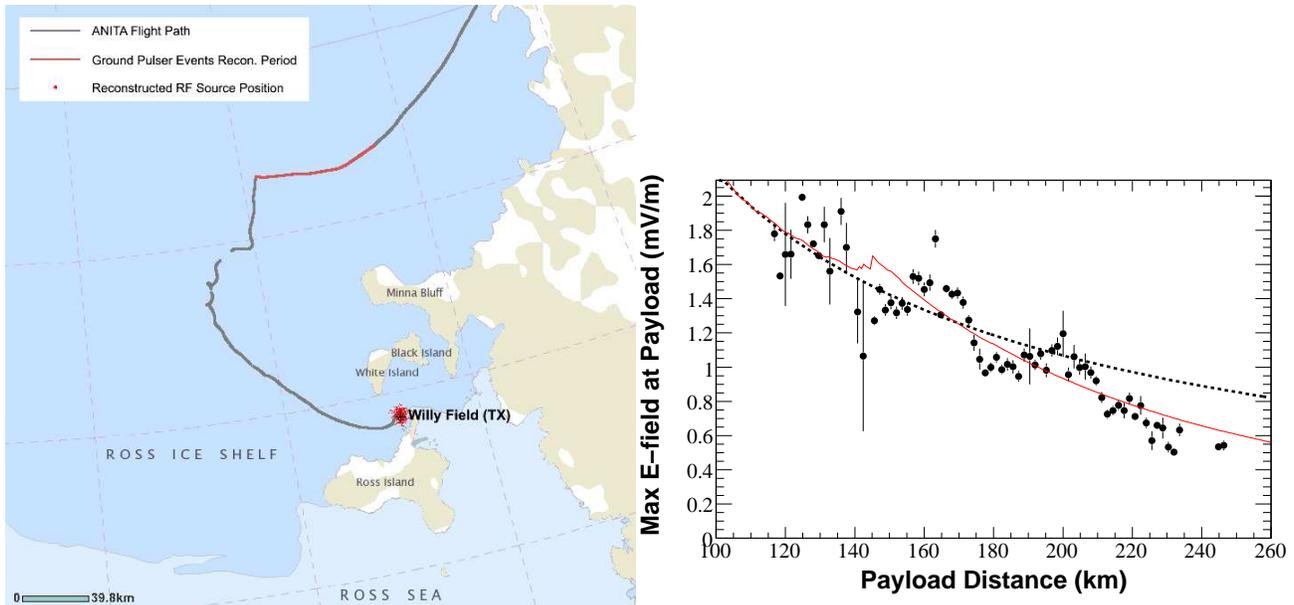,width=3.5in}}
\caption{Left: The flight path of ANITA in the first day after launch, with the 
reconstructed positions of signals from McMurdo ground calibration systems. Red dots 
are projected positions of the reconstructed events, the line is a 
trajectory of the ANITA flight, and the red part of the line is for the flight segment for this reconstruction.
Right: Distance of the ANITA payload from Willy Field versus the maximum E-field of the pulse 
received from the McMurdo ground pulser borehole antenna.  The solid line is a fit to a 1/r curve 
multiplied by an angle-dependent Fresnel factor.  The dashed line is a pure 1/r curve with
arbitrary normalization for reference. (The overall normalization agreed with the
absolute scale to about 10\%).
\label{cal_rf_map}}
\label{fig:gp:dist_vs_max_v}
\end{figure}

\begin{figure}[ht!]
\centerline{\epsfig{figure=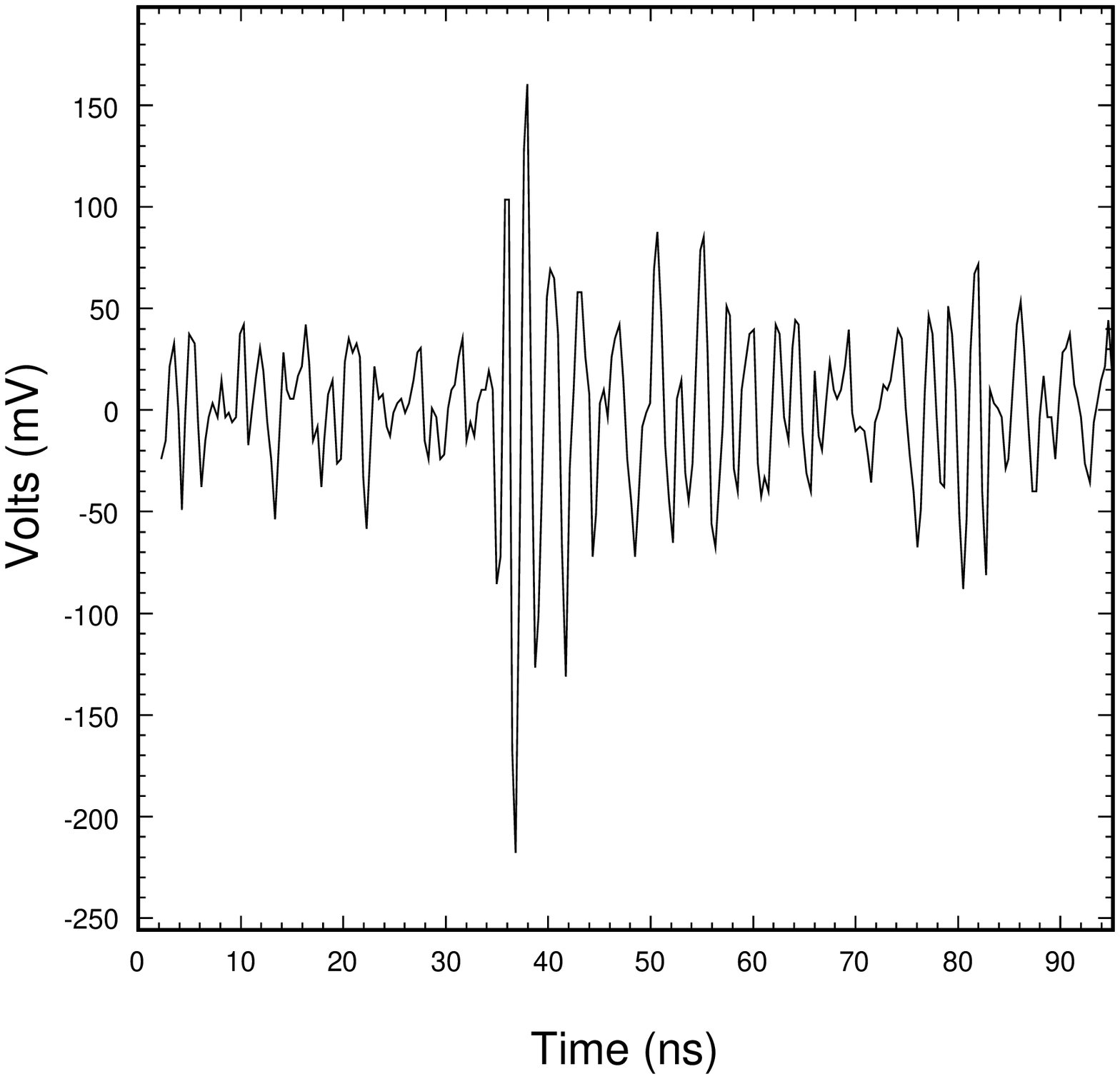,width=3.in}~~~\epsfig{figure=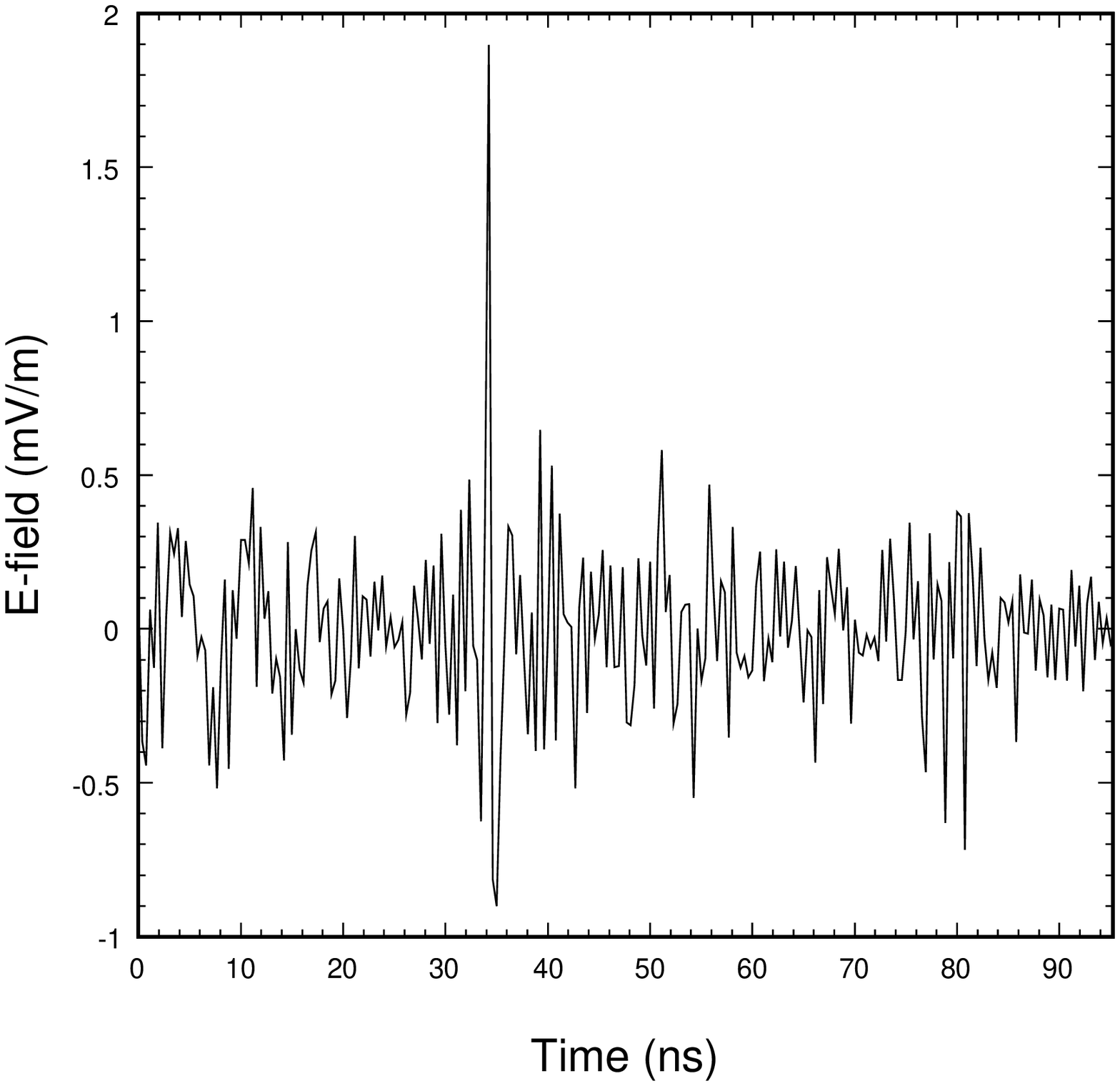,width=3.in}}
\caption{Left: A ground calibration pulse from the McMurdo surface antenna, recorded during the 
2006-2007 ANITA flight.  Right: The same pulse, with instrument gains and group delays removed.}
\label{fig:gp:deconvolution}
\end{figure}

\subsubsection{Uses of Ground Calibration Pulses}
\label{sec:gp:uses}

The ANITA payload successfully recorded tens of thousands of ground pulser signals.  ANITA detected 
signals from the borehole at Williams Field at ranges up to 260 km; Fig.~\ref{cal_rf_map}
shows a portion of the first day's flight path with a segment of the path from which events
were efficiently reconstructed due to a lull in surface noise activity near McMurdo station.  
The surface antennas 
at McMurdo and Taylor Dome could be seen from even further away.  ANITA's flight path
took it no closer than about 250~km to Taylor Dome, and the limited power of the borehole
pulser there, combined with the higher loss in the longer cable needed to accommodate the
much deeper borehole there made it difficult to detect the Taylor Dome borehole pulses,
as most were below the ANITA detection threshold. However, of order 20 of these borehole impulses
were detected, all very close to the instrument hardware threshold.

Figure~\ref{fig:gp:dist_vs_max_v} shows the maximum E-field strength from the McMurdo borehole antenna
at the location of the ANITA instrument, as a function of distance.  More than 10,000 events 
are plotted here.  The line is the predicted peak of the E-field strength, taking into account Fresnel 
factors at the payload's location and the distance of the payload from the transmitting antenna.
This plot illustrates the utility of a calibration source for measuring long-term stability of 
instrument gain.  Changes in ANITA's gain are limited to the $10\%$ level over the day covered 
by the data in Figure~\ref{fig:gp:dist_vs_max_v}, and these constraints on the variation 
include the effects of any variation in surface roughness in the surface Fresnel zone of the
borehole transmitter antenna.

\begin{figure}[htb!]
\centerline{\includegraphics[width=3.3in]{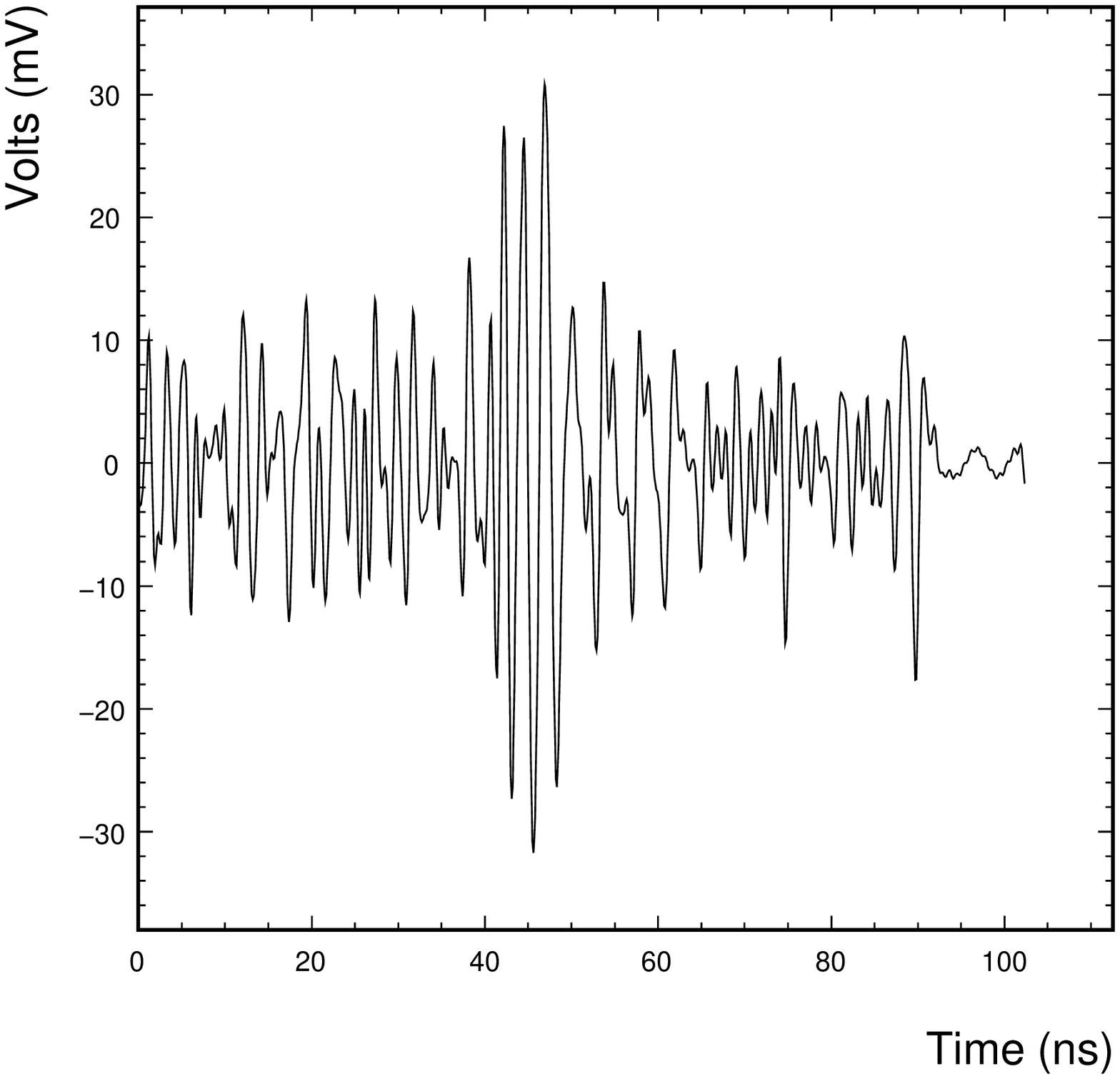}~~\includegraphics[width=3.3in]{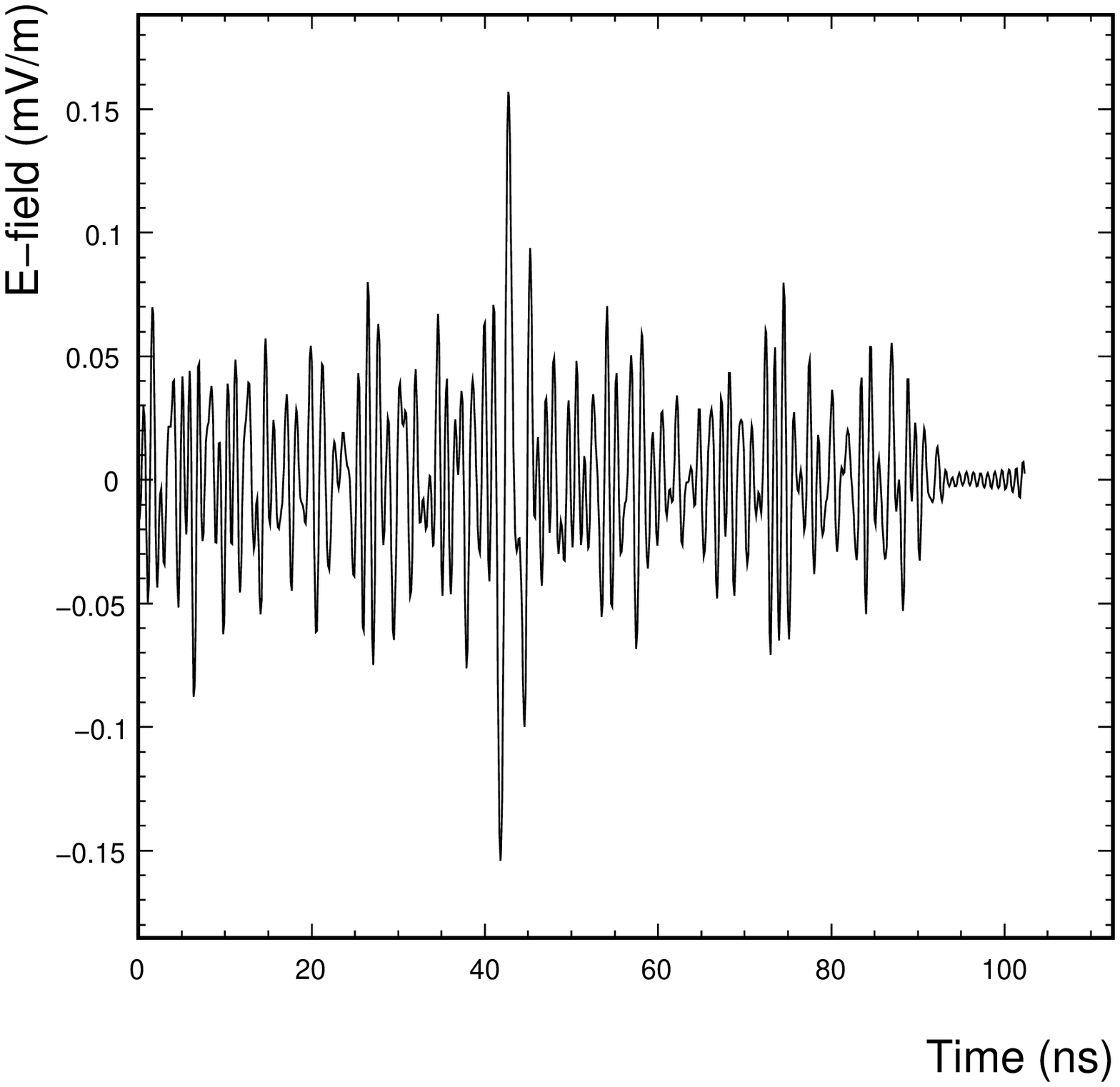}}
\caption{Left: Raw impulse from Taylor dome borehole. Right: Deconvolved impulse from the same waveform data,
showing that the pulse remains highly coherent despite the propagation effects of the 
ice, firn, and surface through which the impulse was transmitted.
\label{TDwfm} }
\end{figure}

We perform a final adjustment of antenna timing using the plane wave
source. We use it to check the impulse response measured in the lab.

The amplifiers, filters, and other components of the flight system introduce frequency-dependent 
gains and group delays in the measured waveforms.  
Using lab measurements of the instrument's RF properties, 
we can deconvolve the instrument's impulse response from our data, retrieving 
the E-field that was incident upon our antennas.  
We perform a final adjustment of antenna timing using the plane wave
source. We use it to check the impulse response measured in the lab.
Once again, the ground calibration
data allows us to test this process, since we know what the incoming signal was.
Figure~\ref{fig:gp:deconvolution} is an example of such a test.  The deconvolved signal
is much sharper.  In fact, the deconvolved signal is the impulse response of an 
ANITA quad-ridged horn antenna -- which is also the result of transmitting an impulse 
such as the one used in the ANITA ground calibration system.

This deconvolution process is also useful in establishing the coherent transmission properties
of the ice, both internally as the pulse must pass through the ice bulk, and
interface transmission as it passes through the ice surface which is potentially rough
on scales that may not be negligible compared to the wavelength of the radiation.
This propagation test was a focus of our calibration system at Taylor Dome.
In Fig.~\ref{TDwfm} we show a pair of waveforms, the left one the raw waveform of a
Taylor Dome pulse, and the right one the deconvolved impulse which is highly coherent.
The fact that this impulse was transmitted of order 150 m (e.g., the slant depth to the surface exit point)
through the ice and firn, along with the
firn surface, with no apparent loss of coherence (which would appear as possible
distortion or splitting of the pulse) provides good evidence that the methodology
employed by ANITA is effective in detection of Askaryan impulse events from
high energy showers.

\subsection{Timing/Pointing analysis}

\section{Angular Reconstruction}
Angular reconstruction of RF signals is a crucial part in the 
ANITA data analysis. It provides powerful rejection of incoherent 
thermal noise events which comprise $\ge$95\%  of the data set, 
anthropogenic RF events from existing bases and field camps, and 
radio Cherenkov events from air showers. If ANITA observes candidate 
neutrino events, angular reconstruction will be important for 
first-order neutrino energy estimation, for providing directional information, 
propagation distance  $(r)$ for $1/r^2$ corrections, and the RF refraction 
angle on the ice-air boundary for the Fresnel correction.

Although ANITA's trigger uses circularly polarized signals to provide
an unbiased sensitivity to both linearly polarized
events, the signals are separately recorded as their horizontally- and vertically-
polarized waveform components. Our Monte-Carlo analysis has indicated that
the neutrino signals of interest tend to be strongly favor vertical over
horizontal polarization, and thus for simplicity we analyze the 
H-pol and V-pol waveforms as separate data sets. In the following
description, the analysis steps are applied in parallel to both the 
polarizations independently. All steps of the analysis are optimized for
either a 10\% fraction of the data, or the full data set with the
payload orientation kept unknown till after the analysis was complete.
This is done to eliminate analyst bias in developing the data cuts. The
separation of the data streams into H-pol and V-pol sets also helps to 
eliminate any selection bias, since the cuts are applied in identical
fashion between the two sets. 

\subsection{Cross Correlation}
Direction reconstruction of the RF signal uses the arrival timing 
difference ($\Delta t$) information between pairs of antennas. $\Delta t$ 
is determined by using a cross-correlation technique between recorded 
waveforms of the antennas. The cross correlation coefficient ($R$) is defined by 

\begin{equation}
R= \frac{{\displaystyle \sum_{t}}(V_i(t)-\bar{V_i})(V_j(t+ \delta t)- 
\bar{V_j})}{\sqrt{{\displaystyle \sum_{t}} (V_i(t)-\bar{V_i})^2} 
\sqrt{ {\displaystyle \sum_{t}}(V_j(t+\delta t)-\bar{V_j})^2}}.
\end{equation}

Here, $V(t)$ is a recorded voltage value at a time bin $t$ and 
$\delta t$ is time delay. The $i$ and $j$ denote channel numbers. For fast 
calculation, the mean voltage $\bar{V}$ is first subtracted to normalize the mean to zero. 
In order to minimize the effects any anthropogenic continuous 
wave interference, we use a limited time window of the signal waveform of 
the antenna $i$; $|{t-t_{\rm peak}}| \le 15~{\rm ns}$, where $t_{\rm peak}$ 
is the time bin in which the maximum peak voltage, $V_{\rm peak}$, exists.  

For a given pair of antennas $i$ and $j$, we search for the time 
lag $\Delta t$ which gives the maximum correlation coefficient, $R_{max}$, 
while varying $\delta t$ within $\pm 25~\rm{ns}$.
In order to obtain more precise time resolution than the 0.38 ns 
sampling time interval of the ANITA digitization system, we use interpolated 
waveforms with a 7.6 ps bin width which is a five-fold oversampling of
the original waveforms.

\subsection{Group Delay Calibration and Time Resolution}
A precise timing measurement using the cross correlation 
technique allows an accurate calibration of the fixed group delay 
for each channels. Although all fixed group delays were also measured 
in the laboratory before the flight, calibration during the 
flight was performed to reduce systematic uncertainties 
such as temperature dependence and possible antenna coordinate changes 
after the launch. We measured all relative group delays 
using the impulse signal transmitted from ground based 
calibration pulser system~\ref{ground_pulser}[reference 
for ground pulser here]. A group delay difference 
($\delta T^{ij}$) between two channels of $i$ and $j$ is 
obtained by comparing observed timing difference $\Delta t_{\rm obs}$ 
and its expectation $\Delta t_{\rm exp}$;
\begin{equation}
\delta T^{i,j}= T_{\rm delay}^{i} - T_{\rm delay}^{j} =   
\Delta t_{\rm obs}^{ij}-\Delta t_{\rm exp}^{ij},
\end{equation}
where the $\Delta t_{\rm exp}^{ij}$ is given by the difference 
of RF propagation times from the coordinate of transmitter to each receiver antenna. 

\begin{figure}[thb!]
\centerline{
\epsfxsize=6in 
\epsfbox{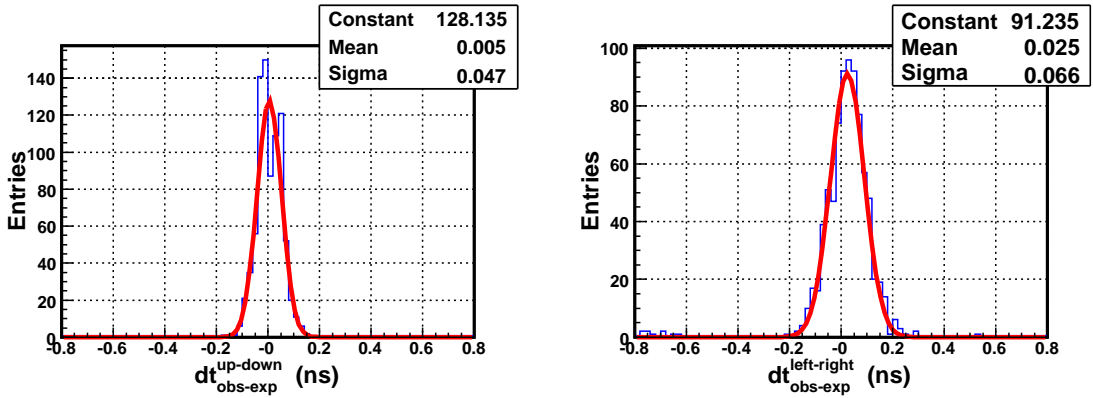}}
\caption[dt distributions]{$dt$ distributions: 
Upper/lower antenna pairs in the same $\Phi$-sector (Left) 
and antenna pairs in different $\Phi$-sectors (Right).}
\label{fig:dt}
\end{figure}

Figure~\ref{fig:dt} shows distributions of time differences 
$dt=\Delta t_{\rm obs}-\Delta t_{\rm exp}$ after applying the 
group delay correction for all channels. Obtained time resolutions 
for the upper/lower antenna pairs in the same $\Phi$-sector is 
$\sigma_{\rm same~\phi}=47~{\rm ps}$.  Because of additional 
jitter due to the intrinsic jitter of time synchronization between different SURF boards 
allocated to different $\Phi$-sectors~\ref{SURF}, a
slightly worse time resolution of $\sigma_{\rm diff~\phi}=66~{\rm ps}$ 
is obtained for antenna pairs in different $\Phi$-sectors.  
 
\begin{figure}[htb!]
\centerline{\includegraphics[width=3.9in]{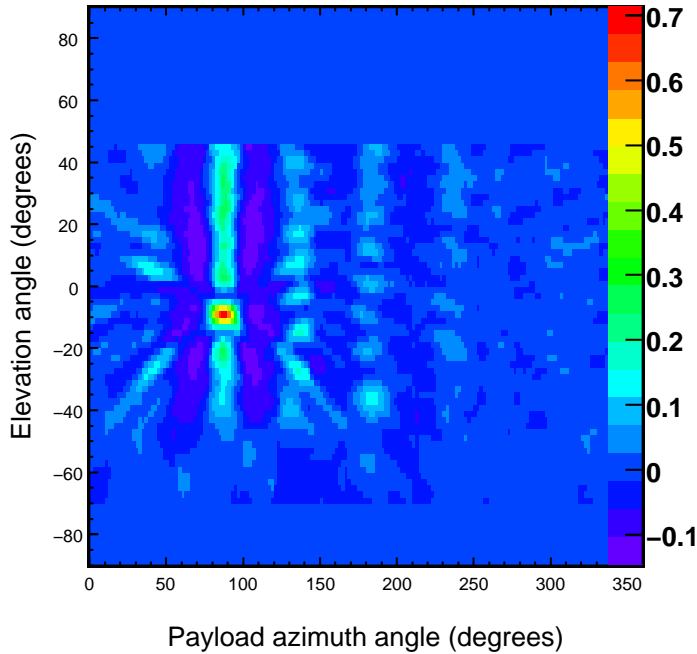}}
\caption{Interferometric image of a single impulsive event from the Taylor Dome ground calibration pulser.
\label{TDimagemap} }
\end{figure}

Once the calibrations are applied, a first order method to
establish the presence of a coherent signal source is to
sum the two-dimensional cross-correlated intensities of all baseline pairs 
into a summed-power interferometric image, as shown in Fig.~\ref{TDimagemap}.
for one of the surface antenna pulses from Taylor Dome. In this 
procedure each pair of antennas gives a fringe across the sky at an
angle corresponding to the baseline direction, and
the fringes sum together with the greatest strength at the source location.
The virtue of this method is that it can also be used to identify the location
of sidelobes in the image, so that later quantitative fitting of the source
location can be tested to ensure that it has not produced a 
misreconstruction at one of the sidelobe locations. 

Such an image is highly analogous to a ``dirty map'' in radio
astronomical usage, and a reduction in sidelobes is possible with
further image processing. However, we have found in general that for this
type of ``pulse-phase interferometry,'' these maps are adequate since
all sources are unresolved for ANITA. We have also studied phase closure
methods as applied to our data, and we find that further work in this area
is justified and may yield even more robust methods for this type of
source imaging.

\subsection{Good event selection and hit cleaning}
For the angular reconstruction, we use antennas in three 
$\Phi$-sectors around the maximum $\Phi$-sector where an average peak 
voltage of the upper and lower antennas is the highest among all $\Phi$-sectors. 
The upper antenna on the maximum$\Phi$-sector is used for a reference channel 
for the cross correlation. Channels having anomalous or outlying timing information 
are not used for the reconstruction. We compare $\Delta t_{\rm obs}$s for 
all possible baselines (15 possible) in the 6 antennas around the maximum $\Phi$-sectors, 
then we regard the channel as an isolated hit if multiple baselines associated 
with this channel have more than 12 ns deviation.

A selection for good events is applied before the angular fit in order to 
speed up analysis and reduce misreconstructions. We require the number of 
good antenna hits $N_{\rm hit} > 5$, the peak voltage to be $V_{\rm peak}>35~mV$, 
and a signal to noise ratio to be ($SNR=V_{peak}/V_{rms}>3.5$) for both of 
upper and lower antennas in the maximum $\Phi$-sector, where the $V_{rms}$ 
is a root mean square of voltage obtained in the first third of the recorded 
waveform which is well-separated from the signal region near the center of
the recorded waveform. Note that the SNR recorded
in the digitizer waveform is generally lower than the SNR of the analog
RF trigger, since there are additional insertion losses associated with the digitizer
itself, as described in a previous section.

\begin{figure}[ht!]
\centerline{
\epsfxsize=3.65in 
\epsfbox{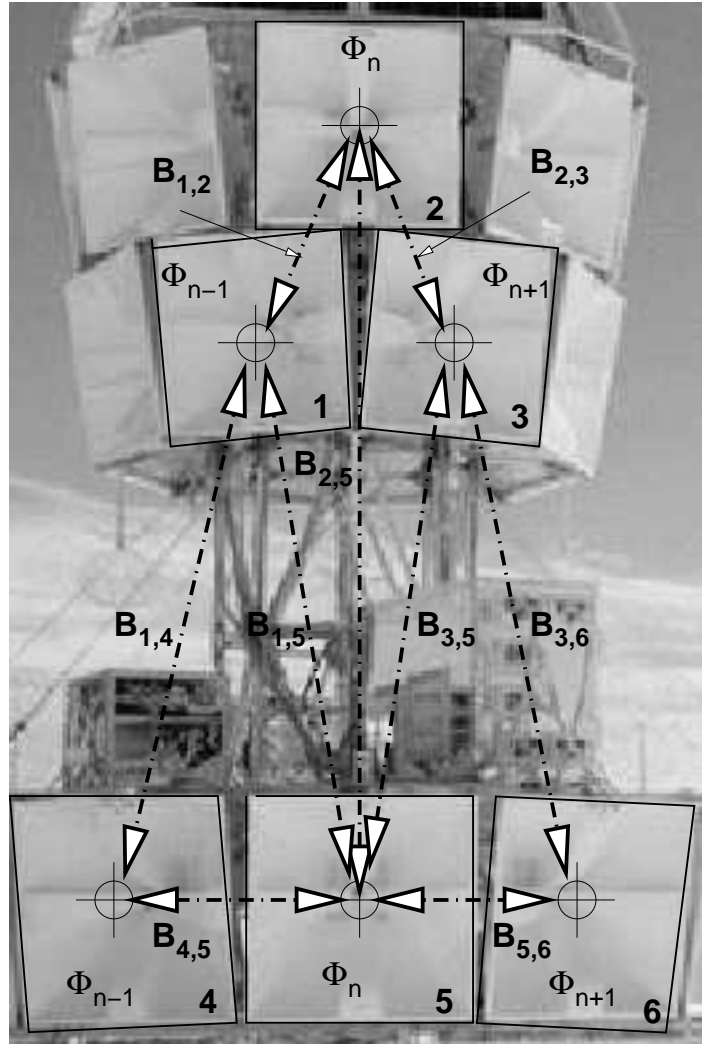}}
\caption[Illustration of basic set of baselines for the angular fit]
{Illustration of basic set of baselines for the angular fit. Not all of
the fifteen possible baselines for 6 antennas are shown.}
\label{fig:baseline}
\end{figure}

\begin{figure}[ht!]
\centerline{
\epsfxsize=6in 
\epsfbox{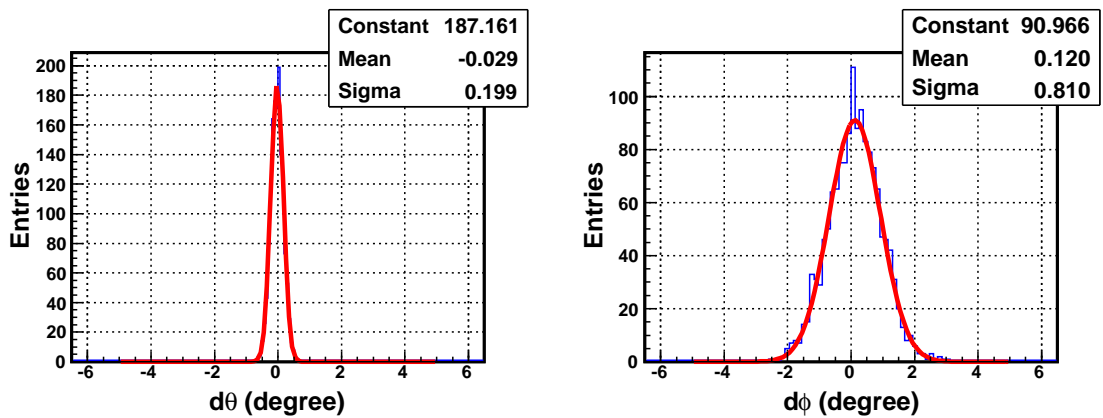}}
\caption[Angular resolution]{Angular resolution: $d \theta$ distribution 
(Left) and $d \phi$ distribution (Right)}
\label{fig:angle_resol}
\end{figure}

\subsection{Angular Fit}
The number of timing baselines $N_{\rm base}$ we use in the fit is $N_{\rm hit}-1$. 
A selection from among the fifteen possible baselines for six antennas
is made in a way to minimize timing uncertainties 
and correlations between the baselines. Figure~\ref{fig:baseline} illustrates 
some of the possible baselines for an event with a phi-sector containing the maximum
possible vertical baseline; either adjacent phi sector would involve a shorter
principal baseline paired with longer adjacent ones. 

We perform a $\chi^2$ fit in order to find the direction of the RF signal. It  minimizes
\begin{equation}
\chi^2 = \sum_{k}^{N_{\rm base}} \left ( \frac{\Delta t_{\rm obs}^{k} - 
\Delta t_{\rm hypo}^{k}(\theta,\phi)}{\sigma_{k}} \right )^2.
\label{eq:chi2}
\end{equation}
Here, the $N_{\rm base}$ is the number of baselines,  
the $\Delta t_{\rm obs}^{k}$ is timing difference of the 
antenna channels participating in baseline $k$, the 
$t_{\rm hypo}^{k}(\theta, \phi)$ is its expectation for 
the RF signal with a given hypothesis of a plane wave 
direction ($\theta, \phi$), and the $\sigma$ is the 
time resolution of the corresponding baseline. 
Since the time resolution is varied depending on the signal 
strength, we use $SNR$ dependent time resolution;
\begin{equation}
\sigma = \sigma^{\rm ch1}_{SNR} \oplus \sigma^{\rm ch2}_{SNR} \oplus \sigma_{\rm sys},
\label{eq:sigma}
\end{equation}
where the $\sigma_{SNR}$ is a single channel timing error caused 
by thermal noise and the $\sigma_{\rm sys}$ is intrinsic system error. 
The ch1 and ch2 denote two channel numbers involved with the baseline. 
A functional form of $\sigma_{SNR}$ is obtained by a Monte Carlo 
simulation (MC) study of a thermal noise effect on time resolution 
of the impulsive signal. We use the measured time resolutions 
for $\sigma_{\rm sys}$, which can be either $\sigma_{\rm same~\phi}$ 
or $\sigma_{\rm diff~\phi}$ depending on the geometry of the baseline.

We perform 10 iterations of fits to reduce misreconstructions caused by 
trapping of the solution in local minima of $\chi^2$.
In each iteration,
initial fit parameters of $(\theta, \phi)$ are uniformly varied within 
$\theta>90^\circ$ and $\phi_{\rm max}-25^\circ < \phi < \phi_{\rm max}+25^\circ$, 
where the $\phi_{\rm max}$ is an azimuth angle of the boresite of the antenna in the maximum 
$\Phi$-sector. We require $\chi^2<4$ for a well reconstructed event.

Figure~\ref{fig:angle_resol} shows angular resolution with the 
ground-based calibration pulsers. The $d \theta$ ($d \phi$) is a 
deviation of zenith (azimuth) angle from expected value. Achieved 
resolutions are $0.2^\circ$ and $0.8^\circ$ respectively. The worse 
resolution of $d \phi$ than  $d \theta$ is due to the shorter length 
of baselines in the $\phi$ direction and the worse time resolution of the inter-$\phi$ baselines.

\subsection{Misreconstruction Rejection}
A misreconstruction is a fit result that deviates from the expected source direction while still
having a good fit quality or low $\chi^2$. Misreconstruction 
is one of the most important potential background sources for the neutrino search 
in the ANITA experiment, because a misreconstructed event of anthropogenic origin
from, for example, a known encampment, might become a neutrino candidate.

The main sources of the misreconstruction are incorrect timing information 
and a fit that becomes trapped in a false local minimum. Incorrect timing means timing with 
an much larger error than the expected $\sigma$. These may be caused by an unresolved 
cycle ambiguity of the signal pulse in the cross correlation procedure. 
In order to reduce this, we compare the timing with it from another 
cross-correlation with different time window of signal waveform; 
$|{t-t_{\rm peak}}| \le 12~{\rm ns}$, then reject the event if the 
difference is greater than 0.5~ns in any of $N_{\rm base}$ baselines.

We also reduce the misreconstruction events using the excellent 
directivity characteristic of the ANITA horn antenna. Because of relatively high
gain of the horn antennas, and corresponding  
narrow width of their angular response, 
the sector $\Phi_{\rm max}$ should be consistent 
with the signal direction. Therefore, 
we require $|\phi - \phi_{\rm max}| < -22.5 ^\circ$, 
where the $22.5^\circ$ is corresponding to the $\Phi$ occupancy of an antenna.

A further requirement for the misreconstruction reduction is 
an internal consistency check with sub-sets of the baselines. We perform 
angular fits not using one of the baselines in the original baseline set. We 
reject events if any of the $N_{\rm base}-1$ of possible subset fits 
provides a space angle result that differs from the original by more than $5^\circ$.

\subsection{Efficiency}

\begin{figure}[ht!]
\centerline{
\epsfxsize=5.5in 
\epsfbox{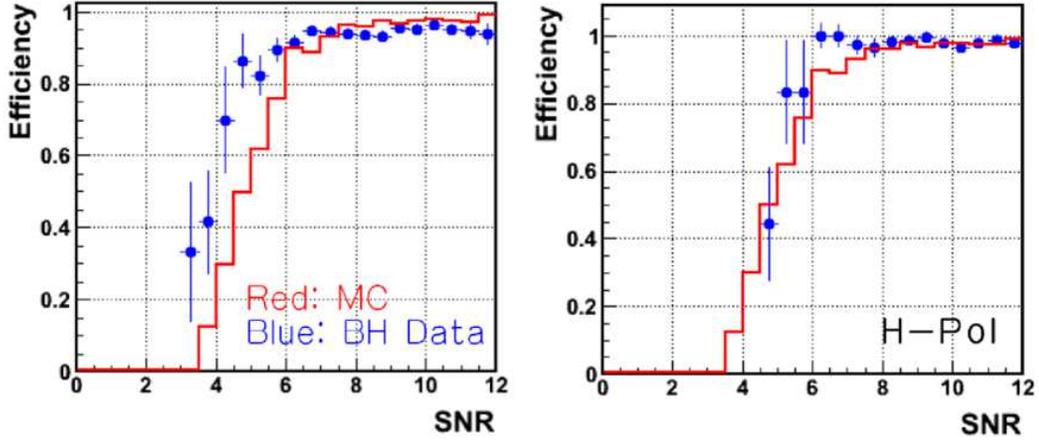}}
\caption[Reconstruction Efficiency vs. $SNR$ ]{Reconstruction Efficiency vs.$SNR$
for both polarizations analyzed independently: 
Dots are efficiency measurement with the ground based calibration pulses and the curves
is for instrument simulation Monte Carlo estimation. Left is for vertical polarization
which has a known SNR bias due to low-level EMI contamination in the event
signals. The right shows the same efficiency for H-pol which is less affected by
EMI contamination.}
\label{fig:eff_vs_snr}
\end{figure}


The final reconstruction efficiency is important to understanding the
overall detector efficiency, and is measured with the ground based calibration data 
and additional simulation analysis. 
Figure~\ref{fig:eff_vs_snr} shows the efficiency measurement as a 
function of $SNR$ for events acceptable fits for both polarizations.
There is a good agreement between the ground based calibration data and the 
MC in $SNR>5.5$ while a discrepancy is found in the lower $SNR$ region for the Vpol events. 
The likely source of this discrepancy is the effect of low-level interference 
from Williams Field where the ground calibration system was used, which causes
an error in the SNR estimate. Due to the preponderance of vertically-polarized
signals in the transmitters near Williams field, Vpol is affected to a greater degree
than H-pol. The global efficiency at higher SNR for both polarizations 
with the ground calibration system is 96\% while a 
misreconstruction-reconstruction rate is less than 0.16\%.

\subsection{RF Projection on the Surface}
As noted above, reconstruction of the RF source position on the surface is critical to identify the 
anthropogenic noise sources associated with known bases. In order to obtain 
more precise position measurements we take into account the surface elevation 
variation based on the Bedmap data.~\cite{bedmap} A simple model is implemented 
for a fast calculation. We find line-sphere intersections assuming the spherical 
shape of the earth, then take one solution having shorter distance to the ANITA payload. 
For an initial calculation, we assume the earth radius $R ^{\rm hypo}_{\rm earth}$ 
is the geoid plus surface elevation of the ANITA payload coordinates (longitude and 
latitude). Then the next iteration uses the geoid and surface elevation of result 
coordinate of the previous iteration. We found the results converge after more 
than 2 iterations.  We found the source position 
resolution at the surface is 2.7 km for 170 km of line-of-sight distance at 37 km of altitude.

\section{Summary}

We have presented the basis for a balloon-borne payload search for ultra-high energy neutrinos
from an altitude of 37~km above the Antarctic Ice sheets. The ANITA payload's first flight
resulted in a large database of impulsive events from locations all across Antarctica.
When these events indicated a ground-based point source, it could be geolocated to
a precision of order $0.3^{\circ} \times 1.0^{\circ}$ in angle, which projects to error ellipses
of several square km at typical distances of several hundred km. We have demonstrated that
operation down to near-thermal-noise levels is achievable, and electromagnetic interference
in Antarctical, while still problematic, is manageable and largely confined to a
relatively small number of camps. Results from the analyses of all these data in the
search for neutrino candidates will be presented separately.

\appendix
\section{Extensive air shower radio and ANITA-1}

Radio Detection of extensive air showers (EAS) has recently received
renewed attention as a method for detection and characterization of EAS.
The radio emission is primarily produced by synchrotron effects in the
geomagnetic field, and so has been termed
geo-synchrotron emission~\cite{FalckeGorham}. This type of emission
dominates over Askaryan emission from the charge excess since all of the
charge, the sum of both positrons and electrons, contributes to it as
the component trajectories curve in opposite directions in the magnetic field.
The emission is highly beamed along the shower axis, 
with a characteristic beamwidth of
a few degrees or less. It is linearly polarized, and 
coherent at frequencies below about 100~MHz, 
and partially coherent above 200~MHz, but these values vary widely with
observation angles and the distance to the shower axis. The partially
coherent emission has been detected up to hundreds of MHz~\cite{Spencer, Fegan68, Fegan69},
with some evidence for a flat radio spectrum above 200~MHz~\cite{Spencer}.
Thus ANITA-1 may be able to detect such events both as direct emission when
the shower axis points almost directly at the payload from above the
horizon, or potentially as reflected events when the emission 
undergoes near-specular reflection off the ice surface. 

\begin{figure}[htb!]
\centering
\scalebox{0.6}[0.6]{\includegraphics{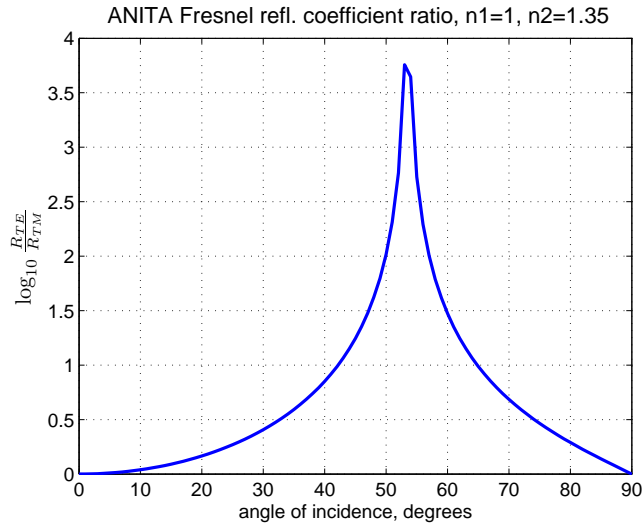}}
\caption{Fresnel coefficient ratio for Te vs TM waves, resulting in
H-polarization for most reflected signals.\label{fresnelfig}}
\end{figure}

In the latter case, such events could be a background for neutrino signals,
if they are not distinct in some way. In fact, such events are likely to be
highly horizontally polarized for two reasons: (1) They are intrinsically stronger
in Hpol because the geomagnetic field is more vertical near the poles, and thus
the field bending of the charges for moderately inclined showers (seen in reflection) 
yields mostly intrinsic Hpol; and (2) the Fresnel reflection coefficient for TE waves
at the surface is always larger than for TM waves, and over most of ANITA's
aperture, it is much larger. Thus even an unpolarized initial signal will
acquire horizontal polarization, and Hpol will tend to be accentuated even more
for signals with some intrinsic Hpol. Figure~\ref{fresnelfig} shows the
base-10 logarithm of ratio of reflected intensity for the TE vs TM wave. It
is evident that it exceeds a factor of 10 over much of ANITA's 
angular range of acceptance.

\subsection{Simulation.}

To simulate the geosynchrotron radio emission for ANITA's band, we made use
of a recent detailed simulation by Huege \& Falcke~\cite{Huege_Falcke05} 
(hereafter HF05) using CORSIKA for the
EAS simulation, married to a standard electromagnetic radiation treatment for the
radio emission. HF05 were mainly interested in the lower frequency bands below
100 MHz, but they did extend their simulations to the partially coherent
region up to about 500~MHz, although none of their parameterizations are
strictly valid there as they were only fit for the coherent regime. But
we use the reported field strengths in the 200-500~MHz regime to create
an {\it ad hoc} parameterization. Based on the given results, the emission
is nearly flat with frequency above 200~MHz in all cases, but varies widely
with the distance away from the shower axis. Energy scaling is not reported
but can be heuristically estimated from the reported results as well.

Because the average sine of the magnetic field inclination angle is close to 1 for all
of Antarctica~\cite{NOAA}, 
and since ANITA's antenna geometry favors showers inclined 
by $45^{\circ}$ or more, we treat all showers as producing predominantly
horizontally polarized signals as seen in ANITA's projection; this is
evidently true on average. The vertically polarized component on average
is less than 1/3 of the Hpol in field strength, and after reflection this
ratio is increased even more. As a result, detection of air showers in the
Vertically polarized component is strongly suppressed, and we will quantify
to what level in a later update.

With these assumptions, we use
\begin{equation}
\mathcal{E}  = 10^{-5} \left ( \frac{E_{shower}}{10^{19}~{\rm eV}} \right )^{0.85} 
\frac{R_0}{R} e^{-\theta/\theta_{0}}~\rho~\langle{\sigma}\rangle~\beta~\zeta~\xi ~~{\rm Volts~m^{-1}~~MHz^{-1}}
\end{equation}
where $\mathcal{E}$ is the field strength at the payload; $E_{shower}$ is the shower
energy in eV; $R_0=6.7$~km is a reference distance related to the effective origin of the
shower radio emission above the ground; $R$ is the distance from the point of reflection
to the payload; $\theta$ and $\theta_0=5.1^{\circ}$ are the angle with respect to projected
shower axis, and a characteristic angle based on the radial distances of
exponential falloff with respect to the shower axis;
$\rho$ is the field reflection Fresnel coefficient; $\langle \sigma\rangle = 0.95$ is the
average sine of the magnetic field inclination with respect to the horizontal, and
$\langle \sigma\rangle$ here thus gives the  
first-order average fractional component of the horizontally polarized field
strength; $\beta$
is the ratio of the Antarctic mean magnetic field strength to the B-field strength
used in the HF05 calculations ($\beta=1.08$ in our case); $\zeta$ is a factor
to account for the fact that the HF05 estimates are in the radio near-field;
and $\xi$ is a factor to account for surface roughness scattering losses.

The value of $\theta_0=5.1^{\circ}$ is about a factor of two higher than estimates
for vertical showers at ground level. HF05's results for inclined showers do however
indicate that showers at 45$^{\circ}$ or more have much flatter exponential falloff than
vertical showers, even after correcting for projection effects, and the value
adopted here is consistent with the simulation data presented in HF05.

For the present results we are using $\zeta=1.2$ as indicated by HF05's scaling with
inclined showers, which tend to move the observer out of the near field.
In general, near field effects reduce the relative field strength as compared 
to the far-field, so $\zeta>1$ is to be expected. We have also assumed
$\xi=0.9$ in reflection--since the average wavelengths are long and the
refractive index contrast low, we expect surface roughness to play a minor role here.
We do not include surface slope variations here; the surface is assumed to be
the locally flat portion of a sphere.
We have assumed flat emission
with frequency based on the reported HF05 results up to 500~MHz and weakly
supported by earlier observations~\cite{Spencer}; above this
we have arbitrarily reduced the field strength by factors of 0.7 and 0.5 for
the MID2 and HIGH bands respectively.

Events are simulated over the entire ANITA field of view (assuming phi symmetry), and
several adjacent phi-sectors are used to form a trigger. A standard 3 of 8
L1 trigger using LCP+RCP is used, with proper Rician noise and signal estimation. The antenna beam
patterns are included. The EAS signal is assumed to be initially completely Hpol.
The exponential off-axis attenuation determined by $\theta_0= tan^{-1}(d/D)$ where
$d$ is the off-axis characteristic distance for 1/e attenuation, and 
$D\sim 6.7$~km is the distance to the shower effective origin. 
The value used in the near-field parameterizations is very conservative
since it is based on vertical showers, where $d=300$~m was indicated. 
The simulations for inclined showers indicate a value of $d=600$~m,
and we have used this for the far-field ANITA case.
Also, these values are energy dependent, but the dependence is currently 
difficult to estimate from the reported results so we have neglected it.

For the UHECR shower rates and absolute normalization, we used the Auger
2007 normalization and the Auger 2007 parameterization of the 
empirical GZK cutoff as observed in the Auger data. The integral flux
above $10^{19}$~eV is $J(>E) = 2.0 \times 10^{-18}$~cm$^{-2}$~s$^{-1}$~sr$^{-1}$.
This spectrum was sampled via a von Neumann method.

\begin{figure}[htb!]
\centering
\scalebox{0.5}[0.5]{\includegraphics{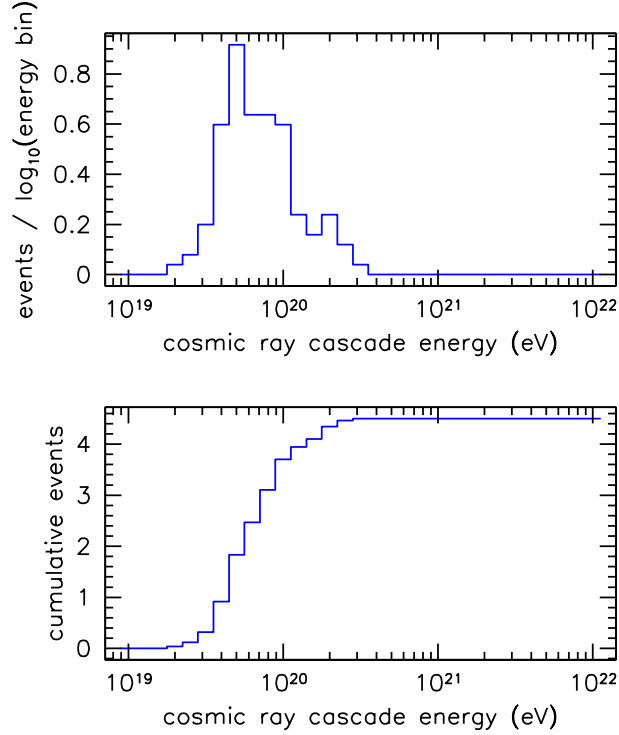}} 
\caption{Top: Differential energy distribution of simulated events that are detected by ANITA-1.
Bottom: Cumulative energy distribution of events, showing that the detectability turns
on in the vicinity of the GZK cutoff.
\label{EAS_fig1}}
\end{figure}

\subsection{Results.}

The simulation predicts a total of $4.5 \pm 0.4$(statistical) detected events for 18 days of cumulative
livetime for the ANITA-1 flight. However, the systematic errors on this result are estimated
to be of order 100\%--that is, our result is also consistent with zero detected events.
This level of uncertainty stems from the high degree of uncertainty on simulation of
the partially coherent signals from EAS in the ANITA band, which is well above the
region where EAS radio measurements are primarily focused. 

Putting aside the issue of the absolute detected flux, we may still investigate the 
relative energy spectral content of the detected events. Figure~\ref{EAS_fig1} shows the energy distribution of
the detected events; here once again the overall flux levels are subject to large
systematic, although the relative fluxes are less sensitive to these same
systematics. The energy threshold is about $4 \times 10^{19}$~eV and shapes the
low-energy edge of the distribution. The GZK cutoff itself shapes the high
energy edge. 

In summary, ANITA's sensitivity was adequate to possibly detect a handful of events from extensive air 
shower radio emission, seen in reflection from the ice surface. The uncertainties in
this prediction are dominated by the uncertainties in the magnitude of partially
coherent emission from EAS radio processes.


\section{Characterization of Surface Roughness and its Effect 
on Neutrino Aperture}

One concern for modeling the detection of neutrinos in the ice 
is the propagation of the radio emissions from
under the ice to the surface, {\it i.e.},
the firn, into the air.  
If the surface of the ice were a smooth boundary, we could
simply apply the proper Fresnel coefficient for each polarization
at the surface corresponding to an index of refraction $n\simeq1.3$ to air $n=1$. 
However, the Antarctic winds and other effects
serve to carve up the surface, making it rough on many length
scales.  

RF energy propagating from the neutrino interaction to
the instrument does not pass through a single point on the surface of the
ice. Rather, all radiation passing through an area given by the first Fresnel zone
on the ice surface can contribute to the received band-limited impulse. This region
is determined by considering the locus of points on the surface where
the pathlength is within a quarter-wavelength of the direct geometric
ray for an idealized smooth surface. Straightforward geometric
considerations for a simple phase-screen perpendicular to the 
direction of propagation between a transmitter and receiver give
a first Fresnel zone radius 
$$R_{F,1} = \sqrt{\frac{cd_1 d_2}{\langle f \rangle (d_1+d_2)}}$$
where $c$ is the velocity of light, $\langle f \rangle$ the mean frequency of
the waves, and $d_1,~d_2$ are the source-to-phase screen and receiver-to-phase screen
distances, respectively.
Within this radius, any sub-wavelength phase disturbance to the propagating waves
will cause constructive or destructive interference to the beam.
Higher order Fresnel zones at larger radii may be important in some cases, but we neglect them
here.

In ANITA's case, the Fresnel zone geometry must be modified to account
for the intervening medium between the neutrino shower ``transmitter'' and
the payload receivers. In addition, projection effects must be considered for
the oblique incident angles of the Cherenkov radiation on the underside
of the ice surface, and the often steep refracted angles of near-tangential
outgoing rays. Accounting for these effects, and the Snell's law divergence
associated with them, the modified first Fresnel zone for ANITA is an 
approximate ellipse with
a major axis diameter, transverse to the plane of incidence and reflection along
the ice surface, of
\begin{equation}
D_{F}^{tr} = 2 \sqrt{\frac{c}{\langle f \rangle}\frac{d_2}{1 + n\frac{d_2}{d_1}}}~
\end{equation}
and a minor axis along-track diameter of
\begin{equation}
D_{F}^{\ell} = \frac{2}{\sin{\theta_i}} 
\sqrt{\frac{c}{\langle f \rangle}\frac{d_2}{1 + \mu\frac{d_2}{d_1}}}~
\end{equation}
where $\mu = n(\cos{\theta_i}/\cos{\theta_r})^2$ and $\theta_i,~\theta_r$ are
the incident and refracted angles with respect to the normal for the
underside of the ice surface.

\begin{figure}
\centerline{\includegraphics[width=3.5in]{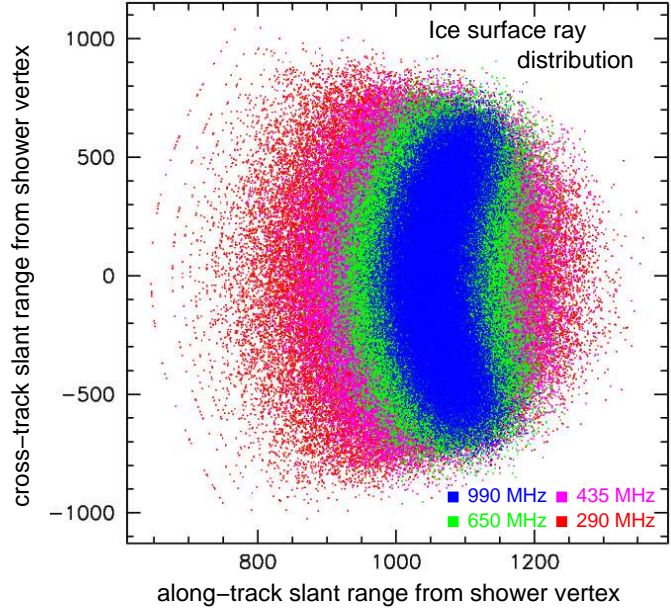}}
\caption{The distribution of ray exit points on the ice surface from a
near-horizontal neutrino event at a depth of about 1~km. The colors
indicate the ray patterns for the center frequencies shown.
\label{ERays} }
\end{figure}\hspace*{5ex}

Thus, for example, for a mean frequency of 300~MHz, an index of refraction
of $n=1.35$ corresponding to the near-surface firn, a 
near-tangential refracted angle of 85\deg, an incident angle on the underside of
the ice surface of 47.5\deg, and distances $d_1=1400$~m, and $d_2=300$~km, 
$D_{F}^{tr}\simeq 50$~m and $D_{F}^{\ell}\simeq 11 $~m, many wavelengths across 
in either dimension. Figure~\ref{ERays} shows a simulation result for radio
Cherenkov emission from a shower similar to this example. Here only rays that
can escape total-internal-reflectance are shown, and the colors show the ray
patters for different frequencies. It is evident that the scale of the emission
at the surface for a typical shower is much larger than the first Fresnel
zone for any of the frequencies of interest to ANITA; thus in almost all
cases the Fresnel zone scale is the relevant scale for surface roughness effects.

Variations
in the surface over the Fresnel zone scale potentially alter the transmission
relative to a smooth surface.  Overall surface tilts on
scales larger than this size are of order 1 degree 
and serve only to tip the refracting surface slightly; the Fresnel
coefficient for a plane wave still applies.  To this end, a distribution
of slopes at random angles is employed in the simulations.  Since
most of our events are close to total-internal-reflection for a horizontal
surface, this variation in slope tends to increase ANITA's overall
aperture.  We 
distinguish this effect from the optically more complicated effect of
roughness on scales within the first Fresnel zone.

Surface ``facets'' on scales of order a wavelength or less will affect the
tranmission properties, and they must be integrated over the scale
of the Fresnel zone to assess their net effect.
In the firn ANITA's wavelengths are  0.2~to 1.2~m.  As described below,
the Antarctic surface has roughness on scales smaller, equal, and
larger than this range.  We are presented with a problem
similar to a  frosted glass effect in the
optical case  where irregularities on the scale of a wavelength can
significantly disturb the angular distribution of 
transmission~\cite{nieto}.
In principle the effect could either increase or decrease
ANITA's sensitivity.  The spreading of the radio Cherenkov cone
will serve to increase (worsen) ANITA's energy threshold.  However, the
spreading of the cone could also greatly improve the solid angle
for neutrino detection by allowing ANITA to see events that
would have been totally internally reflected by a smooth surface.

\subsection{Antarctic Surface Roughness Data}

\hspace*{5ex}
Quantitative studies of the Antarctic surface roughness comes from
two sources:  Over-snow traverses~\cite{furukawa,goodwin} have characterized
surface features in detail, but only in limited areas,
mostly in East Antarctica or near the coasts.   Satellite
altimeter data covers about 80\% of the 
continent but with limited resolution
and scales on the order we care about are difficult to extract.
Over-snow traverses identify roughly three types of features.  
{\it Depositional surface features} or ``snow dunes'' are typically
$50-100$~m long, $10-15$~wide and 1~m high.  {\it Redistribution surfaces}
are usually $20-50$~m long, $2-3$~m wide and $0.3-0.5$~m high.
{\it Erosional surfaces} are 0.3~m in mean height.  The term
{\it sastrugi} in the literature can refer either to erosional
surfaces exclusively, or to any meter-scale feature.   Features larger
than 1~meter tend to align along the direction of the prevailing 
winds.  Features on the centimeter scale are {\it micro-roughness}.

The reader may have a mental picture of the Antarctic surface as
covered with large sastrugi, a picture perhaps biased by the most
interesting photography of the continent.  However, in reality, such
features are rare.  The oversnow traverse of Furukawa~\cite{furukawa}
on the East Antarctic plateau (along a 1000km traverse at
$40^{\circ}$E  from $69^{\circ}$~S to $77^{\circ}$~S ) encountered
features larger than 0.3 meters at most 75 times per kilometer (1 per
13 meters)  and typically 50 times per kilometer (1 per 20 meters),
{\it i.e.}, 5\% of the surface.    The traverse of
Goodwin~\cite{goodwin} provides complementary data that
shows the average feature  size only exceeded 0.5 meters during 7\% of
their 750 kilometer traverse (traverse at $  72^{\circ}$~S latitude
from $112^{\circ}$ to $132^{\circ}$ E).  Only once on their 750~km
journey do they report sastrugi that are 2~meters high.  Hence we may
extract that the fraction of the area of the East Antarctic plateau
covered by even half-meter features to be less than 5\% of the
surface.  Although we do not have traverse data from West Antarctica,
anecdotal evidence from skiers travelling from Patriot Hills to South
Pole indicates that these features are rare there as well.  The
risetime of pulses transmitted by satellite altimeters (Seasat,
ENVISAT, ERS), which is sensitive to the size of surface features,
shows the western plateau to be only about 25\% rougher than the
eastern plateau.

A compilation of the traverse and altimeter data can be
found in Ref.~\cite{griswold1}.  We summarize the data here 
as containing features with 30~cm typical heights and 8~m
spacing or 70~cm typical heights and 13~m spacing.  Hence the
features have vertical scales comparable to a wavelength with
separations of many wavelengths.   Considered as angles,
these roughness features would be of order 2~degrees.

\begin{figure}
\begin{center}
\epsfig{file=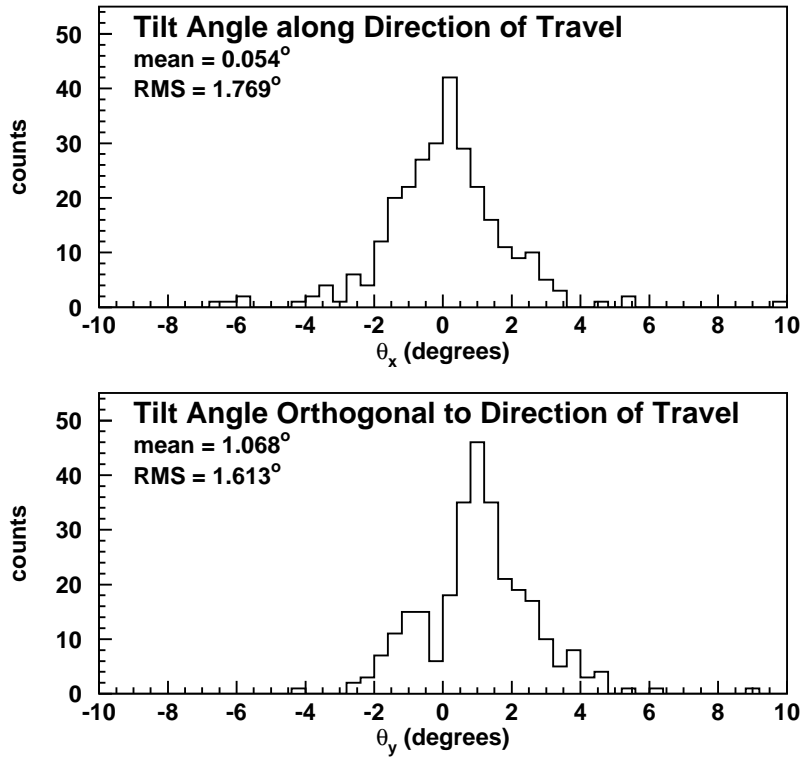,width=4.5in}
\caption{\label{traverseangles} Left: distribution of angles measured
by placing a 0.5~m plate on the surface of the snow at Taylor Dome.  Angles
were measured sequentially along a 100~m traverse in the prevailing wind
direction.   Right: Orthogonal angle measured along the traverse.}
\end{center}
\end{figure}

\begin{figure}
\begin{center}
\epsfig{file=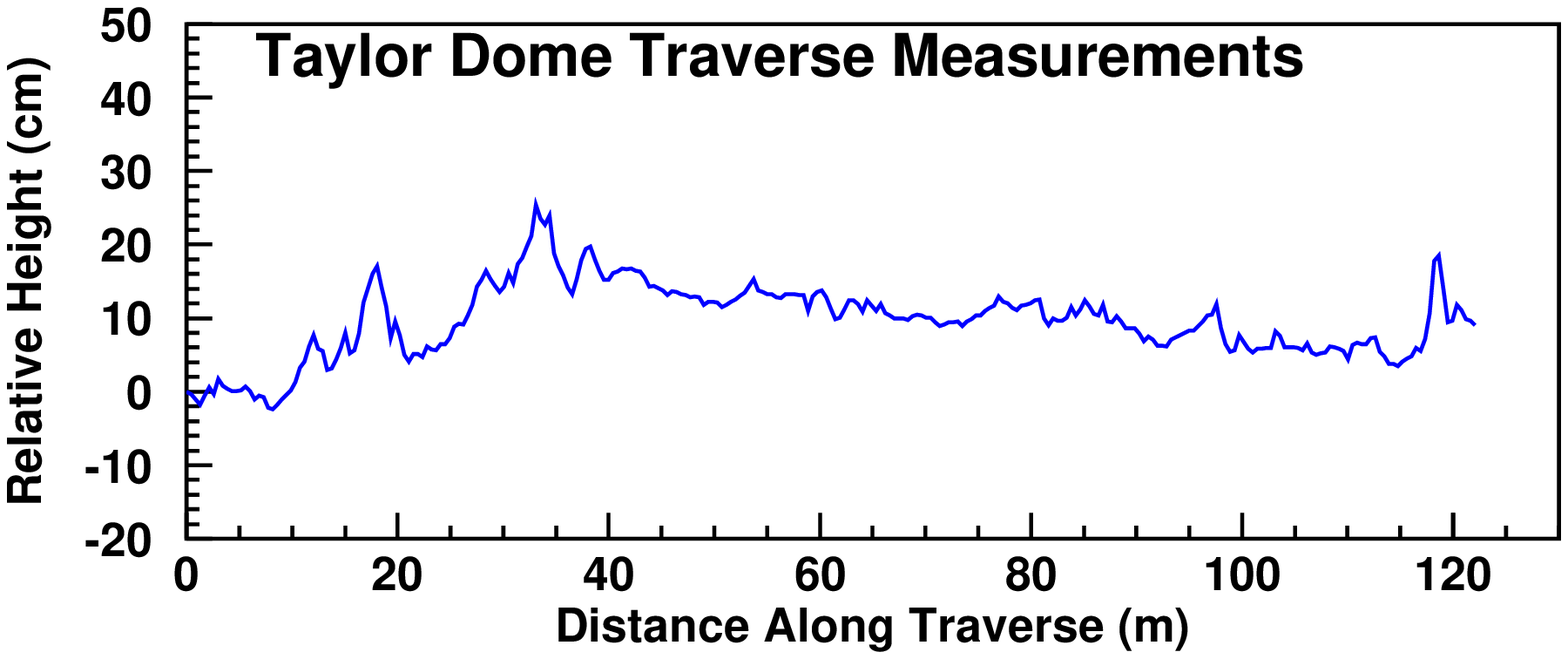,width=4.5in}
\caption{\label{elevationmap} Elevation map created by integrating
the measured angles over the length of the 100~m traverse at Taylor Dome.}
\end{center}
\end{figure}

\begin{figure}
\begin{center}
\epsfig{file=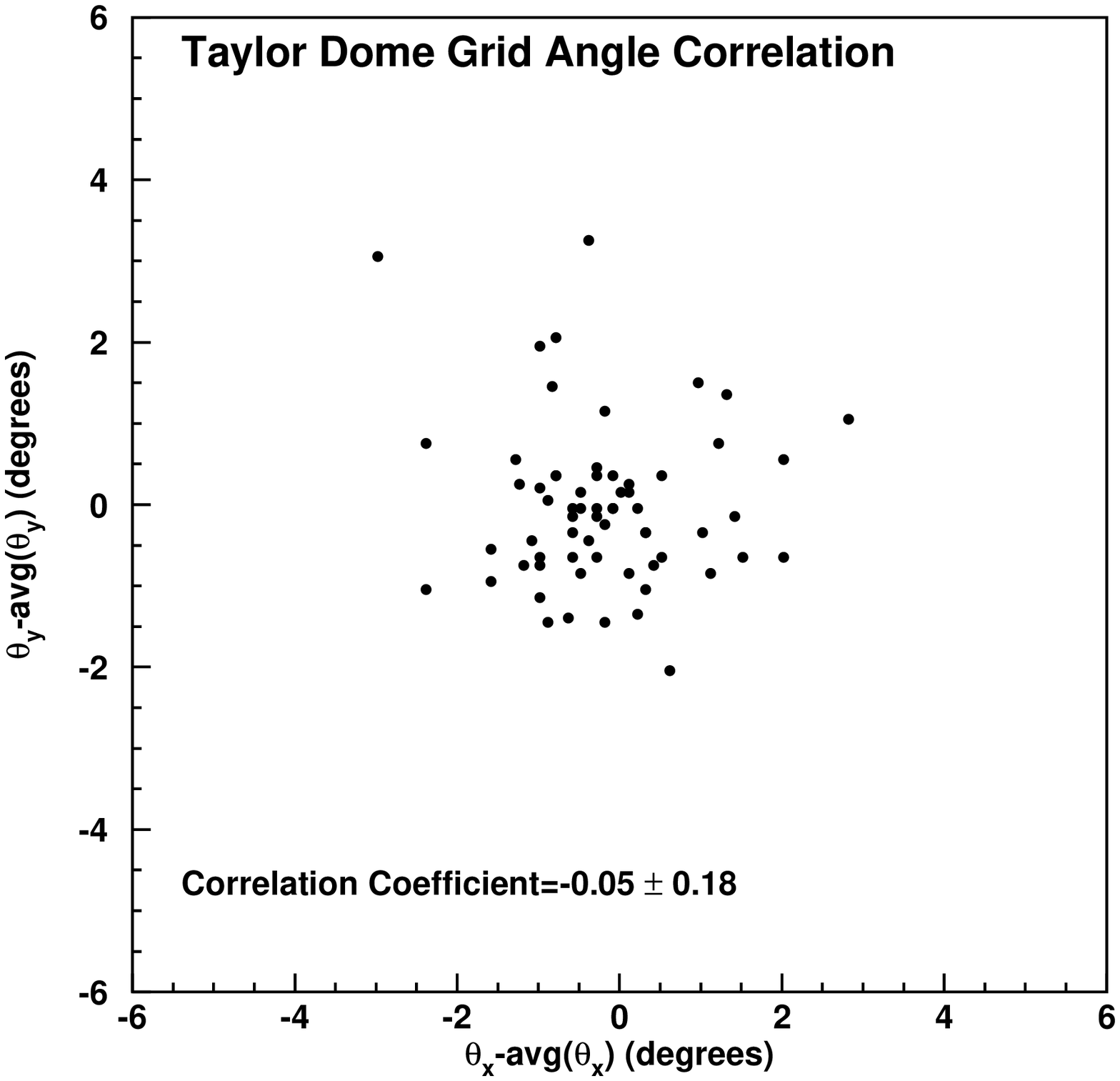,width=3.5in}
\caption{\label{twodimcorrel} Scatter plot of angles measured in a grid taken
with the 0.5~m scale at Taylor Dome.  
No significant correlation between orthogonal
directions was observed.}
\end{center}
\end{figure}

Members of the ANITA group made quantitative measurements 
of roughness features at one location on the Antarctic Plateau,
near Taylor Dome~\cite{romero}.
Hundreds of measurements of angles on a 0.5~m scale with 0.1 degree 
precision are
shown in Figure~\ref{traverseangles}.
The RMS distribution of angles on the 0.5~m
scale had an RMS of 1.7~degrees, consistent
with the traverse data above. 
The ordered angle measurements yield
an elevation map, shown in Figure~\ref{elevationmap}.
We were concerned with correlations between angles 
in orthoganal directions along a traverse.
As shown in Figure~\ref{twodimcorrel},
data taken in a 2-dimensional surface grid
show essentially no correlation of angles 
between measurements taken
in perpendicular directions.   
We made larger scale (100~m)  measurements
at a few points and angles of order 0.3~to 1.1~degrees
were measured, again consistent with the
traverse data. 
An overall slope of the area on the scale of 1~degree
was measured, consistent with the known glacier flow.

\subsection{Optical Modeling}

\hspace*{5ex}
The scale of roughness with horizontal scales of several wavelengths
and vertical scales of order a wavelength is a difficult region
to model mathematically since few electromagnetic approximations
are valid in this regime.   We emulated the situation experimentally and
scaled to the optical regime~\cite{griswold2}.  
The radio transmission was modeled
by a 632~nm laser and incoherent red light source.  The surface roughness
was modeled by a range of diffusers: 400-, 1000-, and 1500-grit diffusers.
We measured the surface 
features of the diffuser with a Veeco atomic force microscope and
found the characteristics of the 1500 grit
diffuser ($200-500$~nm height, $400-4000$~nm transverse feature size)
and 632nm light to closely emulate the
Antarctic surface for the decimeter radiation ANITA detects.

We affixed a half cylinder to the back of the diffuser so
that the incident light would always enter the glass perpendicularly
and not be refracted on its way into the glass. In this way,
all angles could be probed, especially beyond that of 
total-internal reflection if the surface were smooth.
The coherent and
incoherent light sources gave the same results.
We observed, as expected, that there is a wider distribution
of scattered light, but this was still relatively small compared
to the natural width of the Cherenkov cone from the Askarayan effect.
The total integrated power within a few degrees of the specular result 
stayed the same so radiation is not lost, only spread.  We also observed
significant radiation beyond the total internal reflection angle.
To our knowledge, 
this experiment was the first to observe the predicted~\cite{dogariu-boreman}
suppression of
Snell's Law relative to the specular case so that rays are less
bent than a $n=1.5$ to $n=1$ transition.   The effect increased with roughness,
as expected, and deviations as large as 20~degrees were observed.

\begin{figure}
\begin{center}
\epsfig{file=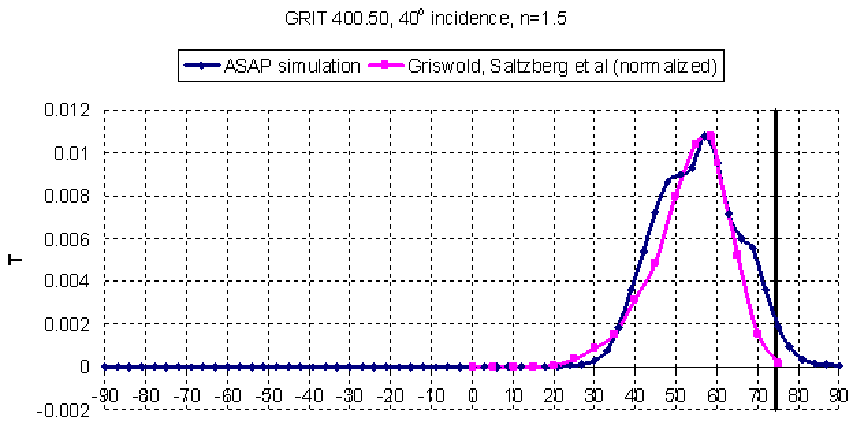,width=4.5in}
\caption{\label{asap_vs_data} A representative comparison of the
optical diffuser intensity data vs the ASAP simulation.  Pink is
data and blue is simulation.  Details
in the text.}
\end{center}
\end{figure}

The simulation employs a simple ``facet model''~\cite{dogariu-boreman} of 
roughness,  performing
ray tracing through small smooth faces on scales smaller than
a wavelength.  We saw in the optical case, this simple geometric model 
reproduces the data extremely well~\cite{dookayka}.
We input the measured atomic force microscope (AFM) surface data 
from the same diffusers we used into the
electromagnetics simulation code Advanced Simulation Analysis
Program (ASAP), developed by the Breault Corporation for simulation
of optical systems.  To test the facet model, we ran the code with
diffractive effects turned off.
All the angular distributions, including intensity and
the suppression effect of Snell's law could
were reproduced to high accuracy 
using the ASAP simulation.  A typical plot of data versus
simulation is shown in Figure~\ref{asap_vs_data}.
The vertical
bar shows the refraction angle expected by Snell's law, which is suppressed
for the rough surface.  Even using only geometric optics,
ASAP successfully reproduces the widths of the
intensity distributions and the observed Snell's Law suppression.
(The shoulders in the simulation are due to finite statistics of the
surface roughness model employed.)
The observed transmission beyond total-internal reflection (TIR) angle was
also measured and simulated.  For
example, the intensity at an angle beyond TIR (70 degrees)
was measured to be about 17\% of the peak transmission, while the simulation
yielded 20\%, demonstrating good agreement.
The success of the
facet model to reproduce the optical data
gives us confidence that the Monte Carlo simulations
which use the same facet model are essentially correct.

The micro-roughness scale is much less than one-tenth of a wavelength
and is ubiquitous.   The effect is difficult to quantify either
for Antarctica or for the ground glass diffusers.  However, by
their construction, sandblasting of glass,
the diffusers surely do have microscale roughness as well, but
the ASAP simulation described the data well without including 
the effect.  So the same is likely true for Antarctica and the decimeter
radiation.

In addition to the optical models we performed
a laboratory test in the radio-frequency domain as well.
During our run using an ice target at the SLAC linear 
accelerator~\cite{slac07}, we took several runs where we placed rows of ice 
20~cm high separated by 50~cm intervals across the surface of the 
ice target.  This was a deliberate overestimate of the antarctic roughness 
extracted from traverse and satellite data to look for some effect.  
We observed
an on-axis reduction of recorded voltage of 17\%.  Off-axis, the effect dropped
to 8\%.  We may infer the effect of roughness on ANITA is minimal,
hurting the threshold by less than of order 10\%.

\subsection{Time-domain concerns}

\hspace*{5ex}
The optics test used a continuous-wave
source rather than an impulse.  It is possible that the emissions
are spread out in time as well.  Since the pathlengths through the
rough surface, even with multipath, are an order of magnitude smaller
than our trigger integration time of 7~ns, we do not expect an effect.
To search for the effect in data, we used the 
in-ice disccone placed at 97~m depth at Taylor Dome during the ANITA
flight.  At a distance of 200~km, we reconstructed approximately 20 
events.  As shown in Figure~\ref{TDwfm} in
the main text, no time spreading is 
observed.  When deconvolved with the known instrument response, the signal is
consistent with a delta function indicating time domain spreading effects,
if there, are much smaller than the sample interval of 400~ps and do not effect
the neutrino detection.  For ANITA-II we plan to measure this effect right
down to the horizon, where the effect may be larger, but most of ANITA events
come from a distance already tested here.

\subsection{Implementation into Simulations}

\hspace*{5ex}
The UH ANITA simulation is based on multiple ray tracing and
could most directly incorporate both the deleterious effect of beam
spreading and advantageous effect of being able to see beyond the
total-internal reflection (TIR) angle.  The
net result is that the extended viewing beyond TIR dominates 
the event rate and boosts the
high energy neutrino acceptance by 50\%.  At lower energies events
are lost due to raising
the detection threshold.    Overall acceptance over a GZK spectrum is
affected by about 10\% so the corresponding uncertainty on the neutrino
flux limits, after including the effects of 
surface roughness to first order, are not significant.




\label{ThermalApp}
\section{Thermal-noise accidental rates for ANITA}
\subsection{The single-antenna trigger}

At the single antenna level (Level 1, or L1), we 
require three-of-eight of the dual-circular-polarization,
four-frequency-sub-bands to trigger in hardware. Each of these bands involves a
coincidence window with an effective time width of $T_1 = 19$~ns.  However,
if the trigger is to survive all software cuts, it must also be {\em coherent}
across all of the sub-bands, that is, it must be in phase, a much tighter requirement
than the hardware trigger. 

For thermal noise, the phases are completely 
uncorrelated, and the probability for thermal noise to produce a coherent impulse
in each sub-band which is also coherent across the bands is just given by the
ratio of the coherence time $\tau_{coh}$ of the full-band impulse to the total coincidence window,
multiplied by the number of independent trials for a coherent signal within the
coincidence window to produce the 3 of 8 trigger combination. 
The number of combinations
of $N$ subbands taken $M$ at a time gives a combinatoric factor of $C=N!/(M!(N-M!))$,
and for $N=8,M=3$ we have $C=56$. For any given combination of $M$ sub-bands that 
gives an L1 trigger, there are
$$\left ( \frac{T_1}{\tau_{coh}} \right )^M$$
combinations of sub-band signals that can trigger, but only 
those for which the relative locations in the sub-bands are identical (ignoring constant
offsets for sub-band differential delays) can yield a coherent
full-band pulse. There are thus ${T_1}/{\tau_{coh}}$ such coherent combinations.
The probability of obtaining one such combination is thus
\begin{equation}
P_1 =  C \left ( \frac{\tau_{coh}}{T_1} \right )^{M-1}~.
\end{equation} ~ 

The coherence time  $\tau_{coh}$ is estimated by requiring that the signal
be phase-coherent across the full-band, and thus that the sub-bands have less than
a $\lambda/8$ phase offset between them for a wavelength at the mean frequency $\bar{f}$
of the band. For $\bar{f}\simeq 600$~MHz as for ANITA, the delay corresponding to a
1/8-wave phase difference is $\tau_{coh}\simeq 200$~ps, 
which gives $P_1 = 6.2 \times 10^{-3}$.

\subsection{The level 2 and 3 trigger: azimuth+elevation constraint}

For the L2 trigger, adjacent antennas must each have an L1 trigger with  5~ns
of overlap for a 30~ns L1 one-shot output, giving an effective coincidence window of
about $T_2 \simeq 40$~ns. The number of possible combinations of pairs of
L1 impulse arrival times that can produce L2 triggers is $T_2/\tau_{az}$ where
$\tau_{az}$ is the azimuthal time resolution for pair of antennas within an
azimuthal ring.
Similar analysis applies to the elevation constraint via the L3 trigger,
which couples the upper and lower rings. The coincidence window for L3 triggers
is the convolution of the L2 windows, and gives  $T_3 \simeq 50$~ns.

However, the analysis requires an additional azimuth constraint in the
other ring, combined with a pair of elevation constraints for adjacent phi-sector
pairs. To estimate the fraction of such combinations of 4-antenna triggers that
add coherently to give a proper direction reconstruction, we
take the ratio of the geometric mean of the elevation and azimuth time resolutions
to the total number of combinations of random impulses that can form in 
the 4 $T_2$ coincidence windows required for an L3. This represents the probability 
that a single direction (the normal incidence plane wave) can be mimicked by
thermal noise:
$$ p_2 = \left ( \frac{\sqrt{\tau_{az}\tau_{el}}}{T_2}\right )^4$$

Multiplying this by the number of independent directions within the
antenna beam, and by the number of phi sectors $N_\phi$
then gives the total probability at the L3 level. 

For a
beam with 
$\Omega = \pi \Delta \theta \Delta \phi \simeq 
\pi(0.3^{\circ})(1.0^{\circ})=0.94$ square degrees, and an
antenna main beam solid angle of $\Omega_{ant} \simeq \pi \Phi^2/4$ for
a half-power beam width $\Phi\simeq 40^{\circ}$, we finally get
\begin{equation}
P_2 = N_{\phi} K\frac{\Omega_{ant}}{\Omega} ~ p_2~
\end{equation}
where $K$ is the fraction of the main antenna beam that is allowed (eg, the
fraction below the horizon). 
For ANITA, the geometric mean of the timing errors is about 60~ps,
and thus $P_2 \simeq 5.4 \times 10^{-8}$, for $K=1$.

\subsection{Combined probability}

Combining the L1 probability and the L2 probability since they are independent for 
thermal noise, the total probability for a thermal noise trigger to produce an
acceptable direction reconstruction is
\begin{equation}
P_{tot} = P_1~P_2 = (6.2 \times 10^{-3}) \times (5.4 \times 10^{-8}) = 3.4 \times 10^{-10}
\end{equation}
or about $0.003$ events expected during ANITA's flight, based on 8.2M triggers.

\end{document}